\documentclass[fleqn,usenatbib,onecolumn]{mnras}

\usepackage{amsmath}	
\usepackage{amssymb}	

\usepackage{newtxtext,newtxmath}
\usepackage[T1]{fontenc}
\usepackage{ae,aecompl}

\usepackage{graphicx}	

\usepackage{color}
\usepackage{ulem}
\usepackage{breqn}
\allowdisplaybreaks

\usepackage[caption=false]{subfig}

\makeatletter
\def\@threej(#1,#2,#3,#4,#5,#6){\Bigl(\begin{smallmatrix}#1&#3&#5\\#2&#4&#6\end{smallmatrix}\Bigr)}
\newcommand{\threej}[3]{\@threej(#1,#2,#3)}
\makeatother

\usepackage{placeins}

\title[Spin characterisation of systematics]{Spin characterisation of systematics in CMB surveys - a comprehensive formalism}

\author[]{
Nialh McCallum,$^{1}$\thanks{E-mail: nialh.mccallum@postgrad.manchester.ac.uk}
Daniel B. Thomas,$^{1}$
Michael L. Brown,$^{1}$
and Nicolas Tessore$^{1}$
\\
$^{1}$Jodrell Bank Centre for Astrophysics, School of Physics \& Astronomy, The University of Manchester, Manchester M13 9PL, UK 
}

\date{Accepted XXX. Received YYY; in original form ZZZ}

\pubyear{2020}

\begin{document}
\label{firstpage}
\pagerange{\pageref{firstpage}--\pageref{lastpage}}
\maketitle

\begin{abstract}
The CMB $B$-mode polarisation signal --- both the primordial gravitational wave signature and the signal sourced by lensing --- is subject to many contaminants from systematic effects. Of particular concern are systematics that result in mixing of signals of different ``spin'',  particularly leakage from the much larger spin-0 intensity signal to the spin-2 polarisation signal. We present a general formalism, which can be applied to arbitrary focal plane setups, that characterises signals in terms of their spin.  We provide general expressions to describe how spin-coupled signals observed by the detectors manifest at map-level, in the harmonic domain, and in the power spectra, focusing on the polarisation spectra --- the signals of interest for upcoming CMB surveys. We demonstrate the presence of a previously unidentified cross-term between the systematic and the intrinsic sky signal in the power spectrum, which in some cases can be the dominant source of contamination. The formalism is not restricted to intensity to polarisation leakage but provides a complete elucidation of all leakage including polarisation mixing, and applies to both full and partial (masked) sky surveys, thus covering space-based, balloon-borne, and ground-based experiments. Using a pair-differenced setup, we demonstrate the formalism by using it to completely characterise the effects of differential gain and pointing systematics, incorporating both intensity leakage and polarisation mixing. We validate our results with full time ordered data simulations. Finally, we show in an Appendix that an extension of simple binning map-making to include additional spin information is capable of removing spin-coupled systematics during the map-making process.
\end{abstract}




\section{Introduction}
Observations of the cosmic microwave background (CMB) anisotropies provide one of the most powerful constraints on the standard cosmological model \citep[for recent reviews see e.g.][]{durrer2015, staggs2018}. One of the key science goals of future CMB experiments (e.g. LiteBIRD, Simons Observatory (SO), \citealt{2018JLTP..193.1048S,2019JCAP...02..056A}) is the observation of the large scale $B$-mode polarisation signal related to primordial gravitational waves. $B$-modes can also arise on smaller scales via gravitational lensing of the higher amplitude $E$-mode signal, which has been observed by a number of experiments (e.g. Polarbear, \citealt{2020ApJ...893...85F}).

The primordial $B$-mode signal is usually characterised by the tensor-to-scalar ratio $r$ . The objective of upcoming Stage IV experiments is to constrain the tensor-to-scalar ratio to an accuracy of $\Delta r \approx 0.001$ \citep{2016arXiv161002743A}. There are a number of difficulties in achieving this goal including removal of foreground contamination, delensing, and instrumental systematics. An unprecedented characterisation of systematics will be required for upcoming surveys in order that any spurious $B$-mode production may be disentangled from the true signal.

CMB instruments typically measure the incoming radiation as a mixture of intensity ($I$) and the two linear polarisations (Stokes parameters $Q$ and $U$). This mixing means that some systematics leak the (much larger) intensity and $E$-mode signals into spurious $B$-mode polarisation, and also that measurements at multiple crossing angles in each pixel are required in order to separate the signals.

Satellite experiments can easily incorporate precession and rotation into their scan strategies, and thus obtain good crossing angle coverage. For ground based surveys, the instrument has much less freedom and sky rotation is relied on to access different observing angles, resulting in limited availability of crossing angles in each patch of the sky. Additional crossing angle coverage can be gained by rotating the boresight of the instrument; this is of particular importance for South Pole based telescopes where sky rotation does not provide additional crossing angles \citep[e.g.][]{2014ApJ...792...62B}.

To aid with separation of the Stokes parameters, surveys may also use techniques such as pair differencing of co-located detector pairs with orthogonal polarisation sensitivity directions, and continuously rotating Half-wave Plates (CHWPs), which can also aid in mitigating systematics \citep[e.g.][]{brown2009, 2018SPIE10708E..48S}. A number of ``spin-coupled'' systematics may manifest in these scenarios, where by spin-coupled we mean systematics that have a spin-dependence and as such are usually reducible by scanning each pixel at multiple crossing angles.\footnote{The spin dependence we refer to is the dependence of the fields with crossing angle $\psi$ such that $e^{in\psi}$ describes a spin-$n$ dependence. Also note that if the systematic has the same spin as the signal of interest, it is irreducible by the scan strategy unless other means are used (such as a CHWP).} For instance, in the pair differencing case, any differences in the fabrication or placement of the detectors in a pair may lead to leakage of the intensity and $E$-mode polarisation signals into the much smaller $B$-mode signal. As shown by several formalisms \citep[e.g.][]{2003PhRvD..67d3004H, 2007MNRAS.376.1767O, PhysRevD.77.083003, 0806.3096, Wallisetal2016}, this leakage appears with different spin dependence in different scenarios, e.g. in the presence of gain, pointing or ellipticity differences between the detectors.

The spin properties of the systematic signals mean that most of them vanish for an ideal scan \citep{2009arXiv0906.1188B}. Additionally, even in the absence of an ideal scan, the spin symmetries may be exploited to remove certain systematics by taking measurements at two particular crossing angles with a specific rotation between them. This was noted before \citep[in e.g.][]{2007MNRAS.376.1767O,PhysRevD.77.083003,0806.3096,2015ApJ...814..110B,2020MNRAS.491.1960T}. For example, differential pointing and gain couple the temperature to polarisation through spin-(1 and 3) and spin-2 scan terms, so pairs of observations within a pixel separated by $180^\circ$ and $90^\circ$ respectively will zero these systematics. This will work for any number of observations in a pixel, as long as they each come in pairs separated by the appropriate angle. It is generally desirable to maximise your crossing angle coverage when designing scan strategies. However, the advantage of these specific angle combinations is that they can have a large impact on certain systematics, even in the cases where there are other limitations on the scan strategy and thus on the crossing angle coverage. More generally, the spin properties of the systematics can be used to design optimal scan strategies such that the spin-coupled systematics are minimised \citep{Wallisetal2016}.

The previous approaches of \cite{2003PhRvD..67d3004H, 2007MNRAS.376.1767O, PhysRevD.77.083003, 0806.3096} and \cite{Wallisetal2016} limited their curved-sky approach to the case of the full sky and used the flat-sky approximation for smaller patches. They additionally focused primarily on pair differenced signals. In this work we extend the full sky approach of \cite{Wallisetal2016} to apply to partial sky experiments (such as ground-based experiments or masked satellite experiments), including mode mixing effects. We note that the reasoning we employ to do this could equally be applied to the other previous works on this topic. We further extend the formalism to general spin-coupled signals seen by a detector, for an arbitrary focal plane setup (such as using a single detector or a pair differencing setup). We include the effects of polarisation mixing in addition to the temperature to polarisation leakage, and we also demonstrate the presence of cross terms between the systematic and intrinsic sky signal at the power spectrum level, which we show can be the dominant contaminant in some cases.

The impact of systematics on the measurement of the polarisation of the CMB has been and continues to be an important consideration for CMB surveys. As such there has been a wide breadth of studies of various systematic effects by many experiments, including satellite experiments such as WMAP \cite[e.g.][]{2011ApJS..192...14J}, {\it Planck} \cite[e.g.][]{2014A&A...571A...7P,2016A&A...594A...7P}, EPIC \cite[e.g.][]{2009arXiv0906.1188B}, and CORE \cite[e.g.][]{2018JCAP...04..022N} and ground-based such as BICEP2 \cite[e.g.][]{2015ApJ...814..110B}, and SO \cite[e.g.][]{2018SPIE10708E..3ZC}. Consequently a number of tools have been developed to deal with systematic effects, including those which are coupled to the scanning strategy. The formalism we present here is akin to that of \cite{2017A&A...598A..25H} which has already considered the inclusion of a number of the effects mentioned for arbitrary scanning strategies. We note here that complimentary to the general weighted power spectrum approach of \cite{2017A&A...598A..25H}, the derivations presented here in section \ref{section:Power Spectrum Derivation} specific to the polarisation spectra, mean we are able to further elucidate the presence of a cross term between the systematic and intrinsic sky signal at the power spectrum level, which we then examine for the specific cases of differential gain and pointing showing it can be the dominant source of systematic.

This paper is organised as follows. In Section \ref{section:Spin Characterisation} we present a general form for incorporating spin-coupled systematics with arbitrary spin at map-level and $a_{\ell m}$ level. In Section \ref{section:Power Spectrum Derivation} we use this formalism to describe the effects of spin-coupled systematics on the polarisation power spectra. In Section \ref{section:Demonstration} we demonstrate the formalism by applying it to the case of differential pointing and gain systematics, using it to describe both intensity to polarisation ($I \rightarrow P$) leakage systematics and polarisation to polarisation ($P \rightarrow P$) mixing for both full and partial sky experiments. In Section \ref{section:Conclusion} we summarise our results.

\section{Spin Characterisation Of Scan Strategies}
\label{section:Spin Characterisation}
During a CMB experiment, each pixel of a map will usually be observed at a variety of crossing angles. If we consider how this signal is coupled with the scan we may write the detected signal within a pixel as a function of the orientation as
\begin{equation}
    S^d(\psi,\Omega) = h(\psi,\Omega) S(\psi,\Omega) \text{,}
    \label{eq:SDTot}
\end{equation}
where $\Omega = (\theta,\phi)$ are the latitude and longitude coordinates of each pixel on the sky, and $\psi$ is the orientation angle of the scan direction of the instrument with respect to North. For some arbitrary scanning strategy we may describe the scan by the real space field
\begin{equation}
h(\psi, \Omega)=\frac{2\pi}{N_{\text{hits}}(\Omega)}\sum_j \delta(\psi-\psi_j(\Omega))\text{,}
\end{equation}
where $\psi_j$ is the $j$th element of a discrete set of crossing angles for the sky pixel in question, and $N_{\text{hits}}(\Omega)$ is the total number of observations of that pixel \footnote{The Dirac delta functions here should not be taken as saying that the signal is infinite at any point, it is just a construction that allows us to represent the discrete observations as a continuous signal in a simple way, that can be easily treated by continuous Fourier analysis. As shown in equations \ref{eq:dtosdictionary}-\ref{eq:dtosdictionary2}, the Fourier transform of the signal written in this way corresponds to the terms in standard map making that occur from averaging the discrete signals.}. By noting that the real space variable $\psi$ is periodic with domain $0-2\pi$, meaning that this is equivalent to a Fourier series with discrete spins, we may decompose the total signal detected within a pixel into the sum of the components of different spins as
\begin{align}
    &S^{d}(\psi, \Omega) = \sum_{k} {}_k\tilde{S}^d(\Omega) e^{-i k \psi}\nonumber
    \\
    &{}_{k}\tilde{S}^{d}(\Omega) = \sum_{k'=-\infty}^{\infty}\tilde{h}_{k-k'}(\Omega){}_{k'}\tilde{S}(\Omega) \text{,}
    \label{eq:spinsyst}
\end{align}
where the second line comes from using equation \ref{eq:SDTot} and the convolution theorem. The symbols topped with a tilde denote the Fourier space quantities. The case $k=2$ relates to the spin-2 polarisation and $k=0$ to the spin-0 temperature signal. More specifically, the spin-2 case of equation \ref{eq:spinsyst} relates to terms encountered in simple-binning polarisation map-making as
\begin{align}
&\langle d_j \cos(2\psi)\rangle ={\Re{}}\left({}_2\tilde{S}^d\right)\nonumber
\\
&\langle d_j \sin(2\psi)\rangle={\Im{}}\left({}_2\tilde{S}^d\right) \text{,}\nonumber 
\end{align} where $d_j$ corresponds to a single data point from the timestream being considered, which has some associated crossing angle $\psi$. We give further details on how equation \ref{eq:spinsyst} relates to the terms used in simple binning map-making algorithms in Appendix~\ref{section:Map-Making}.  The orientation function, $\tilde{h}_{k-k'}(\Omega)$, may be constructed from the scan by averaging within a pixel as
\begin{equation}
    \tilde{h}_{k-k'}(\Omega) = \frac{1}{2\pi}\int d\psi e^{i(k-k')\psi} \frac{2\pi}{N_{\text{hits}}(\Omega)}\sum_j \delta(\psi-\psi_j) = \frac{1}{N_{\text{hits}}(\Omega)} \sum_{j} e^{i(k-k')\psi_j(\Omega)} \text{,}
    \label{eq:orientation function}
\end{equation}
where $\tilde{h}_0$ describes the window function of the survey, while the other terms ($\tilde{h}_1, \tilde{h}_2$ etc.) can be used to characterise the ability of a scan strategy to mitigate certain classes of systematic effects \citep{2009arXiv0906.1188B}.

The $\tilde{h}_{k-k'}$ term couples signals of spin $k'$ to the spin-$k$ quantity of interest or, in the case where $k=k'$, acts as the window function of the survey. In principle the value of $k'$ can go to infinity. However, by examining the spin of the signals being considered, the important $k'$ values for a given experiment may be isolated. One can construct these $\tilde{h}_{k-k'}(\Omega)$ fields with prior knowledge of the scan strategy, and this can be used to develop a fast map-based approach for simulating certain spin-coupled systematics. This will be demonstrated in a forthcoming paper. Here, we concentrate on showing how to use this formalism to analytically model the effects of systematics on the CMB power spectra, in both full and partial sky experiments.

\subsection{Single detector example}
To aid intuition, we present an example here for measurements from a single detector. The detector equation for a CMB experiment corresponding to a single data point from a single detector may be written as
\begin{equation}
S^{\text{SD}}(\psi,\Omega)=I(\Omega)+Q(\Omega)\cos(2\psi)-U(\Omega)\sin(2\psi)+ \sum_{n\geq0}(Z_n^Q(\Omega)\cos(n\psi)-Z_n^U(\Omega)\sin(n\psi))\text{,}
\label{eq:Detector Equation}
\end{equation}
where $I, Q$, and $U$ are the Stokes parameters describing the intensity and polarisation of the on-sky signal (containing CMB and foregrounds). The superscript ${\text{SD}}$ denotes that this corresponds to $S(\psi,\Omega)$ of equation \ref{eq:SDTot} for the case where the timestream considered is from a single detector. The $Z_n^Q$ and $Z_n^U$ are additional spin-$n$ signals that may be attributed to systematics (or, potentially, some additional, unaccounted-for, on-sky signal) where the superscripts $Q$ and $U$ are in analogy to the complex polarisation signal. In the absence of systematics this would reduce to the first three terms of the standard detector equation where $I, Q$ and $U$ are the on-sky fields including CMB and foregrounds. In the presence of a Gaussian symmetric beam (which we apply in this work), these may also be taken to be the beam smoothed quantities. In the case of more complicated instrument beams, further analysis would be required in order to consider effects such as beam ellipticity, though we note that it is possible to write this as part of the systematic term \citep[e.g.][]{2014MNRAS.442.1963W, Wallisetal2016}.

Note that, to be consistent with the previous work of \cite{Wallisetal2016}, we adopt the polarisation convention with a positive $Q$ (and zero $U$) along the North-South axis, and a positive $U$ (and zero $Q$) along the North-West to South-East axis as in \cite{1999astro.ph..5275G}. This is why the polarisation signal $P(\Omega)=Q(\Omega)+iU(\Omega)$ contributes a negative $-U(\Omega)\sin(2\psi)$ term in the detector equation such that the spin-2 signal of interest is $P(\Omega)e^{i2\psi}$ (see equation 14 of \citealt{Wallisetal2016}). However the results of this work are generally applicable and can be trivially rewritten to adhere to other conventions.

We may rewrite equation \ref{eq:Detector Equation} as
\begin{equation}
    S^{\text{SD}}(\psi,\Omega) = I(\Omega) + \frac{1}{2}(P(\Omega)e^{2i\psi} + P^{*}(\Omega)e^{-2i\psi}) + \sum_{n\geq0}\frac{1}{2}(Z_n(\Omega)e^{ni\psi} + Z_n^{*}(\Omega)e^{-ni\psi}) \text{,}
    \label{eq:sky signal}
\end{equation}
where $Z_n(\Omega) = Z_n^Q(\Omega) + iZ_n^U(\Omega)$. We then choose to rewrite the signal according to equation \ref{eq:spinsyst}, which comes from applying the $h(\psi,\Omega)$ field to equation \ref{eq:sky signal} and transforming to Fourier space. As a result, we may write the total signal detected in this single detector setup as
\begin{equation}
    {}_{k}\tilde{S}^{d,\text{SD}}(\Omega) =
    \tilde{h}_{k-0}(\Omega)I(\Omega) + \frac{1}{2}(\tilde{h}_{k-2}(\Omega)P(\Omega) + \tilde{h}_{k+2}(\Omega)P^*(\Omega))
    + \sum_{n\geq0}\frac{1}{2}(\tilde{h}_{k-n}Z_n(\Omega) + \tilde{h}_{k+n}Z_n^{*}(\Omega)) \text{,}
    \label{eq:Single Detector Signal}
\end{equation}
where $I(\Omega)$ contributes to ${}_0\tilde{S}(\Omega)$, $P(\Omega)$ to ${}_2\tilde{S}(\Omega)$, $P^*(\Omega)$ to ${}_{-2}\tilde{S}(\Omega)$, $Z_n(\Omega)$ to ${}_{n}\tilde{S}(\Omega)$, and $Z_n^*(\Omega)$ to ${}_{-n}\tilde{S}(\Omega)$. The systematic terms have been written as general spin-$n$ quantities and could in principle also contribute to the spin-0 and spin $\pm2$ quantities.

This shows simply what signal will be present in the maps if "spin-$k$ only" map-making is applied to a timestream arising from one detector. If other spin terms are included in the map-making then this will separate some of the components from each other as is discussed further in Appendix \ref{section:Map-Making}. Note that equation \ref{eq:Single Detector Signal} would change depending on how the timestream is treated. For example in the case of a pair-differencing experiment, the timestream will be contributed to by both detectors in a pair simultaneously. We shall provide an example of how to apply equation \ref{eq:spinsyst} to a different focal plane setup in Section \ref{section:Demonstration}, but we note that, provided the signal can be written using equation \ref{eq:spinsyst}, then the formalism preesnted here may be applied to it.

\subsection{Deriving Systematics in \texorpdfstring{$\mathbf{a_{\ell m}}$}{TEXT} Space}
\label{section:AlmDerivation}
We restate equation \ref{eq:spinsyst}, which describes the spin-$k$ map-level signal, as the key starting point for the formalism,
\begin{equation}
    {}_{k}\tilde{S}^{d}(\Omega) = \sum_{k'=-\infty}^{\infty}\tilde{h}_{k-k'}(\Omega){}_{k'}\tilde{S}(\Omega) \text{.}
    \label{eq:spinsyst_copy}
\end{equation}
We have shown an example of how this manifests for the single detector timestream. However, the following derivation is general and applies to arbitrary setups as long as they can be decomposed in this form, as the sum of some $\tilde{h}_{k-k'}(\Omega)$ fields that each multiply some ``signal'' that is a function of $\Omega$. As such equations \ref{eq:spinsyst} and \ref{eq:spinsyst_copy} cover different focal plane setups, as we shall demonstrate in Section \ref{section:Demonstration}.

For the spin of interest $k$ we may write the total observed signal as
\begin{equation}
    {}_k\tilde{S}^d_{\ell m} = \int d\Omega {}_kY_{\ell m}^*(\Omega) \sum_{k'} \tilde{h}_{k-k'}(\Omega) \, {}_{k'}\tilde{S}(\Omega) \text{,}
\end{equation}
where the spin of the detected components present in the total signal within a pixel ${}_{k}\tilde{S}^{d}(\Omega)$ of equation \ref{eq:spinsyst} dictate which $k'$ to sum over. In the case of a CMB experiment these components could in principle include the true CMB signal, foregrounds, and systematics of various spin. Expanding each of the fields in harmonic space and using the relationship that ${}_kY^*_{\ell m}(\Omega) = (-1)^{m+k}{}_{-k}Y_{\ell -m}(\Omega)$ we may rewrite the total observed signal as
\begin{align}
\label{eq:Observed Signal}
    {}_k\tilde{S}^d_{\ell m}
    &= \int d\Omega (-1)^{m+k}{}_{-k}Y_{\ell -m}(\Omega) \sum_{\substack{k'\\l_1m_1\\l_2m_2}} {}_{(k-k')}h_{l_{1}m_{1}}{}_{(k-k')}Y_{l_1m_1}(\Omega) {}_{k'}S_{l_{2}m_{2}}{}_{k'}Y_{l_2m_2}(\Omega) \nonumber
    \\
    &= (-1)^{m+k} \, \sum_{\substack{k'\\l_1m_1\\l_2m_2}} {}_{(k-k')}h_{l_1m_1} \, {}_{k'}S_{l_2m_2} \, \sqrt{\tfrac{(2\ell + 1)(2l_1 + 1)(2l_2 + 1)}{4\pi}}
    \threej{\ell,-m}{l_1,m_1}{l_2,m_2} \, \threej{\ell,k}{l_1,-(k-k')}{l_2,-k'} \text{,}
\end{align}
where we have used identities from \cite{Varshalovich:1988ye}, which makes use of the spin-weighted spherical harmonic convolution theorem. We note that this agrees with the work of \cite{2017A&A...598A..25H} and is comparable to equation A.14 of their formalism. For this work, we choose to use a uniform weighting of the map in the harmonic analysis. (In principle it would be straightforward to apply other weightings at this point, e.g. the hitmap could be encoded with an appropriate normalisation.) We have expanded the orientation function fields as
\begin{equation}
    \tilde{h}_{k-k'}(\Omega) = \sum_{\ell m} {}_{(k-k')}h_{\ell m} \, {}_{(k-k')}Y_{\ell m} \text{,}
\end{equation}
where the modes are given by
\begin{equation}
    {}_{(k-k')}h_{\ell m}
    = \int d\Omega {}_{(k-k')}Y_{\ell m}^*(\Omega) \tilde{h}_{(k-k')}(\Omega) = \int d\Omega {}_{(k-k')}Y_{\ell m}^*(\Omega) \frac{1}{N_{\text{hits}}(\Omega)} \sum_{j} e^{i(k-k')\psi_j(\Omega)}\text{.}
    \label{eq:orientation function modes}
\end{equation}
The ${}_0h_{\ell m}$ field describes the window function of the survey, while the other spins can be used when determining particular systematic effects. One subtlety here is that if a pixel is unobserved (or masked), then the orientation function (equation \ref{eq:orientation function}) may be given the value zero. Therefore, the mask and scan strategy are both incorporated into ${}_{(k-k')}h_{\ell m}$ and they are dealt with simultaneously in this formalism. This means that the spherical harmonic terms that we derive later (equations \ref{eq:True Polarisation} and \ref{eq:Spin2 General Systematic alm}) are ``pseudo-$a_{\ell m}$'' terms: systematic contributions modulated by the ${}_{(k-k')}h_{\ell m}$ fields naturally contaminate the pseudo-$a_{\ell m}$ polarisation signal. This process to convert to partial sky should be equally applicable to the full-sky literature on spin characterised systematics such as \cite{PhysRevD.77.083003, 0806.3096, Wallisetal2016} among others.

In principle one could use these equations to consider an observed signal of any spin. It is worth noting that it is thus possible to model spin-coupled effects on the spin-0 field with this formalism, thus it could in principle also be applied to intensity mapping experiments (see Appendix \ref{section:Additional Spectra} for further detail). 

We can write the total observed polarisation signal as a combination of the true sky signal (CMB and foregrounds) ${}_2\tilde{P}_{\ell m}$ (this will usually be a cut-sky signal) and the combined signal from all other sources ${}_2\tilde{\Delta}_{\ell m}$ as
\begin{equation}
    {}_2\tilde{S}^d_{\ell m} = {}_2\hat{P}_{\ell m}
    = {}_2\tilde{P}_{\ell m}
    + {}_2\tilde{\Delta}_{\ell m} \text{,}
    \label{eq: Polarisation Signal}
\end{equation}
where this corresponds to setting $k=2$ in equation \ref{eq:spinsyst}: we are considering the spin-2 signal ${}_2\tilde{S}^d(\Omega)$, which is made up of the true spin-2 sky signal ${}_2\tilde{P}_{\ell m}$ and a spurious term ${}_2\tilde{\Delta}_{\ell m}$. This latter couples signals of arbitrary spin-$k'$ ${}_{k'}\tilde{S}(\Omega)$ to the spin-2 signal of interest through scan coupling terms ${}_{(2-k')}h_{l_1m_1}$. More explicitly, ${}_2\tilde{P}_{\ell m}$ corresponds to the spin-2 map signal $\tilde{h}_0(\Omega) P(\Omega)$, which is the true on-sky polarisation signal multiplied by the survey mask, while the ${}_2\tilde{\Delta}_{\ell m}$ corresponds to the additional spurious terms present in the ${}_2\tilde{S}^d(\Omega)$ signal.

Using equation \ref{eq:Observed Signal}, and isolating the $k'=2$ terms that are only contributed to by the true on-sky spin-2 polarisation signal, we may write the polarisation signal as
\begin{equation}
    {}^{}_2\tilde{P}_{\ell m}
    = \sum_{\substack{l_1m_1\\l_2m_2}} {}_0h_{l_1m_1} \, {}_2P_{l_2m_2} \, (-1)^m \, \sqrt{\tfrac{(2\ell + 1)(2l_1 + 1)(2l_2 + 1)}{4\pi}}
    \threej{\ell,-m}{l_1,m_1}{l_2,m_2} \, \threej{\ell,2}{l_1,0}{l_2,-2} \text{,}
    \label{eq:True Polarisation}
\end{equation}
where ${}_2P_{l_2m_2}$ are the true spin-2 polarisation modes on the full sky. 

The terms from the other sources, such as systematics, may be contributed to by any spin $k'$. These terms contributing to the spin-2 polarisation may be written using equation \ref{eq:Observed Signal} as
\begin{equation}
\label{eq:Spin2 General Systematic alm}
    {}_2\tilde{\Delta}_{\ell m} = (-1)^{m} \, \sum_{\substack{k'\\l_1m_1\\l_2m_2}} {}_{(2-k')}h_{l_1m_1} \, {}_{k'}S_{l_2m_2} \, \sqrt{\tfrac{(2\ell + 1)(2l_1 + 1)(2l_2 + 1)}{4\pi}}
    \threej{\ell,-m}{l_1,m_1}{l_2,m_2} \, \threej{\ell,2}{l_1,-(2-k')}{l_2,-k'}  \text{.}
\end{equation}
The $k'$ which are present will depend on the additional signals contributing at non-negligible levels in a given experiment. In principle the ${}_{k'}S_{l_2m_2}$ could be contributed to by any additional spin-coupled signal. It is likely they would be contributed to by a combination of sources such as the CMB fields which are coupled due to some systematic, or systematic leaked foreground signals \citep[e.g. foreground leakage due to bandpass mismatch,][]{2017JCAP...12..015T,2020MNRAS.491.1960T}. It is worth noting that the spin-0 true temperature signal would also contribute.  However it is standard to remove it with methods such as separation in the map making process or by differencing detector pairs. Imperfections in this separation may contribute a systematic term and these will be explored later.

In summary we have written the general map-space equation \ref{eq:spinsyst} to describe a total observed signal of arbitrary spin-$k$, allowing for contributions from any spin-$k'$ signals through $\tilde{h}_{(k-k')}$ coupling terms. We have propagated this to $a_{\ell m}$-space and written the general equation \ref{eq:Observed Signal} to describe a total observed signal of arbitrary spin-$k$, allowing for contributions from any spin-$k'$ signals through ${}_{(k-k')}h_{l_1m_1}$ coupling terms. We have then applied this to describe the observed spin-2 polarisation signal ${}_2\hat{P}_{\ell m}$ and split this into contributions from the true spin-2 sky signal ${}_2\tilde{P}_{\ell m}$ and all other additional signals ${}_2\tilde{\Delta}_{\ell m}$ which can take arbitrary spin-$k'$ but are coupled to the spin-2 signal by ${}_{(2-k')}h_{l_1m_1}$ coupling terms. Furthermore the presence of the ${}_{(2-k')}h_{l_1m_1}$ coupling terms means the effects of partial sky are included in both equations \ref{eq:True Polarisation} and \ref{eq:Spin2 General Systematic alm}.

\section{Propagating Systematics to Pseudo Power Spectra}
\label{section:Power Spectrum Derivation}
In this section we show the effect that an additive systematic, of the form derived in the previous section, will have on the polarisation pseudo power spectra (see Appendix~\ref{section:Additional Spectra} for the effect on the temperature and cross spectra). The resulting estimated $E-$ and $B-$mode pseudo power spectra may be written in terms of the intrinsic power spectra and systematic bias terms. This derivation is independent of the form of the systematic and experimental set up. We reiterate that in the presence of a Gaussian symmetric beam these equations will also be applicable to the beam convolved quantities simply by scaling by the appropriate $B_{\ell}^2$ factor  (we include this when comparing to our simulations in Section \ref{section:Demonstration}, which feature a Gaussian beam).

Before proceeding, we note that in order to quantify the effect the systematic would have on the reconstructed "full-sky" power spectra (rather than on the pseudo spectra), one would have to follow through the reconstruction methods of e.g. \cite{BrownCastroTaylor}, but include the spurious signal. However, it would be advisable to attempt to remove any systematic  contamination in some way before the reconstruction (for example foreground removal is regularly done in map space,  e.g. \citealt{2018JCAP...04..023R}) but we don't consider this issue further in this work.

It is standard practise to decompose the spin-2 polarisation signal into the gradient containing $E$-mode and curl containing $B$-mode contributions. Following the convention in \cite{durrer_2008} we write the decomposition as
\begin{align}
    &\tilde{B}_{\ell m} = \frac{-i}{2}({}^{}_2\tilde{P}_{\ell m} -{}^{}_{-2}\tilde{P}_{\ell m})
    \\
    &\tilde{E}_{\ell m} = \frac{1}{2}({}^{}_2\tilde{P}_{\ell m} +{}^{}_{-2}\tilde{P}_{\ell m}) \text{,}\nonumber
\end{align}
for the true sky signal, and
\begin{align}
    &\tilde{\Delta}^{B}_{\ell m} = \frac{-i}{2}({}^{}_2\tilde{\Delta}_{\ell m} -{}^{}_{-2}\tilde{\Delta}_{\ell m})
    \\
    &\tilde{\Delta}^{E}_{\ell m} = \frac{1}{2}({}^{}_2\tilde{\Delta}_{\ell m} +{}^{}_{-2}\tilde{\Delta}_{\ell m}) \text{,}\nonumber
\end{align}
for the systematic signal. Applying these to equations \ref{eq:True Polarisation} and \ref{eq:Spin2 General Systematic alm}, we can express the effect of the systematics directly on the total polarisation pseudo-$a_{\ell m}$s as
\begin{equation}
    \hat{X}_{\ell m} = \tilde{X}_{\ell m}+\tilde{\Delta}^{X}_{\ell m} \text{,}
\end{equation}
where $X \in \{E,B\}$. The hat signifies contributions from both the true sky signal and systematic are present, and the tilde denotes it is the pseudo quantities i.e. including effects of partial sky. Note again that the presence of the ${}_0h_{l_1m_1}$ and ${}_{2-k'}h_{l_1m_1}$ terms in equations \ref{eq:True Polarisation} and \ref{eq:Spin2 General Systematic alm} mean that unless the entire sky is observed and there is no mask, these terms are all ``pseudo'' (i.e. partial sky) quantities.

\subsection{\texorpdfstring{$\mathbf{B}$}{TEXT}-mode}
The $B$-mode pseudo power spectrum, may be written as a combination of the true and systematic signals as
\label{section:ClBBSyst}
\begin{dmath}
\tilde{C}_\ell^{\hat{B}\hat{B}}=\frac{1}{\left(2\ell+1 \right)}\sum_{m}\langle\left(\tilde{B}_{\ell m}+\tilde{\Delta}^{B}_{\ell m} \right) \left(\tilde{B}_{\ell m}+\tilde{\Delta}^{B}_{\ell m} \right)^*\rangle \text{,}
\end{dmath}
where the angle brackets denote the average over CMB realisations. This can then be rewritten as a combination of the true $B$-mode spectrum\footnote{The on-sky signal would include both the CMB signal and foregrounds. Fully describing this on-sky signal using a power spectrum requires the fields to be Gaussian. This is a fair assumption for the CMB signal. However it would not truly capture the effects of the foreground signals, which can be highly non-Gaussian. It is likely that an attempt at foreground removal will have been made before this stage and, as such, equations \ref{eq:ClBB} and \ref{eq:ClEE} should still hold in this case. However we note that systematic leakage of foregrounds could still be present in the ${}_2\tilde{\Delta}_{\ell m}$ terms.} and systematic contribution (where we rewrite the systematic in terms of the spin-2 signals) as
\begin{equation}
    \tilde{C}_{\ell}^{\hat{B}\hat{B}}
    = \tilde{C}_{\ell}^{\tilde{B}\tilde{B}}
    + \frac{1}{(2\ell + 1)} \sum_{m} \Bigl\{
        \bigl\langle \frac{1}{4}({}_2\tilde{\Delta}_{\ell m} - {}_{-2}\tilde{\Delta}_{\ell m})({}_2\tilde{\Delta}_{\ell m} - {}_{-2}\tilde{\Delta}_{\ell m})^* - \frac{i}{2}({}_2\tilde{\Delta}_{\ell m} - {}_{-2}\tilde{\Delta}_{\ell m})\tilde{B}^*_{\ell m} + \frac{i}{2}\tilde{B}_{\ell m}({}_2\tilde{\Delta}_{\ell m} - {}_{-2}\tilde{\Delta}_{\ell m})^* \rangle
    \Bigr\} \text{,}
\end{equation}
which simply becomes
\begin{equation}
    \tilde{C}_{\ell}^{\hat{B}\hat{B}}
    = \tilde{C}_{\ell}^{\tilde{B}\tilde{B}}
    + \frac{1}{2 \, (2\ell + 1)} \sum_{m} \Bigl\{
        \bigl\langle{}_2\tilde{\Delta}_{\ell m} \, {}^{}_2\tilde{\Delta}_{\ell m}^*\bigr\rangle
        - (-1)^m \, {\Re{}} \bigl\langle{}_2\tilde{\Delta}_{\ell m} \, {}_2\tilde{\Delta}_{\ell -m}\bigr\rangle + 2 {\Re{}} \langle i \tilde{B}_{\ell m} ({}^{}_2\tilde{\Delta}_{\ell m} -{}^{}_{-2}\tilde{\Delta}_{\ell m})^*\rangle
    \Bigr\} \text{,}
\end{equation}
and we may use the relations $\sum_{m} \tilde{X}_{\ell m} \, {}_{-2}\tilde{\Delta}_{\ell m}^* = \sum_{m} (-1)^m \, \tilde{X}_{\ell m} \, {}_2\tilde{\Delta}_{\ell -m} = \sum_{m} (-1)^m \, \tilde{X}_{\ell -m} \, {}_2\tilde{\Delta}_{\ell m} = \sum_{m} \tilde{X}_{\ell m}^* \, {}_2\tilde{\Delta}_{\ell m}$ where $X \in \{E,B\}$ to rewrite this as
\begin{equation}
    \tilde{C}_{\ell}^{\hat{B}\hat{B}}
    = \tilde{C}_{\ell}^{\tilde{B}\tilde{B}}
    + \frac{1}{2 \, (2\ell + 1)} \sum_{m} \Bigl\{
        \bigl\langle{}_2\tilde{\Delta}_{\ell m} \, {}^{}_2\tilde{\Delta}_{\ell m}^*\bigr\rangle
        - (-1)^m \, {\Re{}} \bigl\langle{}_2\tilde{\Delta}_{\ell m} \, {}_2\tilde{\Delta}_{\ell -m}\bigr\rangle - 4 {\Im{}} \langle  \tilde{B}_{l m} {}^{}_2\tilde{\Delta}_{\ell m}^*\rangle
    \Bigr\} \text{,}
    \label{eq:ClBB}
\end{equation}
noting the presence of a cross-term between the true signal and the systematic.

\subsection{\texorpdfstring{$\mathbf{E}$}{TEXT}-mode}
\label{section:ClEESyst}
We now follow the same approach for the $E$-mode. The $E$-mode power spectrum, may be written as a combination of the true and systematic signals as
\begin{dmath}
\tilde{C}_\ell^{\hat{E}\hat{E}}=\frac{1}{\left(2\ell+1 \right)}\sum_{m}\langle\left(\tilde{E}_{\ell m}+\tilde{\Delta}^{E}_{\ell m} \right) \left(\tilde{E}_{\ell m}+\tilde{\Delta}^{E}_{\ell m} \right)^*\rangle  \text{.}
\end{dmath}
This can then be rewritten as a combination of the true $E$-mode spectrum\footnotemark[\value{footnote}] and systematic contribution as
\begin{equation}
    \tilde{C}_{\ell}^{\hat{E}\hat{E}}
    = \tilde{C}_{\ell}^{\tilde{E}\tilde{E}}
    + \frac{1}{(2\ell + 1)} \sum_{m} \Bigl\{
        \bigl\langle \frac{1}{4}({}_2\tilde{\Delta}_{\ell m} + {}_{-2}\tilde{\Delta}_{\ell m})({}_2\tilde{\Delta}_{\ell m} + {}_{-2}\tilde{\Delta}_{\ell m})^* + \frac{1}{2}({}_2\tilde{\Delta}_{\ell m} + {}_{-2}\tilde{\Delta}_{\ell m})\tilde{E}^*_{\ell m} + \frac{1}{2}\tilde{E}_{\ell m}({}_2\tilde{\Delta}_{\ell m} + {}_{-2}\tilde{\Delta}_{\ell m})^* \rangle
    \Bigr\} \text{,}
\end{equation}
which simply becomes
\begin{equation}
    \tilde{C}_{\ell}^{\hat{E}\hat{E}}
    = \tilde{C}_{\ell}^{\tilde{E}\tilde{E}}
    + \frac{1}{2 \, (2\ell + 1)} \sum_{m} \Bigl\{
        \bigl\langle{}_2\tilde{\Delta}_{\ell m} \, {}^{}_2\tilde{\Delta}_{\ell m}^*\bigr\rangle
        + (-1)^m \, {\Re{}} \bigl\langle{}_2\tilde{\Delta}_{\ell m} \, {}_2\tilde{\Delta}_{\ell -m}\bigr\rangle + 2 {\Re{}} \bigl\langle \tilde{E}_{\ell m} ({}^{}_2\tilde{\Delta}_{\ell m} +{}^{}_{-2}\tilde{\Delta}_{\ell m})^*\bigl\rangle
    \Bigr\} \text{,}
\end{equation}
which can be simplified, using similar relationships to the previous section, as
\begin{equation}
    \tilde{C}_{\ell}^{\hat{E}\hat{E}}
    = \tilde{C}_{\ell}^{\tilde{E}\tilde{E}}
    + \frac{1}{2 \, (2\ell + 1)} \sum_{m} \Bigl\{ \bigl\langle {}_2\tilde{\Delta}_{\ell m} \, {}_2\tilde{\Delta}_{\ell m}^*\bigr\rangle
        + (-1)^m \, {\Re{}} \bigl\langle {}_2\tilde{\Delta}_{\ell m} \, {}_2\tilde{\Delta}_{\ell -m}\bigr\rangle
        + 4 \, {\Re{}} \langle \tilde{E}_{\ell m} \, {}_2\tilde{\Delta}_{\ell m}^* \rangle
    \Bigr\} \text{,}
    \label{eq:ClEE}
\end{equation}
again noting the presence of a cross-term between the true signal and the systematic. This term can in fact dominate for some systematics, such as in the differential gain example which we provide as a demonstration later. It is therefore an important term to include in any analysis.

These equations are independent of the form of systematic and experimental set up and rather show the effect of an additive systematic on $BB$ and $EE$ pseudo-spectra in full generality. We have shown that a cros-term between the intrinsic signal and systematic signal exists for both $BB$ and $EE$ spectra (and we also show the presence of similar terms in the other four spectra of interest in Appendix \ref{section:Additional Spectra}). Any appreciable difference between the effect of systematics on $BB$ and $EE$ will usually be driven by the difference in the cross-terms between the intrinsic signal and the systematic between the two cases. It is worth noting again here that these equations apply generally to full or partial sky analyses as the mask is incorporated within the systematic term along with the scan strategy.

\subsection{Expansion of the terms}
\label{section:Power Spectrum Equations}
In order to use equations \ref{eq:ClBB} and \ref{eq:ClEE} we shall expand each of the terms here to include the general spin-2 spin-coupled systematic of equation \ref{eq:Spin2 General Systematic alm}.
\subsubsection{First Term}
\begin{equation}
\begin{split}
    \frac{1}{2 \, (2\ell + 1)} \sum_{m}  \bigl\langle {}_2\tilde{\Delta}_{\ell m} \, {}_2\tilde{\Delta}_{\ell m}^*\bigr\rangle
    &= \frac{1}{2 \, (2\ell + 1)}\sum_{\substack{m\\l_1m_1\\l_1'm_1'\\l_2m_2}} \bigg\langle \bigg(\sum_{k'}{}_{2-k'}h_{l_1m_1} \threej{\ell,2}{l_1,-(2-k')}{l_2,k'} \, {}_{k'}S_{l_2m_2}\bigg)
    \bigg\{\sum_{k''}{}_{2-k''}h^{*}_{l_1'm_1'} \threej{\ell,2}{l_1',-(2-k'')}{l_2,k''} \, {}_{k''}S^{*}_{l_2m_2}\bigg\}
    \\
    & \tfrac{(2\ell + 1)\sqrt{(2l_1 + 1)(2l_1' + 1)}(2l_2 + 1)}{4\pi} \, \threej{\ell,-m}{l_1,m_1}{l_2,m_2}\threej{\ell,-m}{l_1',m_1'}{l_2,m_2} \bigg\rangle \text{,}
\end{split}
\end{equation}
but using $\sum_{mm_{2}}\threej{\ell,-m}{l_1',m_1'}{l_{2},m_{2}}\threej{\ell,-m}{l_1,m_1}{l_{2},m_{2}} = \frac{1}{2l_1 + 1}\delta_{l_1'l_1}\delta_{m_1'm_1}$ and defining the spectrum of the scan fields as 
\begin{equation}
    \tilde{C}_{l_1}^{{}_{k'}h{}_{k''}h} = \frac{1}{2\ell +1}\sum_{m_1} {}_{k'}h^{}_{l_1m_1} {}_{k''}h^{*}_{l_1m_1} \text{,}
    \label{eq:Scan Spectrum}
\end{equation}
we may write this as
\begin{equation}
\begin{split}
\label{eq:First Term}
    \frac{1}{2 \, (2\ell + 1)} \sum_{m}  \bigl\langle {}_2\tilde{\Delta}_{\ell m} \, {}_2\tilde{\Delta}_{\ell m}^*\bigr\rangle
    &= \frac{1}{8\pi}\sum_{\substack{l_1\\l_2m_2}} (2l_2 + 1)
    \sum_{k'k''}\tilde{C}^{{}_{2-k'}h{}_{2-k''}h}_{l_1} \threej{\ell,2}{l_1,-(2-k')}{l_2,-k'} \threej{\ell,2}{l_1,-(2-k'')}{l_2,-k''} \, \bigg\langle ({}_{k'}S_{l_2m_2}) ({}_{k''}S^{*}_{l_2m_2}) \bigg\rangle  \text{.}
\end{split}
\end{equation}
When applying this to realistic systematics later we will see it is possible to further simplify some systematics as they are often proportional to the on-sky polarisation ${}_{2}P_{l_2m_2}$ or intensity $I_{l_2m_2}$ signals. This means we can rewrite the terms containing the CMB sky signal as $\langle X_{l_2m_2}Y_{l_2m_2}^{*} \rangle = C_{l_2}^{XY}$ where $X,Y \in \{T,E,B\}$ since the angle brackets denote the average over CMB realisations in this case (note the $C_{l_2}^{XY}$s without tildes are the full sky power spectra). However, the foreground signal leaked by systematics may be highly non-Gaussian and as such this simplification would not be possible for them. This means the full sums over $l_2,m_2$ in equations \ref{eq:First Term}, \ref{eq:Second Term}, and \ref{eq:Cross Term} must be applied when considering systematics leaking foregrounds into an observed map.

\subsubsection{Second Term}
Following similar procedures to the first term we find the contribution at power spectrum level to the second term of equation \ref{eq:ClBB} to be
\begin{equation}
\begin{split}
\label{eq:Second Term}
    \frac{1}{2 \, (2\ell + 1)} \sum_{m}  (-1)^m {\Re{}} \bigl\langle {}_2\tilde{\Delta}_{\ell m} \, {}_2\tilde{\Delta}_{\ell -m}^*\bigr\rangle
    &= \frac{1}{8\pi}\sum_{\substack{l_1\\l_2m_2}} (2l_2 + 1) {\Re{}}  
    \sum_{k'k''}\tilde{C}^{{}_{2-k'}h{}_{-(2-k'')}h}_{l_1} \threej{\ell,2}{l_1,-(2-k')}{l_2,-k'} \, \threej{\ell,-2}{l_1,(2-k'')}{l_2,k''} \bigg\langle ({}_{k'}S_{l_2m_2}) ({}_{-k''}S^{*}_{l_2m_2}) \bigg\rangle  \text{.}
\end{split}
\end{equation}

\subsubsection{Third Term ($E$-mode)}
Since we are considering the cross term between the systematic and sky signal here, we must include some extra steps. Using
\begin{equation}
\begin{split}
    \tilde{E}_{\ell m}
    = \sum_{l'm'} \bigl\{
        E_{l'm'} \, W^+_{ll'mm'}
        + B_{l'm'} \, W^-_{ll'mm'}
    \bigr\} \text{,}
\end{split}
\end{equation}
we may write the third term of equation \ref{eq:ClEE} as
\begin{equation}
\begin{split}
    \frac{1}{2 \, (2\ell + 1)} \sum_{m} \bigl\langle \tilde{E}_{\ell m} \, {}^{}_2\tilde{\Delta}_{\ell m}^* \bigr\rangle
    &= \frac{1}{2 \, (2\ell + 1)} \sum_{\substack{k'm\\l_1m_1\\l_2m_2}} \bigl\langle
    \bigl\{ E_{l_2m_2} \, W^+_{ll_2mm_2} + B_{l_2m_2} \, W^-_{ll_2mm_2} \bigr\} \,
    \bigg\{{}_{(2-k')}h^{*}_{l_1m_1} \threej{\ell,2}{l_1,-(2-k')}{l_2,-k'} \, {}_{k'}S^{*}_{l_2m_2}
    \bigg\}
    \\
    & (-1)^m\sqrt{\tfrac{(2\ell + 1)(2l_1 + 1)(2l_2 + 1)}{4\pi}} \, \threej{\ell,-m}{l_1,m_1}{l_2,m_2} \rangle  \text{.}
\end{split}
\end{equation}
The windows due to a mask as in \cite{BrownCastroTaylor} are given by
\begin{equation}
    {}_sw^{m m_{2}}_{l l_{2}} = \int d\Omega {}_sY_{l_{2}m_{2}}(\Omega) w_p(\Omega) {}_sY_{\ell m}^{*}(\Omega).
\end{equation}
By expanding the mask as a spin zero field  $\sum_{l_1'm_1'} {}_0h_{l_1'm_1'} Y_{l_1'm_1'}(\Omega)$,
we can rewrite this as
\begin{equation}
    {}_sw^{m m_{2}}_{l l_{2}} = \int d\Omega {}_sY_{l_{2}m_{2}}(\Omega) \sum_{l_1'm_1'} {}_0h_{l_1'm_1'} Y_{l_1'm_1'}(\Omega) {}_sY_{\ell m}^{*}(\Omega)
    = \sum_{l_1'm_1'} {}_0h_{l_1'm_1'} (-1)^{m+s} \sqrt{\frac{(2\ell +1)(2l_{2}+1)(2l_1'+1)}{4\pi}} \threej{l_{2},m_{2}}{l_1',m_1'}{\ell,-m} \threej{l_{2},s}{l_1',0}{\ell,-s} \text{,}
\end{equation}
where ${}_sY_{\ell m}^{*}(\Omega) = (-1)^{m+s}{}_{-s}Y_{\ell -m}(\Omega)$. We can use this to write out the mask terms as

\begin{align}
    W^+_{ll_2mm_2} &= \frac{1}{2}({}_2w^{m m_{2}}_{l l_{2}} + {}_{-2}w^{m m_{2}}_{l l_{2}})
    = \frac{1}{2} \sum_{l_1'm_1'} {}_0h_{l_1'm_1'} \sqrt{\frac{(2\ell +1)(2l_{2}+1)(2l_1'+1)}{4\pi}} (-1)^m\threej{l_{2},m_{2}}{l_1',m_1'}{\ell,-m} \bigg(\threej{l_{2},2}{l_1',0}{\ell,-2}
    + \threej{l_{2},-2}{l_1',0}{\ell,2}\bigg)
    \nonumber
    \\
    W^-_{ll_2mm_2} &= \frac{i}{2}({}_2w^{m m_{2}}_{l l_{2}} - {}_{-2}w^{m m_{2}}_{l l_{2}})
    = \frac{i}{2} \sum_{l_1'm_1'} {}_0h_{l_1'm_1'} \sqrt{\frac{(2\ell +1)(2l_{2}+1)(2l_1'+1)}{4\pi}} (-1)^m\threej{l_{2},m_{2}}{l_1',m_1'}{\ell,-m} \bigg(\threej{l_{2},2}{l_1',0}{\ell,-2}
    - \threej{l_{2},-2}{l_1',0}{\ell,2}\bigg)  \text{.}
    \label{equation:Windows}
\end{align}
This therefore means we can write
\begin{equation}
\begin{split}
    \frac{1}{2 \, (2\ell + 1)} \sum_{m} \bigl\langle \tilde{E}_{\ell m} \, {}^{}_2\tilde{\Delta}_{\ell m}^* \bigr\rangle
    &= \sum_{\substack{k'm\\l_1m_1\\l_1'm_1'\\l_2m_2}} \bigl\langle
    \bigl\{ E_{l_2m_2} \, \bigg(\threej{l_{2},2}{l_1',0}{\ell,-2}
    + \threej{l_{2},-2}{l_1',0}{\ell,2}\bigg) + iB_{l_2m_2} \, \bigg(\threej{l_{2},2}{l_1',0}{\ell,-2}
    - \threej{l_{2},-2}{l_1',0}{\ell,2}\bigg) \bigr\} \, {}_0h_{l_1'm_1'}
    \\
    &\bigg\{\frac{1}{2} {}_{(2-k')}h^{*}_{l_1m_1} \threej{\ell,2}{l_1,-(2-k')}{l_2,-k'} \, {}_{k'}S^{*}_{l_2m_2}\bigg\}
    \tfrac{\sqrt{(2l_1 + 1)(2l_1' + 1)}(2l_2 + 1)}{16\pi} \, \threej{\ell,-m}{l_1,m_1}{l_2,m_2}\threej{l_{2},m_{2}}{l_1',m_1'}{\ell,-m} \rangle.
\end{split}
\label{eq:Cross Term Mask}
\end{equation}
Using equation \ref{eq:Scan Spectrum} to define the spectrum of the scan fields we may write the cross-term contribution as
\begin{equation}
\begin{split}
    \frac{1}{2 \, (2\ell + 1)} \sum_{m} \bigl\langle \tilde{E}_{\ell m} \, {}^{}_2\tilde{\Delta}_{\ell m}^* \bigr\rangle
    &= \sum_{\substack{k'l_1\\l_2m_2}} \tfrac{(2l_1 + 1)(2l_2 + 1)}{16\pi} (-1)^{\ell+l_1+l_2}
    \bigg\{\tilde{C}_{l_1}^{{}_{0}h{}_{(2-k')}h} \threej{\ell,2}{l_1,-(2-k')}{l_2,-k'} \, \bigg\{\langle E_{l_2m_2} {}_{k'}S^{*}_{l_2m_2}\rangle\, \bigg(\threej{l_{2},2}{l_1,0}{\ell,-2}
    + \threej{l_{2},-2}{l_1,0}{\ell,2}\bigg)
    \\
    &+ i\langle B_{l_2m_2} {}_{k'}S^{*}_{l_2m_2}\rangle \, \bigg(\threej{l_{2},2}{l_1,0}{\ell,-2}
    - \threej{l_{2},-2}{l_1,0}{\ell,2}\bigg) \bigg\}  \text{.}
\end{split}
\label{eq:Cross Term}
\end{equation}
The cross term can be the dominant source of systematic but depending on its spin-dependence can also vanish in certain circumstances --- see Appendix \ref{section:Cross Term} for further detail. Additionally flipping the lower row signs of one of the Wigner-3j symbols using the factor of $(-1)^{\ell+l_1+l_2}$ the further simplification can be made that
\small
\begin{equation}
    \frac{1}{2 \, (2\ell + 1)} \sum_{m} \bigl\langle \tilde{E}_{\ell m} \, {}^{}_2\tilde{\Delta}_{\ell m}^*
    \bigr\rangle = \bigg\{
    \begin{array}{l l}
    \sum_{\substack{k'l_1\\l_2m_2}} \tfrac{(2l_1 + 1)(2l_2 + 1)}{16\pi}
    \tilde{C}_{l_1}^{{}_{0}h{}_{(2-k')}h} \threej{\ell,2}{l_1,-(2-k')}{l_2,-k'} \, \langle E_{l_2m_2} {}_{k'}S^{*}_{l_2m_2}\rangle\, \bigg((-1)^{\ell+l_1+l_2} \threej{l_{2},2}{l_1,0}{\ell,-2}
    + \threej{l_{2},2}{l_1,0}{\ell,-2}\bigg)
    & \text{for } \ell+l_1+l_2 = \text{even.}\\
    \sum_{\substack{k'l_1\\l_2m_2}} \tfrac{(2l_1 + 1)(2l_2 + 1)}{16\pi}
    \tilde{C}_{l_1}^{{}_{0}h{}_{(2-k')}h} \threej{\ell,2}{l_1,-(2-k')}{l_2,-k'} \, i\langle B_{l_2m_2} {}_{k'}S^{*}_{l_2m_2}\rangle \, \bigg((-1)^{\ell+l_1+l_2}\threej{l_{2},2}{l_1,0}{\ell,-2}
    - \threej{l_{2},2}{l_1,0}{\ell,-2}\bigg)
    & \text{for } \ell+l_1+l_2 = \text{odd.}\\
    \end{array}
\end{equation}
\normalsize

\subsubsection{Third Term ($B$-mode)}
Using
\begin{equation}
\begin{split}
    \tilde{B}_{\ell m}
    = \sum_{l'm'} \bigl\{
        B_{l'm'} \, W^+_{\ell l'mm'}
        - E_{l'm'} \, W^-_{\ell l'mm'}
    \bigr\} \text{,}
\end{split}
\end{equation}
we may write the third term of equation \ref{eq:ClBB} as
\begin{equation}
\begin{split}
    \frac{1}{2 \, (2\ell + 1)} \sum_{m} \bigl\langle \tilde{B}_{\ell m} \, {}^{}_2\tilde{\Delta}_{\ell m}^* \bigr\rangle
    &= \frac{1}{2 \, (2\ell + 1)}  \sum_{\substack{k'm\\l_1m_1\\l_2m_2}} \bigl\langle
    \bigl\{ B_{l_2m_2} \, W^+_{\ell l_2mm_2} - E_{l_2m_2} \, W^-_{\ell l_2mm_2} \bigr\} \,
    \bigg\{{}_{(2-k')}h^{*}_{l_1m_1} \threej{\ell,2}{l_1,-(2-k')}{l_2,-k'} \, {}_{k'}S^{*}_{l_2m_2}
    \bigg\}
    \\
    & (-1)^m\sqrt{\tfrac{(2\ell + 1)(2l_1 + 1)(2l_2 + 1)}{4\pi}} \, \threej{\ell,-m}{l_1,m_1}{l_2,m_2} \rangle  \text{.}
\end{split}
\end{equation}
Using equation \ref{equation:Windows} we can write this as
\begin{equation}
\begin{split}
    \frac{1}{2 \, (2\ell + 1)} \sum_{m} \bigl\langle  \tilde{B}_{\ell m} \, {}^{}_2\tilde{\Delta}_{\ell m}^* \bigr\rangle
    &= \sum_{\substack{k'm\\l_1m_1\\l_1'm_1'\\l_2m_2}} \bigl\langle
    \bigl\{ B_{l_2m_2} \, \bigg(\threej{l_{2},2}{l_1',0}{\ell,-2}
    + \threej{l_{2},-2}{l_1',0}{\ell,2}\bigg) - iE_{l_2m_2} \, \bigg(\threej{l_{2},2}{l_1',0}{\ell,-2}
    - \threej{l_{2},-2}{l_1',0}{\ell,2}\bigg) \bigr\} \, {}_0h_{l_1'm_1'}
    \\
    &\bigg\{\frac{1}{2} {}_{(2-k')}h^{*}_{l_1m_1} \threej{\ell,2}{l_1,-(2-k')}{l_2,-k'} \, {}_{k'}S^{*}_{l_2m_2}\bigg\}
    \tfrac{\sqrt{(2l_1 + 1)(2l_1' + 1)}(2l_2 + 1)}{16\pi} \, \threej{\ell,-m}{l_1,m_1}{l_2,m_2}\threej{l_{2},m_{2}}{l_1',m_1'}{\ell,-m} \rangle  \text{.}
\end{split}
\label{eq:Cross Term Mask BB}
\end{equation}
Using equation \ref{eq:Scan Spectrum} to define the spectrum of the scan fields we may write the cross-term contribution as
\begin{equation}
\begin{split}
    \frac{1}{2 \, (2\ell + 1)} \sum_{m} \bigl\langle \tilde{B}_{\ell m} \, {}^{}_2\tilde{\Delta}_{\ell m}^* \bigr\rangle
    &=  \sum_{\substack{k'l_1\\l_2m_2}} \tfrac{(2l_1 + 1)(2l_2 + 1)}{16\pi} (-1)^{\ell+l_1+l_2}
    \bigg\{\tilde{C}_{l_1}^{{}_{0}h{}_{(2-k')}h} \threej{\ell,2}{l_1,-(2-k')}{l_2,-k'} \, \bigg\{\langle B_{l_2m_2} {}_{k'}S^{*}_{l_2m_2}\rangle\, \bigg(\threej{l_{2},2}{l_1,0}{\ell,-2}
    + \threej{l_{2},-2}{l_1,0}{\ell,2}\bigg)
    \\
    &- i\langle E_{l_2m_2} {}_{k'}S^{*}_{l_2m_2}\rangle \, \bigg(\threej{l_{2},2}{l_1,0}{\ell,-2}
    - \threej{l_{2},-2}{l_1,0}{\ell,2}\bigg) \bigg\}  \text{.}
\end{split}
\label{eq:Cross Term BB}
\end{equation}
Again, it can be shown (see Appendix \ref{section:Cross Term}) that for many systematics this cross-term will be negligible, Additionally flipping the lower row signs of one of the Wigner-3j symbols using the factor of $(-1)^{\ell+l_1+l_2}$ the further simplification can be made that
\small
\begin{equation}
    \frac{1}{2 \, (2\ell + 1)} \sum_{m} \bigl\langle  \tilde{B}_{\ell m} \, {}^{}_2\tilde{\Delta}_{\ell m}^*
    \bigr\rangle = \bigg\{
    \begin{array}{l l}
    \sum_{\substack{k'l_1\\l_2m_2}} \tfrac{(2l_1 + 1)(2l_2 + 1)}{16\pi}
    \tilde{C}_{l_1}^{{}_{0}h{}_{(2-k')}h} \threej{\ell,2}{l_1,-(2-k')}{l_2,-k'} \, \langle B_{l_2m_2} {}_{k'}S^{*}_{l_2m_2}\rangle\, \bigg((-1)^{\ell+l_1+l_2}\threej{l_{2},2}{l_1,0}{\ell,-2}
    + \threej{l_{2},2}{l_1,0}{\ell,-2}\bigg)
    & \text{for } \ell+l_1+l_2 = \text{even.}\\
    \sum_{\substack{k'l_1\\l_2m_2}} \tfrac{(2l_1 + 1)(2l_2 + 1)}{16\pi}
    \tilde{C}_{l_1}^{{}_{0}h{}_{(2-k')}h} \threej{\ell,2}{l_1,-(2-k')}{l_2,-k'} \, \bigg\{-i\langle E_{l_2m_2} {}_{k'}S^{*}_{l_2m_2}\rangle \, \bigg((-1)^{\ell+l_1+l_2}\threej{l_{2},2}{l_1,0}{\ell,-2}
    - \threej{l_{2},2}{l_1,0}{\ell,-2}\bigg)\bigg\}
    & \text{for } \ell+l_1+l_2 = \text{odd.}\\
    \end{array}
\end{equation}
\normalsize

\section{Demonstration}
\label{section:Demonstration}
In order to demonstrate the utility of the formalism, we now provide two illustrative examples: the gain and pointing mismatches between detectors in a situation where detectors are differenced. Although the formalism is not restricted to differencing (simply requiring that the systematic signal can be written as in equation \ref{eq:spinsyst}), this choice allows us to see how the additional considerations of partial sky, polarisation mixing, and the cross terms identified above, contribute to the analysis in a situation that has been treated in their absence before \cite[e.g.][]{Wallisetal2016}. We shall consider the case without foregrounds in this section for clarity, but as noted earlier it is also possible to apply this formalism to systematics that leak foregrounds.

\subsection{Focal Plane Setup and Pair Differencing}
When considering how to use equation \ref{eq:spinsyst} to write the observed signal we must take into account how the timestream is being treated. Each element that contributes its own distinct timestream measurement $d_i$ to the map-making process (further detail on map-making is available in Appendix~\ref{section:Map-Making}) will have its own distinct crossing angle associated with it. As such, each distinct element will also have a distinct $h(\psi, \Omega)$ field associated. We shall describe this in the context of a pair-differencing experiment in this section.

We use ``pair differencing'' to describe experiments where the detectors are arranged in pairs with orthogonal polarisation sensitivity and their signals are differenced (in order to nominally remove the temperature signal), so that the signal for detector pair $i$ is given by
\begin{equation}
d_i=\frac{1}{2}\left(d^A_i-d^B_i \right) \text{,}
\end{equation}
where $i$ denotes the pair the detector belongs to, and $A$/$B$ denote which detector in the pair we are considering. Any differences in the responses of the detectors in a pair may result in both temperature leakage and polarisation mixing. Characterising these systematics is important given the magnitude of the temperature spectrum in comparison to the polarisation. The polarisation mixing has been largely ignored due to its relative small size to the temperature. However given the unprecedented sensitivities expected in future experiments this should be carefully checked.

In this context it is the differenced signal that contributes to the timestream (as opposed to individual timestreams for each detector): each of these differenced signals $d_i$ is treated as a single entry for the map-making process outlined further in Appendix~\ref{section:Map-Making}. As such, each pair of detectors shares a single set of crossing angles, and the coupling of the differenced signals is described by a distinct $h(\psi, \Omega)$ field for each pair.

We will consider two sets of co-located pairs of detectors, oriented $\pi/4$ radians apart to allow simple simultaneous calculation of both $Q$ and $U$ polarisation signals. 
Since there are now two detector pairs, each pair will have a distinct associated crossing angle and will contribute distinctly to the timestream. We can write the total observed signal, using equation \ref{eq:SDTot}, as
\begin{equation}
    S^{d}(\psi,\Omega)
    = h(\psi,\Omega) S_1(\psi,\Omega) + h'(\psi,\Omega) S_2(\psi,\Omega) \text{,}
\end{equation}
where the subscript 1 and 2 indicate which pair of detectors the signal applies to, and the second detector pair is rotated by $\pi/4$ with respect to the first giving a distinct scan-coupling term
\begin{equation}
    h'(\psi, \Omega)
    = \sum_{k} \tilde{h}_k(\Omega) \, e^{-i \, k \, \pi/4}\text{.}
\end{equation}
We thus write the total signal in Fourier space as a convolution according to equation \ref{eq:spinsyst} as
\begin{equation}
    {}_{2}\tilde{S}^{d} = \sum_{k'=-\infty}^{\infty}\tilde{h}_{2-k'}{}_{k'}\tilde{S}^{1} + \tilde{h}_{2-k'}e^{-i\pi (2-k') / 4}{}_{k'}\tilde{S}^{2} = \sum_{k'=-\infty}^{\infty}\tilde{h}_{2-k'}{}_{k'}\tilde{S} \text{,}
    \label{eq:DiffSpinSyst}
\end{equation}
in which we have selected $k=2$ since we are interested in the spin-2 polarisation signal and in this case the signal may be written as ${}_{k'}\tilde{S} = {}_{k'}\tilde{S}^{1} + e^{-i\pi (2-k') / 4}{}_{k'}\tilde{S}^{2}$ where the superscript 1 and 2 indicate the pair of detectors. This shows the signal for a two detector pair differencing setup can be written according to equation \ref{eq:spinsyst}. This can be extended to arbitrary focal plane setups by following the method we employed above.

As an example if we consider a sky signal as in equation \ref{eq:Single Detector Signal}, but in the absence of systematics, then a single detector would see a $k=2$ signal of
\begin{equation}
    {}_{2}\tilde{S}^{d,\text{single}}(\Omega) = \tilde{h}_{2}(\Omega)I(\Omega) + \tilde{h}_{0}(\Omega)P(\Omega) + \tilde{h}_{4}(\Omega)P^*(\Omega)  \text{,}
\end{equation}
while the signal for the two detector pair setup would be
\begin{equation}
\begin{split}
    {}_{2}\tilde{S}^{d}(\Omega) &= \frac{1}{2}\tilde{h}_{0}(\Omega)({}_{2}\tilde{S}^{1} + {}_{2}\tilde{S}^{2}) + \frac{1}{2}\tilde{h}_{4}(\Omega)({}_{-2}\tilde{S}^{1} + e^{-i\pi}{}_{-2}\tilde{S}^{2})
    =\frac{1}{2}\tilde{h}_{0}(\Omega)(P + P) + \frac{1}{2}\tilde{h}_{4}(\Omega)(P^* - P^*) = \tilde{h}_{0}(\Omega)P(\Omega) \text{.}
\end{split}
\end{equation}
This shows that, in the absence of systematics, this setup will simply give the pure polarisation signal. This is the on-sky polarisation signal multiplied by the window function, so this corresponds to the $\tilde{P}$ contribution to $\hat{P}$ in equation \ref{eq: Polarisation Signal}. Any other terms present in the detector equation would contribute to the $\Delta$ term.

\subsection{Differential Gain}
\label{sec:diff_gain}
Differential gain arises from a miscalibration of the detectors and results in two effects: temperature leakage into the polarisation signal, and the direct amplification of the polarisation signals.

We first write the signal in a single detector as
\begin{equation}
    d_{i} = (1+g_{i}) (I(\Omega) + \frac{1}{2}(P(\Omega)e^{2i\psi} + P^{*}(\Omega)e^{-2i\psi})) \text{,}
    \label{eq:Gain Single Detector}
\end{equation}
where $g_{i}$ is some offset due to gain miscalibration. The pair-differenced signal is then given by
\begin{equation}
\begin{split}
    S_i &= \frac{1}{2}[d^A_i - d^B_i]
    = \frac{1}{2} \bigg[(g^{A}_i-g^{B}_i)I(\Omega) + \frac{1}{2}\Big(P(\Omega)[(1+g^A_i)e^{2i\psi} - (1+g^B_i)e^{2\psi+\pi}] 
    + P^*(\Omega)[(1+g^A_i)e^{-2i\psi} - (1+g^B_i)e^{-2i(\psi+\pi)}]\Big)\bigg]
    \\
    &=\frac{1}{2}\bigg[(g^A_i-g^B_i)I(\Omega) + \frac{1}{2}\Big(P(\Omega)[2+g^A_i+g^B_i]e^{2i\psi} + P^*(\Omega)[2+g^A_i+g^B_i]e^{-2i\psi}\Big)\bigg]\text{.}
\end{split}
\end{equation}
The spurious signal for a single detector pair is thus
\begin{equation}
\delta d^g_i=\frac{1}{2}(g^A_i-g^B_i)I(\Omega) + \frac{g^A_i+g^B_i}{2}\frac{1}{2} (P(\Omega)e^{2i\psi} + P^*(\Omega)e^{-2i\psi}) \text{,}
\end{equation}
which contains both leakage from the temperature signal and amplification of the polarisation signals. Since these signals are of spin-0 and spin-2 respectively, a "3x3" map-making\footnote{The map-making we refer to here is a naive binning method where a simple matrix inversion is performed in each pixel --- Appendix \ref{section:Map-Making} examines this in more detail. For the rest of this work we shall adopt a map-making terminology based on the matrix size to describe this naive approach. i.e. 2x2 refers to solving for just the $Q$ and $U$ Stokes parameters with the $I$ row removed, 3x3 solves for all three Stokes parmameters ($I$, $Q$ and $U$), 5x5 includes an additional two rows to solve for spin-1 fields etc.} approach would separate them, thus ensuring only a gain amplification persists, rather than any temperature-to-polarisation leakage from differential gain. However one approach regularly used in pair differencing experiments is to remove the spin-0 row from the map-making process. In this scenario the spin-0 temperature signal from gain mismatch will still leak into the polarisation and would need to be accounted for in some other way (e.g. \cite{2015ApJ...814..110B} use a template fitting method). This is explored further in Appendix \ref{section:2x2 vs 3x3}.

For the two sets of detector pairs we may use equation \ref{eq:DiffSpinSyst} to write the total spin-2 signal including differential gain as
\begin{equation}
{}_2\tilde{S}^{d}(\Omega) = \tilde{h}_0(\Omega) P(\Omega)
    + \frac{1}{2} \tilde{h}_2(\Omega) (\delta g_1 - i\delta g_2) \, I(\Omega)
    + \frac{1}{4}\tilde{h}_0(\Omega)(g^A_1 + g^B_1 + g^A_2 + g^B_2) \, P(\Omega)
    + \frac{1}{4}\tilde{h}_4(\Omega)(g^A_1 + g^B_1 - g^A_2 - g^B_2) \, P^*(\Omega) \text{,}
    \label{eq:Differential Gain}
\end{equation}
where we define $\delta g_i = g^A_i-g^B_i$. Comparing to equation \ref{eq:spinsyst}, the signals in this case are given by
\begin{align}
{}_0\tilde{S} &= {}_0\tilde{S}^1 + e^{-i\pi(2-0)/4}{}_0\tilde{S}^2 = \frac{1}{2}(\delta g_1 -i\delta g_2) I(\Omega)\nonumber
\\
{}_2\tilde{S} &= {}_2\tilde{S}^1 + e^{-i\pi(2-2)/4}{}_2\tilde{S}^2 =\left[1 + \frac{1}{4}(g^A_1 + g^B_1 + g^A_2 + g^B_2)\right]\nonumber
P(\Omega)
\\
{}_{-2}\tilde{S} &= {}_{-2}\tilde{S}^1 + e^{-i\pi(2+2)/4}{}_{-2}\tilde{S}^2 =\left[\frac{1}{4}(g^A_1 + g^B_1 - g^A_2 - g^B_2)\right] \, P^*(\Omega).
\end{align}
Inserting this into equation \ref{eq:Spin2 General Systematic alm}, the differential gain systematic contribution can be written in harmonic space as
\begin{equation}
\begin{split}
    {}_2\tilde{\Delta}_{\ell m}^{g} = \sum_{\substack{l_1m_1\\l_2m_2}}  \, \bigg( \frac{1}{2} {}_2h_{l_1m_1} \, (\delta g_1 - i \, \delta g_2) \, \threej{\ell,2}{l_1,-2}{l_2,0} \, I_{l_2m_2}
    + \frac{1}{4} {}_0h_{l_1m_1} \, (g_1^A + g_1^B + g_2^A + g_2^B) \, \threej{\ell,2}{l_1,0}{l_2,-2} \, {}_2P_{l_2m_2}
    \\
    + \frac{1}{4} {}_4h_{l_1m_1} \, (g_1^A + g_1^B - g_2^A - g_2^B) \, \threej{\ell,2}{l_1,-4}{l_2,2} \, {}_{-2}P_{l_2m_2}\bigg)
    (-1)^m \sqrt{\tfrac{(2\ell + 1)(2l_1 + 1)(2l_2 + 1)}{4\pi}} \, \threej{\ell,-m}{l_1,m_1}{l_2,m_2}.
\end{split}
\end{equation}
We may then use this to look at the effect the systematic would have on the observed power spectra in the case where the signals are not removed by other means. In principle you could examine the spins of the systematic present in equation \ref{eq:Differential Gain} and use this to inform choices of map-making --- we examine this more in Appendix~\ref{section:2x2 vs 3x3}. In the following example, we shall assume that the spin-2 content of the observed data is identified as the on-sky polarisation signal, where a 2x2 map-making approach has been employed.

Note again that since we are considering a pure CMB sky signal in the absence of foregrounds for this demonstration we may use the Gaussianity of the CMB fields to write them as $\langle X_{l_2m_2}Y_{l_2m_2}^{*} \rangle = C_{l_2}^{XY}$ where $X,Y \in \{T,E,B\}$, also using the fact that $C_{l_2}^{XY} = C_{l_2}^{YX}$ (note $I$ and $T$ are used interchangeably here, the temperature spectrum comes from the intensity signal). 

Following the process outlined generally in Section \ref{section:Power Spectrum Equations} and defining ${}_2P_{\ell m} = E_{\ell m} + i B_{\ell m}$ \citep{durrer_2008} \footnote{This is a different convention to that used by {\sevensize HEALPIX} so cross spectra ($TE$, $EB$ etc.) would be need multiplying by -1 if using {\sevensize HEALPIX}.} we find the following systematic terms due to differential gain:
\small
\begin{equation}
\begin{split}
    \frac{1}{2 \, (2\ell + 1)} \sum_{m} \bigl\langle {}_2\tilde{\Delta}_{\ell m} \, {}_2\tilde{\Delta}_{\ell m}^* \bigr\rangle
    &= \sum_{l_1l_2} \tfrac{(2l_1+1)(2l_2 + 1)}{32\pi} \,
    \bigg(\tilde{C}_{l_1}^{{}_2h{}_2h} \, \bigl|\delta g_1 - i \, \delta g_2\bigr|^2 \, \threej{\ell,2}{l_1,-2}{l_2,0}^2 \, C_{l_2}^{TT}
    \\
    &+ \tilde{C}_{l_1}^{{}_2h{}_0h} \, (g_1^A + g_1^B + g_2^A + g_2^B)  \, \threej{\ell,2}{l_1,-2}{l_2,0} \threej{\ell,2}{l_1,0}{l_2,-2} \, (\delta g_1 C_{l_2}^{TE} - \delta g_2 C_{l_2}^{TB})
    \\
    &+\tilde{C}_{l_1}^{{}_2h{}_4h} \, (g_1^A + g_1^B - g_2^A - g_2^B) \, \threej{\ell,2}{l_1,-2}{l_2,0} \threej{\ell,2}{l_1,-4}{l_2,2} \, (\delta g_1 C_{l_2}^{TE} + \delta g_2 C_{l_2}^{TB})
    \\
    &+ \frac{1}{4} \tilde{C}_{l_1}^{{}_0h{}_0h} \, (g_1^A + g_1^B + g_2^A + g_2^B)^2 \, \threej{\ell,2}{l_1,0}{l_2,-2}^2\, (C_{l_2}^{EE} + C_{l_2}^{BB})
    \\
    &+ \frac{1}{2} \tilde{C}_{l_1}^{{}_0h{}_4h} \, (g_1^A + g_1^B + g_2^A + g_2^B) \, (g_1^A + g_1^B - g_2^A - g_2^B) \, \threej{\ell,2}{l_1,0}{l_2,-2} \threej{\ell,2}{l_1,-4}{l_2,2} \, (C_{l_2}^{EE} - C_{l_2}^{BB})
    \\
    &+ \frac{1}{4} \tilde{C}_{l_1}^{{}_4h{}_4h} \, (g_1^A + g_1^B - g_2^A - g_2^B)^2 \, \threej{\ell,2}{l_1,-4}{l_2,2}^2 \, (C_{l_2}^{EE} + C_{l_2}^{BB})\bigg)
    \label{eq:Term 1 Gain}
\end{split}
\end{equation}

\begin{equation}
\begin{split}
    \frac{1}{2 \, (2\ell + 1)} \sum_{m} (-1)^m \, \Re{} \bigl\langle{}_2\tilde{\Delta}_{\ell m} \, {}_{2}\tilde{\Delta}_{\ell -m}\bigr\rangle
    &= \sum_{l_1l_2} \Re{} \tfrac{(2l_1+1)(2l_2 + 1)}{32\pi} \,
    \bigg(\tilde{C}_{l_1}^{{}_2h{}_{-2}h} \, (\delta g_1 - i \, \delta g_2)^2 \, \threej{\ell,2}{l_1,-2}{l_2,0} \threej{\ell,-2}{l_1,2}{l_2,0} \, C_{l_2}^{TT}
    \\
    &+ \frac{1}{2} \tilde{C}_{l_1}^{{}_0h{}_{-2}h} \, (g_1^A + g_1^B + g_2^A + g_2^B) \, (\delta g_1 - i \, \delta g_2) \, \threej{\ell,-2}{l_1,2}{l_2,0} \threej{\ell,2}{l_1,0}{l_2,-2} \, (C_{l_2}^{TE} + i C_{l_2}^{TB})
    \\
    &+ \frac{1}{2} \tilde{C}_{l_1}^{{}_4h{}_{-2}h} \, (g_1^A + g_1^B - g_2^A - g_2^B) \, (\delta g_1 - i \, \delta g_2) \, \threej{\ell,-2}{l_1,2}{l_2,0} \threej{\ell,2}{l_1,-4}{l_2,2} \, (C_{l_2}^{TE} - i C_{l_2}^{TB})
    \\
    & + \frac{1}{2} \tilde{C}_{l_1}^{{}_2h{}_0h} \, (\delta g_1 - i \, \delta g_2)
    \, (g_1^A + g_1^B + g_2^A + g_2^B) \, \threej{\ell,-2}{l_1,0}{l_2,2} \threej{\ell,2}{l_1,-2}{l_2,0} \, (C_{l_2}^{TE} + i C_{l_2}^{TB})
    \\
    &+ \frac{1}{4} \tilde{C}_{l_1}^{{}_0h{}_0h} \, (g_1^A + g_1^B + g_2^A + g_2^B)^2 \, \threej{\ell,2}{l_1,0}{l_2,-2} \threej{\ell,-2}{l_1,0}{l_2,2}\, (C_{l_2}^{EE} - C_{l_2}^{BB} +2i C_{l_2}^{EB} )
    \\
    &+ \frac{1}{4} \tilde{C}_{l_1}^{{}_4h{}_0h} \, (g_1^A + g_1^B - g_2^A - g_2^B) \, (g_1^A + g_1^B + g_2^A + g_2^B) \, \threej{\ell,-2}{l_1,0}{l_2,2} \threej{\ell,2}{l_1,-4}{l_2,2} \, (C_{l_2}^{EE} + C_{l_2}^{BB})
    \\
    & + \frac{1}{2} \tilde{C}_{l_1}^{{}_2h{}_{-4}h} \, (\delta g_1 - i \, \delta g_2)
    \, (g_1^A + g_1^B - g_2^A - g_2^B) \, \threej{\ell,-2}{l_1,4}{l_2,-2} \threej{\ell,2}{l_1,-2}{l_2,0} \, (C_{l_2}^{TE} - i C_{l_2}^{TB})
    \\
    &+ \frac{1}{4} \tilde{C}_{l_1}^{{}_0h{}_{-4}h} \, (g_1^A + g_1^B + g_2^A + g_2^B) \, (g_1^A + g_1^B - g_2^A - g_2^B) \, \threej{\ell,-2}{l_1,4}{l_2,-2} \threej{\ell,2}{l_1,0}{l_2,-2} \, (C_{l_2}^{EE} + C_{l_2}^{BB})
    \\
    &+ \frac{1}{4} \tilde{C}_{l_1}^{{}_4h{}_{-4}h} \, (g_1^A + g_1^B - g_2^A - g_2^B)^2 \, \threej{\ell,2}{l_1,-4}{l_2,2} \threej{\ell,-2}{l_1,4}{l_2,-2} \, (C_{l_2}^{EE} - C_{l_2}^{BB} - 2i C_{l_2}^{EB})\bigg)
    \label{eq:Term 2 Gain}
\end{split}
\end{equation}

\begin{equation}
\begin{split}
    \frac{1}{2 \, (2\ell + 1)} \sum_{m} \Re{} \bigl\langle \tilde{E}_{\ell m} \, {}^{}_2\tilde{\Delta}_{\ell m}^* \bigr\rangle
    &= \sum_{\substack{l_1l_2}} \Re{} \tfrac{(2l_1 + 1)(2l_2 + 1)}{16\pi} (-1)^{\ell+l_1+l_2}
    \bigg\{\frac{1}{2} \tilde{C}_{l_1}^{{}_{0}h{}_{2}h} \, (\delta g_1 + i \, \delta g_2) \, \threej{\ell,2}{l_1,-2}{l_2,0} \, \bigg\{ C_{l_2}^{TE} \, \bigg(\threej{l_{2},2}{l_1,0}{\ell,-2}
    + \threej{l_{2},-2}{l_1,0}{\ell,2}\bigg) + iC_{l_2}^{TB} \, \bigg(\threej{l_{2},2}{l_1,0}{\ell,-2}
    - \threej{l_{2},-2}{l_1,0}{\ell,2}\bigg) \bigg\}
    \\
    &+\frac{1}{4} \tilde{C}_{l_1}^{{}_{0}h{}_{0}h} \, (g_1^A + g_1^B + g_2^A + g_2^B) \, \threej{\ell,2}{l_1,0}{l_2,-2} \,
    \bigg\{ (C_{l_2}^{EE}-iC_{l_2}^{EB}) \, \bigg(\threej{l_{2},2}{l_1,0}{\ell,-2}
    + \threej{l_{2},-2}{l_1,0}{\ell,2}\bigg) + (iC_{l_2}^{EB}+C_{l_2}^{BB}) \, \bigg(\threej{l_{2},2}{l_1,0}{\ell,-2}
    - \threej{l_{2},-2}{l_1,0}{\ell,2}\bigg)\bigg\}
    \\
    &+\frac{1}{4} \tilde{C}_{l_1}^{{}_{0}h{}_{4}h} \, (g_1^A + g_1^B - g_2^A - g_2^B) \, \threej{\ell,2}{l_1,-4}{l_2,2} \,
    \bigg\{ (C_{l_2}^{EE} + iC_{l_2}^{EB}) \, \bigg(\threej{l_{2},2}{l_1,0}{\ell,-2}
    + \threej{l_{2},-2}{l_1,0}{\ell,2}\bigg) + (iC_{l_2}^{EB} -C_{l_2}^{BB}) \, \bigg(\threej{l_{2},2}{l_1,0}{\ell,-2}
    - \threej{l_{2},-2}{l_1,0}{\ell,2}\bigg) \bigg\}
    \label{eq:Term 3 Gain}
\end{split}
\end{equation}

\begin{equation}
\begin{split}
    \frac{1}{2 \, (2\ell + 1)} \sum_{m} {\Im{}} \bigl\langle \tilde{B}_{\ell m} \, {}^{}_2\tilde{\Delta}_{\ell m}^* \bigr\rangle
    &= \sum_{\substack{l_1l_2}} {\Im{}} \tfrac{(2l_1 + 1)(2l_2 + 1)}{16\pi} (-1)^{\ell+l_1+l_2}
    \bigg[\frac{1}{2} \tilde{C}_{l_1}^{{}_{0}h{}_{2}h} \, (\delta g_1 + i \, \delta g_2) \, \threej{\ell,2}{l_1,-2}{l_2,0} \, \bigg\{ C_{l_2}^{TB} \, \bigg(\threej{l_{2},2}{l_1,0}{\ell,-2}
    + \threej{l_{2},-2}{l_1,0}{\ell,2}\bigg) - iC_{l_2}^{TE} \, \bigg(\threej{l_{2},2}{l_1,0}{\ell,-2}
    - \threej{l_{2},-2}{l_1,0}{\ell,2}\bigg) \bigg\}
    \\
    &+\frac{1}{4} \tilde{C}_{l_1}^{{}_{0}h{}_{0}h} \, (g_1^A + g_1^B + g_2^A + g_2^B) \, \threej{\ell,2}{l_1,0}{l_2,-2} \,
    \bigg\{ (C_{l_2}^{BE}-iC_{l_2}^{BB}) \, \bigg(\threej{l_{2},2}{l_1,0}{\ell,-2}
    + \threej{l_{2},-2}{l_1,0}{\ell,2}\bigg) - (iC_{l_2}^{EE}+C_{l_2}^{EB}) \, \bigg(\threej{l_{2},2}{l_1,0}{\ell,-2}
    - \threej{l_{2},-2}{l_1,0}{\ell,2}\bigg)\bigg\}
    \\
    &+\frac{1}{4} \tilde{C}_{l_1}^{{}_{0}h{}_{4}h} \, (g_1^A + g_1^B - g_2^A - g_2^B) \, \threej{\ell,2}{l_1,-4}{l_2,2} \,
    \bigg\{ (C_{l_2}^{BE} + iC_{l_2}^{BB}) \, \bigg(\threej{l_{2},2}{l_1,0}{\ell,-2}
    + \threej{l_{2},-2}{l_1,0}{\ell,2}\bigg) - (iC_{l_2}^{EE} -C_{l_2}^{EB}) \, \bigg(\threej{l_{2},2}{l_1,0}{\ell,-2}
    - \threej{l_{2},-2}{l_1,0}{\ell,2}\bigg) \bigg\}\bigg] \text{.}
    \label{eq:Term 3 BB Gain}
\end{split}
\end{equation}
\normalsize

Equations~\ref{eq:Term 1 Gain}--\ref{eq:Term 3 BB Gain} demonstrate the use of our formalism in providing a complete characterisation of the effects of differential gain for the experimental setup of two detector pairs. This shows how simple it is to fully characterise the effects of spin-coupled systematics using just their spin properties and the formalism developed in Sections~\ref{section:Spin Characterisation} and \ref{section:Power Spectrum Derivation}. We note again that the procedure can in principle be applied to any focal plane layout for systematics of appropriate form.

\subsection{Differential Pointing}
\label{sec:diff_point}
Differential pointing refers to an incorrect alignment of the beams of the two detectors by an angle $\rho_i$ in a direction on the sky given by the angle $\chi_i$ with respect to the orientation of the telescope with respect to North, $\psi$; see \cite{Wallisetal2016} for further details. The differenced signal observed by a pair of detectors may be written as (where we may assume flat sky co-ordinates $\{x,y \}$ due to the deviation being small)
\begin{equation}
\begin{split}
S_i &= \frac{1}{2}[d^A_i - d^B_i]
    \\
    &=\frac{1}{2}[ I(x,y)-I(x-\rho_i\sin(\psi +\chi_i) ,y-\rho_i \cos (\psi+\chi_i) )
    + \frac{1}{2}(P(x,y)e^{2i\psi}-P(x-\rho_i \sin (\psi+\chi_i),y-\rho_i\cos(\psi +\chi_i))e^{2i(\psi+\pi/2)})
    \\
    &+ \frac{1}{2}(P^*(x,y)e^{-2i\psi}-P^*(x-\rho_i \sin (\psi+\chi_i),y-\rho_i\cos(\psi +\chi_i))e^{-2i(\psi+\pi/2)})]
    \\
    &= \frac{1}{2} [P e^{2i\psi} + P^*e^{-2i\psi}] +\frac{1}{4}[(\frac{\partial I}{\partial y}-i\frac{\partial I}{\partial x})\rho_ie^{i(\psi+\chi_i)}
    +(\frac{\partial I}{\partial y}+i\frac{\partial I}{\partial x})\rho_ie^{-i(\psi+\chi_i)}]
    \\
    &-\frac{1}{8}[(\frac{\partial P}{\partial y} - i \frac{\partial P}{\partial x}) \rho_i e^{i(3\psi+\chi_i)}
    +(\frac{\partial P}{\partial y} + i \frac{\partial P}{\partial x}) \rho_i e^{i(\psi-\chi_i)}]
    -\frac{1}{8}[(\frac{\partial P^*}{\partial y} - i \frac{\partial P^*}{\partial x}) \rho_i e^{-i(\psi-\chi_i)}
    +(\frac{\partial P^*}{\partial y} + i \frac{\partial P^*}{\partial x}) \rho_i e^{-i(3\psi+\chi_i)}] \text{,}
\end{split}
\end{equation}
where in the final step we have approximated using a Taylor expansion. Using the spin raising operator $\eth = \frac{\partial }{\partial y}-i\frac{\partial }{\partial x}$ and its conjugate (the spin lowering operator, denoted by a bar), we can write the signal seen by a detector pair experiencing differential pointing as
\begin{equation}
    S_i = \frac{1}{2} \, e^{2 i \, \psi} \, P
    + \frac{1}{2} \, e^{-2 i \, \psi} \, P^*
    + \frac{\rho}{4} \, e^{i \, (\psi + \chi)} \, \eth I
    + \frac{\rho}{4} \, e^{-i \, (\psi + \chi)} \, \bar{\eth} I
    - \frac{\rho}{8} \, e^{i \, (3\psi + \chi)} \, \eth P
    - \frac{\rho}{8} \, e^{i \, (\psi - \chi)} \, \bar{\eth} P
    - \frac{\rho}{8} \, e^{i \, (-\psi + \chi)} \, \eth P^*
    - \frac{\rho}{8} \, e^{-i \, (3\psi + \chi)} \, \bar{\eth} P^* \text{,}
    \label{eq: Differential Pointing Single Measurement}
\end{equation}
where both the temperature and polarisation coupling have been included. For the two sets of detector pairs we may use equation \ref{eq:DiffSpinSyst} to write the total spin-2 signal including differential pointing as
\begin{equation}
\begin{split}
{}_2\tilde{S}^d(\Omega) &=  \tilde{h}_0 P(\Omega)
    + \frac{1}{4}\tilde{h}_1(\rho_1 \, e^{i \, \chi_1} + \rho_2 \, e^{i \, (\chi_2 - \pi/4)}) \, \eth I(\Omega)
    + \frac{1}{4}\tilde{h}_3(\rho_1 \, e^{-i \, \chi_1}
    + \rho_2 \, e^{-i \, (\chi_2 + 3\pi/4)}) \bar{\eth} I(\Omega)
    \\
    &- \frac{1}{8}\tilde{h}_{1} (\rho_1 \, e^{-i \, \chi_1}
    + \rho_2 \, e^{-i \, (\chi_2 + \pi/4)}) \, \bar{\eth} P(\Omega)
    - \frac{1}{8}\tilde{h}_{3}(\rho_1 e^{i \, \chi_1} + \rho_2 e^{i \, (\chi_2 - 3\pi/4)}) \, \eth P^*(\Omega)
    \\
    &- \frac{1}{8}\tilde{h}_{5} (\rho_1 \, e^{-i \, \chi_1} + \rho_2 \, e^{-i \, (\chi_2 + 5\pi/4)}) \, \bar{\eth} P^*(\Omega)
    - \frac{1}{8}\tilde{h}_{-1}(\rho_1 \, e^{i \, \chi_1} +\rho_2 \, e^{i \, (\chi_2 + \pi/4)}) \, \eth P(\Omega).
    \label{eq:Differential Pointing}
\end{split}
\end{equation}
The spins of the systematic terms here are not 0 or 2 and as such are not included in the standard map-making procedures, so will generally leak into the polarisation signal. While it is possible to solve for them as part of the map-making step (as we demonstrate in Appendix \ref{section:5x5 and 7x7}) in the following power spectrum example we shall assume a simple 2x2 map-making approach is employed.

Examining equation \ref{eq:Differential Pointing} the differential pointing systematic contribution may then be written in harmonic space according to the general prescription of equation \ref{eq:Spin2 General Systematic alm} as
\begin{equation}
\begin{split}
    {}_2\tilde{\Delta}_{\ell m}^{p} &= \sum_{\substack{l_1m_1\\l_2m_2}}\Bigg[
    \bigg(\frac{1}{4} \, {}_1h_{l_1m_1} \, \sqrt{l_2 \, (l_2+1)} \, \bigl(\rho_1 \, e^{i \, \chi_1} + \rho_2 \, e^{i \, (\chi_2 - \pi/4)}\bigr) \, \threej{\ell,2}{l_1,-1}{l_2,-1}
    - \frac{1}{4} \, {}_3h_{l_1m_1} \, \sqrt{l_2 \, (l_2+1)} \, \bigl(\rho_1 \, e^{-i \, \chi_1} - \rho_2 \, e^{-i \, (\chi_2 - \pi/4)}\bigr) \, \threej{\ell,2}{l_1,-3}{l_2,1}\bigg)
    \\
    & I_{l_2m_2} (-1)^m \sqrt{\tfrac{(2\ell + 1)(2l_1 + 1)(2l_2 + 1)}{4\pi}} \, \threej{\ell,-m}{l_1,m_1}{l_2,m_2}
    \\
    &+\bigg(
    \frac{1}{8} \, {}_1h_{l_1m_1} \, \sqrt{(l_2+2) \, (l_2-2+1)}
    \, \bigl(\rho_1 \, e^{-i \, \chi_1} + \rho_2 \, e^{-i
    \, (\chi_2 + \pi/4)}\bigr) \, \threej{\ell,2}{l_1,-1}{l_2,-1}{}_{2}P_{l_2m_2}
    \\
    &+\frac{1}{8} \, {}_5h_{l_1m_1} \, \sqrt{(l_2+2) \, (l_2-2+1)}
    \, \bigl(\rho_1 \, e^{-i \, \chi_1} - \rho_2 \, e^{-i
    \, (\chi_2 + \pi/4)}\bigr) \, \threej{\ell,2}{l_1,-5}{l_2,3}{}_{-2}P_{l_2m_2}
    \\
    &- \frac{1}{8} \, {}_{3}h_{l_1m_1} \, \sqrt{(l_2-2) \, (l_2+2+1)} \, \bigl(\rho_1 \, e^{i \, \chi_1} - \rho_2 \, e^{i \, (\chi_2 + \pi/4)}\bigr) \,
    \threej{\ell,2}{l_1,-3}{l_2,1}{}_{-2}P_{l_2m_2}
    \\
    &- \frac{1}{8} \, {}_{-1}h_{l_1m_1} \, \sqrt{(l_2-2) \, (l_2+2+1)} \, \bigl(\rho_1 \, e^{i \, \chi_1} + \rho_2 \, e^{i \, (\chi_2 + \pi/4)}\bigr) \,
    \threej{\ell,2}{l_1,1}{l_2,-3}{}_2P_{l_2m_2}\bigg)
    (-1)^m \sqrt{\tfrac{(2\ell + 1)(2l_1 + 1)(2l_2 + 1)}{4\pi}} \, \threej{\ell,-m}{l_1,m_1}{l_2,m_2}
    \Bigg] \text{,}
\end{split}
\end{equation}
where the effect of spin raising and lowering operators on the spectral representation have been included as $\eth ({}_{s}Y_{\ell m}) = +\sqrt{(l-s) \, (l+s+1)} \, {}_{s+1}Y_{\ell m}$ and $\bar{\eth} ({}_{s}Y_{\ell m})= -\sqrt{(l+s) \, (l-s+1)} \, {}_{s-1}Y_{\ell m}$. We may then use this to look at the effect the pointing systematic would have on the observed power spectra. Following the process described in Section \ref{section:Power Spectrum Equations}, we can derive the systematic terms due to differential pointing. The resulting expressions are provided in Appendix~\ref{section:Pointing Terms} due to space considerations.

\subsubsection{Lensing Issues}
It is known that the CMB lensing field can be represented as derivatives of the polarisation field, and thus that (differential) pointing effects can appear similar to lensing effects \citep[see e.g.][]{2009PhRvD..79l3002S,2010PhRvD..81f3512Y}. The advantage over these previous works of the comprehensive expansion employed in equation \ref{eq:Differential Pointing}, which includes leakage from the derivatives of both the temperature and the polarisation, is that we can explicitly see this correspondence. This allows us to analytically calculate the value of the "lensing-like" contribution caused by the differential pointing error for a given scan strategy.

As a result, surveys that employ pair differencing could suffer from a systematic bias to the off diagonal terms used by lensing quadratic estimators, which could result in an overestimate of the lensing amplitude. Furthermore, this differential pointing error can produce a spurious lensing curl field, which is generally expected to be zero. Interestingly, \cite{2018arXiv180706210P} reports a $4.3\sigma$ discrepancy in one of the bins of their curl reconstruction band-power amplitudes, which could potentially be explained in this way.

Our characterisation of the differential pointing including the coupling to the scan strategy provides a potential avenue to break the degeneracy between the spurious signal and the true lensing signal: knowledge of the scan strategy, and the expected size of the pointing signal, could be used to debias the lensing signal and reconstruct both signals correctly. In addition, this differential pointing systematic also contributes to spurious signals of different spin to the lensing signal, so it may be possible to quantify the level of the pointing systematic separately using these additional signals. This could then be used to remove the contamination during lensing reconstruction. We leave it to further work to explore these possibilities in detail, but point them out here as a benefit of extending the formalism to include the effects of polarisation leakage.

\subsection{Verification with simulations}
In this section we use time-ordered data (TOD) simulations to generate data using realistic CMB scan strategy information. Since the formalism begins from the detector equation \ref{eq:Detector Equation} and propagates the effect of systematics through the map, $a_{\ell m}$, and power spectrum levels, the full TOD simulations provide a solid verification that the analytic results match realistic spin-coupled systematics. To match sections \ref{sec:diff_gain} and \ref{sec:diff_point}, the simulations use two detector pairs, oriented at $45^{\circ}$ with respect to each other allowing simultaneous measurement of the $Q$ and $U$ Stokes parameters.

The TOD simulation code takes input maps of $I$, $Q$ and $U$ fields created using the {\sevensize SYNFAST} routine of the {\sevensize HEALPIX} package \citep{2005ApJ...622..759G}. The model CMB power spectra used to create the maps were those corresponding to the standard best-fitting 6-parameter $\Lambda$CDM model to the 2015 {\it Planck} results \citep{2016A&A...594A..13P}, specified by the following cosmological parameter values: $H_{0} = 67.3$, $\Omega_{b} = 0.0480$, $\Omega_{cdm} = 0.261$, $\tau = 0.066$, $n_{s} = 0.968$, $A_{s} = 2.19\times10^{-9}$. The input spectra were generated using the Boltzmann code CLASS \citep{2011JCAP...07..034B}. Primordial $B$-modes were not included ($r = 0.0$) but the input maps did include $B$-modes induced by gravitational lensing (approximated as Gaussian). The maximum multipole included when creating the input CMB maps was $\ell_{\rm max} = 4000$ at a resolution of $N_{\rm side} = 2048$, corresponding to a pixel size of 1.7 arcmin. The maps were also convolved with a Gaussian beam of Full Width at Half Maximum (FWHM) of 7 arcmin. 

Systematics were injected at the detector time stream level: gain offsets were included using equation \ref{eq:Gain Single Detector} and differential pointing was included using equation \ref{eq: Differential Pointing Single Measurement}. The {\sevensize HEALPIX} Fortran package {\sevensize SYNFAST} was used to generate the first derivatives of the temperature and polarisation fields \citep{2005ApJ...622..759G} to include the differential pointing systematic into the TOD. This is a similar approach to the way lensing effects are included in simulations in \cite{2013JCAP...09..001N}. The gain simulations used a setup where each pair of detectors experiences the same differential gain, corresponding to parameters $g^A_1 = g^A_2 = 0$ and $g^B_1 = g^B_2 = 0.01$, which models a 1\% differential gain. The third term of equation \ref{eq:Differential Gain} is consequently zeroed. The differential pointing simulations used a systematic level set to $\rho_1 = \rho_2 = 0.1$ arcmin and $\chi_1 = \chi_2 = 0.0$ radians. These levels of systematics are indicative of differential systematics seen in recent CMB surveys \citep[e.g.][]{2015ApJ...814..110B,1403.2369}.

We present results for different scan strategies, representative of a satellite survey both full sky and in the presence of a galactic mask, and both "deep" and "wide" ground based surveys. The input signal is a pure CMB sky, with no foregrounds or noise included; these effects are beyond the scope of these demonstrations but should not change the conclusions about the formalism.

\subsubsection{Satellite Mission}
Satellite scan strategies can be designed such that crossing angle coverage is maximised, as they suffer from fewer restrictions than ground based surveys do. This should limit the impact of any differential systematics whose effects vanish for an ideal scan. Here we present the results for an EPIC-like satellite survey, for the full sky and in the presence of a Galactic mask. The scan strategy of EPIC has been designed to optimise crossing angle coverage and is defined by its boresight angle ($50^\circ$), precession angle ($45^\circ$), spin period (1 min), and precession period (3 hrs). For more details see \cite{2008arXiv0805.4207B}.

\subsubsection{Ground Based Surveys}
\label{section:Ground Based}
Ground based scan strategies are far more restricted in the available crossing angles coverage due to their position on the Earth limiting their pointing options. This will result in less suppression of the differential systematics compared to that of a satellite scan. There are ways of achieving similar suppression through use of boresight rotation and/or the inclusion of a continuously rotating halfwave plate \citep{2020MNRAS.491.1960T,2018SPIE10708E..41S}.

\cite{2020MNRAS.491.1960T} demonstrated it is possible to incorporate boresight rotation of a ground-based telescope into a scan strategy to reduce certain systematics in the case of pair differencing experiments. The form of the $\tilde{h}_{k-k'}$ scan coupling terms of equation \ref{eq:orientation function} means this process can be used to zero $\tilde{h}_{k-k'}$ terms by taking observations at specific pairs of crossing angles, e.g. the $\tilde{h}_1$ term is zeroed if two observations that have crossing angles 180 degrees apart are taken within the same pixel. We note that the expansion of the formalism which we present in this paper implies that this is true of any spin-coupled signals that manifest in this way --- for single detector observations, a full focal plane, or differenced pairs --- provided it is possible to describe them using equation \ref{eq:sky signal}. As such we have shown that the use of boresight rotation to suppress systematics, as demonstrated in \cite{2020MNRAS.491.1960T}, is generally applicable to any focal plane setup and is not specific to the detector differencing setup considered by \cite{2020MNRAS.491.1960T}.

For the ground-based simulations, we have modelled ``deep" and ``wide" scan strategies using a ``synthetic scan" approach as was previously used in \cite{2020MNRAS.491.1960T}, where scan parameters can effectively be averaged within a pixel in order to speed up the TOD process. Deep surveys are designed to target the primordial $B$-mode signal which is expected to peak at relatively large angular scales of a few degrees, corresponding to multipoles $\ell \approx 80$. The wide surveys have several observational targets such as SZ, neutrinos masses, and the lensing $B$-mode power spectrum. The lensing signal peaks on angular scales of a few arcmin, corresponding to multipoles $\ell \approx 1000$. As such for the ground-based surveys in this work we use the deep survey to consider the effects of differential gain as it has a strong low-$\ell$ component, and we use the wide survey to consider the effects of differential pointing as it has a strong high-$\ell$ component \citep{2020MNRAS.491.1960T}.

The synthetic scan procedure vastly speeds up the simulation pipeline and gives comparable results to a full TOD simulation, capturing the important aspects of the scan for this work, when using simple pixel by pixel map-making as is done here. The framework uses two parameters per sky pixel to specify a scan, which may be used to quickly construct full-sky simulated maps for different scan strategies. The two parameters are the number of distinct crossing angles in each pixel, $N_{\psi}$, and the range of crossing angles, $R$. We use the values $N_{\psi} = 2$ and $R = 0.169$ radians for the wide survey and $N_{\psi} = 4$ and $R = 0.719$ radians for the deep survey setting the same value in each pixel for simplicity. These are the appropriate values for $N_{\rm side} = 2048$ but they would be slightly different for other resolutions. Allowing variation of these parameters with sky location would provide a more realistic distribution. However, we have verified using a restricted set of simulations that, for the scans considered in this work, this increase in complexity does not change the conclusions. We have also checked that the quantitative nature of the results, for the scan strategies and systematics considered here, is unaffected by the choice to use the synthetic scan approach. Further details on the use of the framework along with the sky cuts adopted for the ground-based simulations are available in \citep{2020MNRAS.491.1960T} and a detailed exploration will be provided in a forthcoming publication.

\begin{figure}
\centering
\subfloat[Results of the full sky EPIC scan strategy simulation.]{\includegraphics[width=0.4\textwidth]{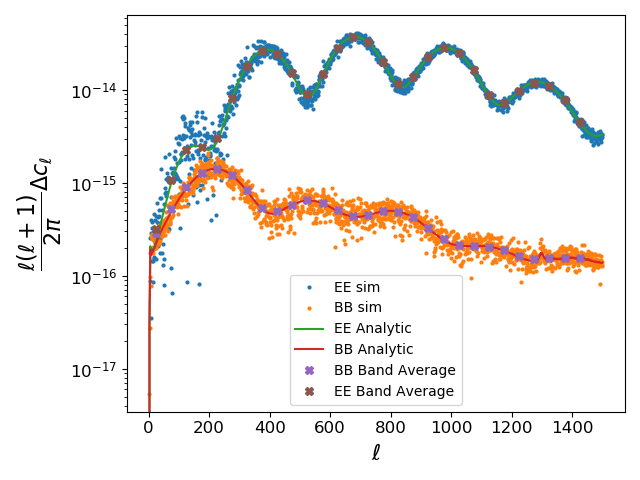}}
\subfloat[Results of the EPIC scan strategy simulation in the presence of a galactic mask.]{\includegraphics[width=0.4\textwidth]{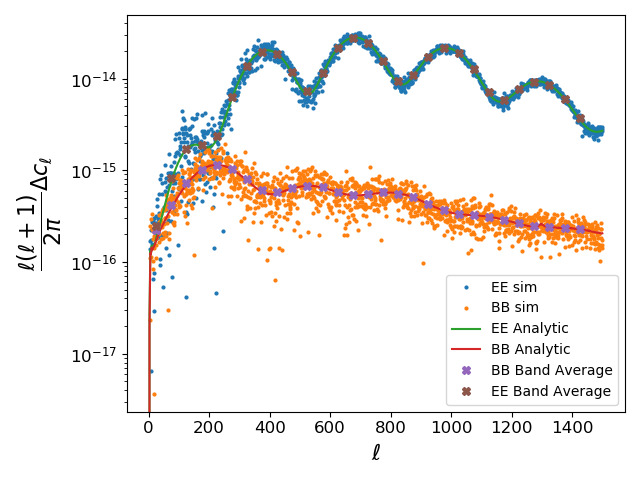}}

\subfloat[Results of a Deep ground based scan strategy simulation.]{\includegraphics[width=0.4\textwidth]{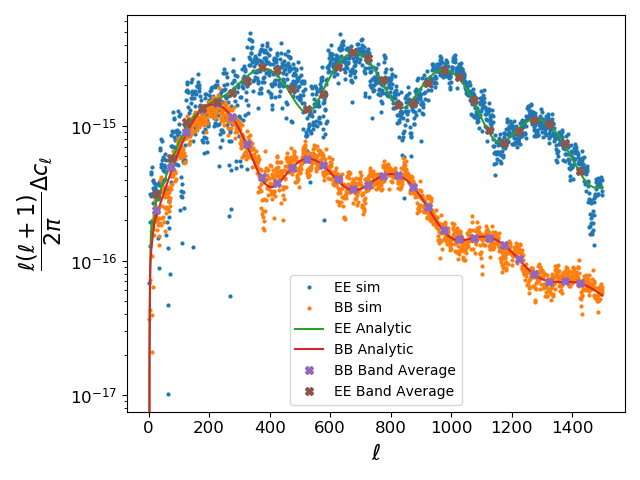}}
\caption{The effect of the differential gain on the $E$ and $B$ mode polarisation spectra. The blue and orange dots show the systematic spectrum output from from a single realisation of the simulation for the $E$-mode and $B$-mode respectively. These have been isolated by plotting the residual between the simulation output including the systematic and a simulation output including no systematic. The green ($E$-mode) and red ($B$-mode) lines show the predicted systematic terms using equations \ref{eq:Term 1 Gain}--\ref{eq:Term 3 Gain}. The predictions clearly work well on the full sky, in the presence of a mask, and for a ground based partial sky experiment, only beginning to deviate from the simulation towards the beam scale. To further emphasise the agreement of the analytic predictions with the simulation, band averages have been overplotted as purple and brown crosses, which are sourced from the average of 50 realisations and have been smoothed into bins of width $\Delta \ell = 50$; the band averages are clearly in close agreement with the analytic predictions. The difference between the leakage into the $E$ and $B$ mode spectra is driven by the large cross-term between the systematic and the sky signal, which is present in the $E$-mode for differential gain. We note the presence of a slight upward spike at $\ell\approx1300$ in the $B$-mode spectra of panels (a) and (b) which corresponds to structure present in the EPIC scan strategy (further detail available in appendix \ref{section:Importance of Terms}).}
 \label{fig:Differential Gain}
\end{figure}

\begin{figure}
\centering
\subfloat[Results of the full sky EPIC scan strategy simulation.]{\includegraphics[width=0.4\textwidth]{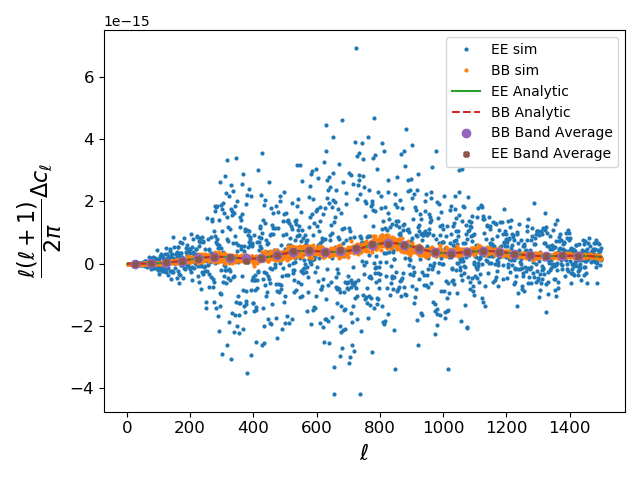}}
\subfloat[Results of the EPIC scan strategy simulation in the presence of a galactic mask.]{\includegraphics[width=0.4\textwidth]{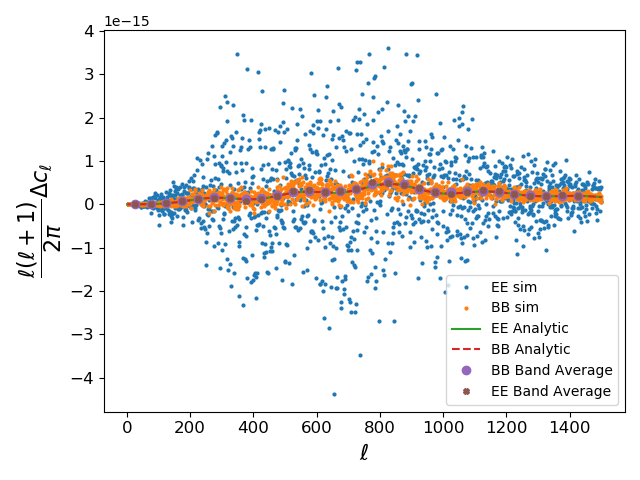}}

\subfloat[Results of a Wide ground based scan strategy simulation.]{\includegraphics[width=0.4\textwidth]{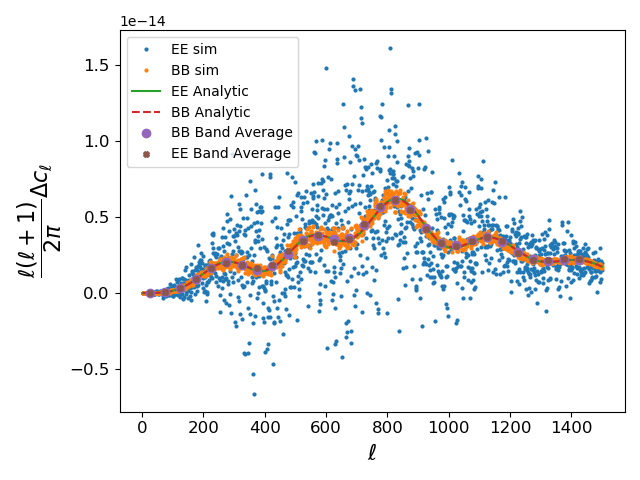}}
\caption{The effect of the differential pointing on the $E$ and $B$ mode polarisation spectra. The blue and orange dots show the systematic spectrum output from the simulation for the $E$-mode and $B$-mode respectively. These have been isolated by plotting the residual between the simulation output including the systematic and a simulation output including no systematic. The green solid line ($E$-mode) and red dashed line ($B$-mode) show the predicted systematic terms using equations \ref{eq:Term 1 Pointing}--\ref{eq:Term 3 Pointing}, the green line sits almost exactly below the red line and as such is somewhat obscured. The predictions clearly work well on the full sky, in the presence of a mask, and for a ground based partial sky experiment. To further emphasise the agreement of the analytic predictions with the simulation, band averages have been overplotted as brown crosses ($E$-mode) and purple circles  ($B$-mode), which are sourced from the average of 50 realisations and have been smoothed into bins of width $\Delta \ell = 50$; the band averages are clearly in close agreement with the analytic predictions. The limited crossing angle coverage in the case of the ground-based survey means the systematic contribution to the polarisation power spectra is larger, but the analytic formalism still captures its behaviour well, showing that it is applicable in cases of both limited and extensive crossing angle coverage. The large scatter on the $E$-mode is caused by the presence of the cross-term between the sky signal and the systematic.}
 \label{fig:Differential Pointing}
\end{figure}

\subsubsection{Results \& discussion}
The output of the simulation pipeline consists of Stokes polarisation maps recovered using a 2x2 simple binning map-making scheme to solve for $Q$ and $U$ (see Appendix \ref{section:Map-Making} for more details). As the simulations do not include noise, this map-making algorithm performs as well as would be achieved with a more optimal map-making scheme.

In order to isolate the systematic signal to compare with the analytic forms, we present results in terms of the residual pseudo power spectra, which is constructed as follows. Two cases of the TOD simulations are run using the same sky realisation, one with the relevant systematic and one in the absence of systematics. The {\sevensize HEALPIX} routine {\sevensize ANAFAST} is then applied to the output $Q$ and $U$ maps to estimate the pseudo power spectra. Note that as the input maps were beam-smoothed then these spectra are convolved with the beam, and we choose not to deconvolve this beam smoothing from the estimated power spectra. The power spectra recovered from the simulation with no systematic are then subtracted from those of the simulation with the systematic present in order to isolate the spurious signal arising from the systematic effect. This allows direct comparison to the analytic forms.

Figures \ref{fig:Differential Gain} and \ref{fig:Differential Pointing} show that the analytic predictions of the power spectra arising from the prescription of Section \ref{section:Power Spectrum Equations} (i.e. the expressions in equations \ref{eq:Term 1 Gain}--\ref{eq:Term 3 BB Gain} and in Appendix \ref{section:Pointing Terms}) clearly agree with the simulation outputs. This shows that the formalism is robust for both limited and extensive angle coverage and is thus applicable in general to CMB experiments including space based, balloon-borne, and ground-based surveys. We have demonstrated that the $\tilde{h}_{k-k'}$ scan fields of equation \ref{eq:orientation function} may be used to accurately predict the effect of systematics on the observed polarisation spectra. These quantities can thus be used to rank scan strategies according to their effectiveness in mitigating systematic effects \citep{Wallisetal2016}. This provides an additional simple selection criteria to minimise when designing scan strategies. It is common knowledge that scans should be set up such that they maximise crossing angle coverage and these $h_{k-k'}$ provide a way of quantifying the degree to which a scan has achieved this. Furthermore it is possible to use them to examine the effect that spin-coupled terms will have on the power spectra \citep{Wallisetal2016}.

There is clear variation of the $E$-mode sourced by the gain systematic in figure \ref{fig:Differential Gain} with sky area. This is in part due to the decrease in (pseudo-)power associated with a smaller survey area, but it also has a contribution from the terms in equation \ref{eq:Term 3 Gain}. The contributions to this cross-term between the sky signal and systematic are mostly zero on the full sky (see Appendix \ref{section:Cross Term}). However for the partial sky cases, particularly evident in the deep scan case which has a substantial mask, all of the cross-terms will contribute to the systematic. This causes a noticeable difference in structure.

\begin{figure}
\centering
\includegraphics[width=5.0in]{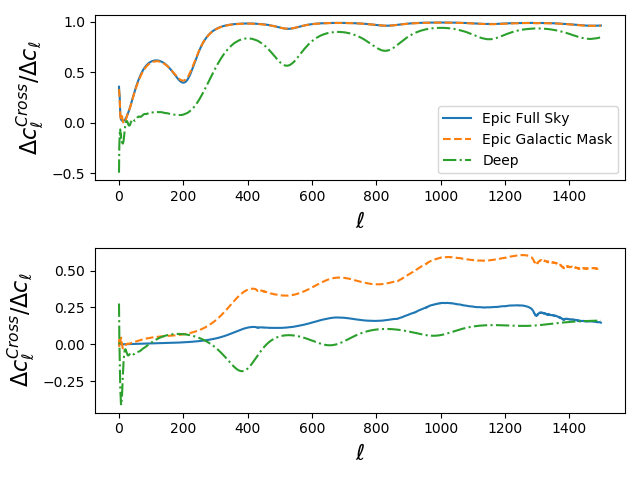}
\caption{Top panel: The ratio of $\Delta C^{Cross}_{\ell} = \frac{1}{2 \, (2\ell + 1)} \sum_{m} 4 \Re{} \bigl\langle \tilde{E}_{\ell m} \, {}^{}_2\tilde{\Delta}_{\ell m}^* \bigr\rangle$ and the total systematic contribution $\Delta C_{\ell}$ for the $E$-mode for the differential gain systematic. Bottom panel: The ratio of $\Delta C^{Cross}_{\ell} = \frac{1}{2 \, (2\ell + 1)} \sum_{m} -4 {\Im{}} \bigl\langle \tilde{B}_{\ell m} \, {}^{}_2\tilde{\Delta}_{\ell m}^* \bigr\rangle$ and the total systematic contribution $\Delta C_{\ell}$ for the $B$-mode for the differential gain systematic. The blue (solid), orange (dashed), and green (dash-dotted) lines show the cases for the EPIC full sky, EPIC in the presence of a galactic mask, and the Deep surveys respectively. Both panels show that the effect of the cross-terms in both the $E$-mode and $B$-mode cases is highly dependent on the survey, including the $\ell$ range they are likely to dominate over. Furthermore, these figures show that the cross-term between the systematic and intrinsic signal can be the dominant source of contamination --- this is evident for the $E$-mode where for all surveys considered it is the dominant source (ratio $>0.5$) of contamination for $\ell \gtrsim 300$. Perhaps more importantly for upcoming CMB surveys, the cross-term can also be the dominant source of contamination for the $B$-mode as is shown in the lower panel for the case of the EPIC survey in the presence of a galactic mask (where it becomes the dominant source of contamination at $\ell \gtrsim 1000$). We note the presence of a slight spike at $\ell\approx1300$ in both the EPIC full sky and EPIC in the presence of a galactic mask lines in the lower panel; this corresponds to structure present in the scanning strategy of the EPIC satellite (further detail available in appendix \ref{section:Importance of Terms}).}
\label{fig:Cross Term Ratios}
\end{figure}

The substantial difference between the $E$-mode and $B$-mode spectra for the differential gain systematic in figure \ref{fig:Differential Gain} is driven by the difference between the cross-terms of equations \ref{eq:Term 3 Gain} and \ref{eq:Term 3 BB Gain}. Having calculated the contributions separately, we see that for the $E$-mode the temperature leakage term present in equation \ref{eq:Term 1 Gain} dominates at low-$\ell$. Beyond this, the dominant contaminant is the direct amplification of the $E$-mode by the gain which is present in the cross-term, as is clear from figure \ref{fig:Cross Term Ratios}, where the cross-term is the dominant source of contamination to the $E$-mode for $\ell \gtrsim 300$ for all the surveys considered in this work. However, the equivalent term in the $B$-mode cross-term is proportional to the on-sky $B$-mode signal, which is a much weaker signal compared to the $E$-mode. As such, the $B$-mode systematic is dominated by the temperature leakage term present in equation \ref{eq:Term 1 Gain} up to higher $\ell$ than the $E$-mode, although the cross-term does still become comparable, and can in fact dominate, at higher-$\ell$. This is shown in the lower panel of figure \ref{fig:Cross Term Ratios}, where the cross-term is the dominant source of contamination at $\ell \gtrsim 1000$ for the EPIC survey in the presence of a galactic mask, while also contributing at non-negligible levels for all surveys at parts of the $\ell$ range of interest. The demonstration of the existence of these cross-terms between the systematic and the intrinsic signal of equations \ref{eq:Cross Term} and \ref{eq:Cross Term BB}, combined with this numerical demonstration that it is not negligible in the case of differential gain, shows that it is an important contaminant to consider for the accurate recovery of both $E$-mode and $B$-mode power spectra with upcoming surveys.

In figure \ref{fig:Differential Pointing} we see that in the case of differential pointing the spurious $B$-mode and $E$-mode signals are larger for the ground-based case. This is a consequence of the design of the wide survey. It covers a larger sky fraction than that of the deep strategy and as a consequence each pixel in the map is visited fewer times with a smaller range of crossing angles. As such the systematic is not suppressed as well by the wide ground-based scan as it is by the EPIC satellite scan strategy, which has been designed to have extensive crossing angle coverage.

In figure \ref{fig:Differential Pointing} there is a large scatter on the $E$-mode, which is caused by excess variance that is introduced by the cross-term between the sky signal and the systematic. The scatter is a consequence of cosmic variance of the quantities contributing to the cross-term. The expectation value of the cross-term is, in fact, analytically zero on the full sky, as shown in Appendix \ref{section:Cross Term}, and is small in comparison to the other terms in the partial sky case due to no scan auto-spectra appearing in the cross-term for differential pointing (equation \ref{eq:Term 3 Pointing}; this is explained further in Appendix \ref{section:Importance of Terms}). However the large scatter persists in both cases. When averaging over many CMB realisations, the scatter decreases as the terms tend towards zero in the case of full sky, and towards a negligible amplitude for the partial sky case. The scatter from the systematic is in fact smaller than the cosmic variance of the true $E$-mode unless the pointing offset is unrealistically large. The effect of this scatter is in fact captured when applying the map-based part of this formalism using equation \ref{eq:spinsyst}. We will show that this technique can be utilised to provide a fast map-based simulation of systematics that capture the effects of the scan in forthcoming work.

\FloatBarrier
\section{Conclusion}
\label{section:Conclusion}
We have presented a general formalism using the spin characterisation of spin-coupled signals to describe the effect of systematics on CMB surveys, including polarisation mixing in addition to leakage of the intensity signal. We have shown for the first time that the formalism is applicable to single detector observations and a full focal plane, in addition to pair differencing setups, provided that the spin-coupled signals can be written in the form of equation \ref{eq:spinsyst}. We have shown how to apply the formalism to both full and partial sky CMB surveys, and we have used TOD simulations  to demonstrate that  the formalism accurately captures the behaviour of spin-coupled signals up to the beam-scale.

We have provided generalised equations to describe spin-coupled signals of arbitrary spin at both map-level and harmonic level, and we propagate these through to the polarisation power spectra. Equations \ref{eq:spinsyst} and \ref{eq:Observed Signal} describe the map-based and $a_{\ell m}$-based forms of the observed signals, and equation \ref{eq:Spin2 General Systematic alm} describes the spin-2 spin-coupled systematic signal that provides the basis for considering the effects of systematics on the polarisation power spectra. However we note that the formalism could equally be applied to examine the effects on any of the power spectra since it can be applied generally to any spin signal. This also means it could in principle be used to consider effects of spin-coupled systematics on intensity mapping experiments (see Appendix \ref{section:Additional Spectra} for further detail).

The effects of a systematic on the $B$-mode and $E$-mode power spectra are given in equations \ref{eq:ClBB} and \ref{eq:ClEE} respectively, where the spurious terms can be calculated for signals of arbitrary spin using equations \ref{eq:First Term}, \ref{eq:Second Term}, \ref{eq:Cross Term}, and \ref{eq:Cross Term BB}. We point out for the first time an additional contribution of systematics to the $B$-mode and $E$-mode spectra stemming from the cross-term between the systematic and the intrinsic sky signal, which can be the dominant source of contamination in some cases. In the case of full sky surveys this cross-term will analytically disappear if the spin of the systematic signal does not match the polarisation, but can leak a large additional scatter into the spectra due to a contribution from cosmic variance. Furthermore the cross-term can survive and can even be the dominant source of contamination as long as the spin of the systematic leaked signal matches the spin-2 polarisation for a full sky survey.

We illustrate this formalism in section \ref{section:Demonstration} with the complete characterisation of differential gain and pointing systematics, validated by TOD simulations. We build on previous work to incorporate both intensity leakage and mixing between the polarisation signals, showing the effects of the cross-term, and demonstrating the ability to predict systematics in both partial and full sky cases. A particularly interesting result is the demonstration that polarisation mixing from differential pointing closely imitates the lensing signal. This is an important contaminant to consider given the effect it could have on lensing estimation, which we shall examine more in future work.

One application of this formalism is to optimise scan strategy design for space-based CMB experiments. This was done for full sky space-based surveys in \cite{Wallisetal2016}. Our extension of the formalism to be applicable to partial sky means this can now be extended to aid scan strategy design for balloon-borne and ground-based experiments. Another application of elucidating systematics in terms of their spin properties is shown in Appendix \ref{section:Map-Making}, where we demonstrate that a straightforward extension of simple binning map-making can be effective in removing spin-coupled systematics during the map-making process (a similar approach in the Fourier domain is described in \citealt{2015MNRAS.453.2058W}). The inclusion of additional spins in the map-making equation successfully separates spin-coupled systematics from the signals of interest. We have shown this to be the case for differential gain by extending the 2x2 map making to include a spin-0 row to capture the temperature leakage. Additionally we have shown this for the spin-1 differential pointing systematic that leaks the derivative of the temperature field into the polarisation, and the spin-1 and 3 systematics that leak the derivative of the polarisation fields, where the systematics were successfully isolated from the signals of interest by including appropriate spins at the map-making stage.

We reiterate that the generalisation of this approach means that the partial boresight rotation results of \cite{2020MNRAS.491.1960T} will apply even when there is no detector differencing. Scan coupling terms that appear in specific pairs of crossing angles (e.g. the $\tilde{h}_1$ term is zeroed if a pair observations which have crossing angles 180 degrees apart) will zero their associated spin-coupled systematic.

Finally, we point out that equation \ref{eq:spinsyst} provides a map-based description of the effect of spin-coupled systematics on the observed signal. Our formalism thus provides a path to fast map-based simulations of systematics that capture the effects of a full TOD simulation including the structure of the scan. In a forthcoming paper we will lay out this framework and show that with prior knowledge of the scan strategy the salient features of spin-coupled systematics can be reproduced.

\section*{Acknowledgements}
NM is supported by a STFC studentship. DBT acknowledges support from Science and Technology Facilities Council (STFC) grants ST/P000649/1 and ST/T000414/1. MLB acknowledges support from STFC grant ST/T007222/1.\\
\textbf{Data Availability:} The data underlying this article will be shared on reasonable request to the corresponding author.
The reader can be provided with the appropriate scanning strategy maps i.e. $\cos(k \psi)$ and $\sin(k \psi)$ to be able to replicate the results. The codes to calculate the contaminated power spectra are written in python and make use of {\sevensize HEALPY} \citep{Zonca2019} and {\sevensize SHTOOLS} \citep{2018GC007529} when calculating the spin-weighted spherical harmonics and Wigner-3j symbols.

\bibliographystyle{mnras}
\bibliography{Formalism}

\FloatBarrier
\appendix

\section{Insights Into Map Making}
\label{section:Map-Making}
We may parameterise some arbitrary signal using the detector equation as 
\begin{equation}
    S(\psi,\Omega) = \sum_{s\geq0} \alpha_s\cos(s\psi) - \beta_s\sin(s\psi).
\end{equation}
Any spin $s$ that this includes may be constrained during simple binned map making by the inclusion of spin $s$ rows in the map making equation, as we will explore below; we will first connect the spin decomposition formalism to standard map making. We can rewrite this signal as
\begin{equation}
    S(\psi,\Omega) = \sum_{s\geq0} \frac{1}{2}[(\alpha_s+i\beta_s)(\Omega)e^{si\psi} + (\alpha_s-i\beta_s)(\Omega)e^{-si\psi}].
\end{equation}
The total signal of spin $k$ detected in a pixel in Fourier space may then be written as a convolution using equation \ref{eq:spinsyst} as
\begin{equation}
    {}_{k}\tilde{S}^{d}(\Omega) = \sum_{s\geq0}\tilde{h}_{k-s}(\Omega){}_{s}\tilde{S}(\Omega) + \tilde{h}_{k+s}(\Omega) {}_{-s}\tilde{S}(\Omega) \text{,}
\end{equation}
where ${}_{s}\tilde{S} = \frac{1}{2}(\alpha_s+i\beta_s)$ and ${}_{-s}\tilde{S} = \frac{1}{2}(\alpha_s-i\beta_s)$. As such, including $\sin(s)$ and $\cos(s)$ terms in the map making solves for both $\pm s$ terms of the Fourier series representation.

For the simple case of no systematics, the sky signal is given by equation \ref{eq:sky signal}, with spin-0 and spin-2 fields present related to the $I$, $Q$, and $U$ signals that we wish to isolate. The simple binning approach to map making reconstructs these fields using \cite[e.g.][]{brown2009}
\begin{equation}
\begin{pmatrix}I\\Q\\U\end{pmatrix}
=
\begin{pmatrix}1&\langle \cos(2\psi_{j}) \rangle&\langle \sin(2\psi_{j}) \rangle\\\langle \cos(2\psi_{j}) \rangle&\langle \cos^{2}(2\psi_{j}) \rangle&\langle \cos(2\psi_{j})\sin(2\psi_{j}) \rangle\\\langle \sin(2\psi_{j}) \rangle&\langle \sin(2\psi_{j})\cos(2\psi_{j}) \rangle&\langle \sin^{2}(2\psi_{j}) \rangle\end{pmatrix}^{-1}
\begin{pmatrix}\langle d_{j} \rangle\\\langle d_{j}\cos(2\psi_{j}) \rangle\\\langle d_{j}\sin(2\psi_{j}) \rangle\end{pmatrix} \text{,}
\label{eq:3x3 map making}
\end{equation}
where the angle brackets $\langle \rangle$ denote an average over the $d_j$ measurements in a pixel, each of which has an associated angle $\psi_j$. The terms on the right hand side of this equation can be related to our approach using
\begin{align}
&\langle d_j \rangle=\frac{1}{N}\sum_j d(\psi_j)=_0\tilde{S}^d\\
&\langle d_j \cos(k\psi)\rangle ={\Re{}}\left({}_k\tilde{S}^d\right)\label{eq:dtosdictionary}\\
&\langle d_j \sin(k\psi)\rangle={\Im{}}\left({}_k\tilde{S}^d\right) \text{,}
\label{eq:dtosdictionary2}
\end{align}
showing why $_2\tilde{S}^d$ is the important quantity to consider when evaluating the effect of systematics on polarisation maps. In particular, the vector of $\langle d_j \cos(k\psi) \rangle$ and $\langle d_j \sin(k\psi) \rangle$ terms on the right hand side (RHS) of the map-making equation (e.g. equation \ref{eq:3x3 map making}) defines which spins are being looked for in the time ordered data. The vector of quantities on the left hand side (LHS) defines the list of signals that are being extracted by taking into account the knowledge of what the survey did. In this sense, the vector on the RHS maps to the signal ${}_k\tilde{S}^d$ and that on the LHS maps to ${}_{k'}\tilde{S}$ of equation \ref{eq:spinsyst}. Therefore, we can use the spin decomposition in the formalism to choose additional terms for inclusion in the standard approach to map making, showing another benefit of representing systematics using this approach.

As an example, consider a sky signal that is made up of the standard Stokes parameters and some spin-1 systematic. The spin zero temperature would contribute a term $\alpha_0 = I$, the spin-2 polarisation $P=Q+iU$ would contribute terms $\alpha_2 = Q$ and $\beta_2 = U$, and the spin-1 contaminant $Z_1=Z^Q_1+iZ^U_1$ would contribute terms $\alpha_1 = Z^Q_1$ and $\beta_1 = Z^U_1$ giving
\begin{equation}
    S(\psi,\Omega) =\frac{1}{2}[Ie^{0i\psi} + Ie^{-0i\psi} + (Z^Q_1+iZ^U_1)(\Omega)e^{i\psi} + (Z^Q_1-iZ^U_1)(\Omega)e^{-i\psi} + (Q+iU)(\Omega)e^{2i\psi} + (Q-iU)(\Omega)e^{-2i\psi}].
\end{equation}
The total detected spin-$k$ signal may then be expressed as
\begin{equation}
    {}_{k}\tilde{S}^{d} = \tilde{h}_{k}I + \frac{1}{2}[\tilde{h}_{k-1}(Z^Q_1+iZ^U_1) + \tilde{h}_{k+1}(Z^Q_1-iZ^U_1) + \tilde{h}_{k-2}(Q+iU) + \tilde{h}_{k+2}(Q-iU)].
\end{equation}
Examining this equation we see that the non-$h$ fields are of spin-0, $\pm1$, and $\pm2$. This means that the simple binning approach to map-making can be extended to include spin-1 ($\cos \psi$ and $\sin \psi$) terms, in addition to the usual spin-0 and spin-2 terms, in order to solve for all the contributions to the signal. We will demonstrate how this works in the rest of this appendix.

Note that a further benefit of the formalism here is that, by using prior scan information to calculate the scan fields $\tilde{h}_{k\pm s}$, one can check how well suppressed the systematics coupled by these terms should be. This aids evaluation of which terms are likely to be numerically important to include in the map making process \citep{2015MNRAS.453.2058W,Wallisetal2016}.

\subsection{Removing Spin-Coupled Systematics With Map Making}
\label{section:5x5 and 7x7}
Although the spin 0 and 2 signals are usually separated in the map making process, any signals of other spin that are present will still contaminate the temperature and polarisation signals. These signals could be sourced in a variety of ways but we will concentrate on spin-coupled systematics in this work. The systematic signals of other spin may be solved for during binned map making by simply adding additional rows which correspond to their spin. Similar attempts to reconstruct spin coupled systematics and remove them from the polarisation signal by their inclusion in the map-making process have been made in \cite{2015MNRAS.453.2058W} \footnote{Note that the presence of polarisation leakage due to a differential pointing systematic as in equation \ref{eq:Differential Pointing} would mean additional $\tilde{h}_{k \pm s}$ terms would need to be included in the work of \cite{2015MNRAS.453.2058W}, where only the temperature leakage terms were considered.}. In this case they use the Fourier series components, reconstructing each of the $\tilde{h}_{k \pm s}$. Here, we instead explore a real space approach that produces similar results through including spin-s rows in the simple binned map making.

We note that the incorporation of additional rows will make the process of inverting the matrix more costly. There will also be an associated noise penalty which will increase the errors on the final observed power spectra \citep{2015MNRAS.453.2058W}. However we leave this to further work to explore and rather use this section to point out that it is possible to reconstruct spin-coupled systematics by including the appropriate spin in the map making process. In addition, both methods require the inversion of a matrix that needs to be well conditioned. We do not examine this issue in detail here, but note that it can be achieved with a scan strategy that includes sufficient crossing angle coverage.

Both this approach, and that in \cite{2015MNRAS.453.2058W}, require a ``stable'' spin-coupled systematic, by which we mean a systematic field that is a well defined function of only the crossing angle within each pixel. Any additional dependence of the systematic on time, local temperature, or any other variables, will introduce an apparent ``scatter'' into the systematic when it is written as a function of crossing angle only within each pixel, meaning that the systematic will no longer be a one-to-one function of the crossing angle, and potentially breaking the simple spin decomposition as a result. If this apparent scatter of the systematic varies too much within a pixel, then the resulting failure of the simple spin decomposition will prevent the additional map-making terms from reconstructing and removing the contaminating signal. However, if the variation within a pixel is small, then there will still be a strong spin-$s$ component to the field that can still be reconstructed and removed. For example, if the apparent scatter comes from an RA and Dec dependence (or equivalently an azimuth and elevation dependence for a ground based survey), it would naturally be small within a pixel. 

We note that allowing for apparent scatter in each pixel potentially increases the number and realism of systematics that the formalism presented here can apply to. For example, it may be possible to model an absolute pointing error (for some cases) as a scatter around a spin-1 signal (which models the stable differential pointing well). We explore the sensitivity of the map-making approach to the apparent scatter in appendix \ref{section:Unstable Systematics}, but leave a thorough exploration to further work.

\subsubsection{Spin-1 Systematics}
\label{section:spin1mapmaking}
The map making of equation \ref{eq:3x3 map making} is easily extended to solve for spin-1 signals by adding in $Z^Q_1$ and $Z^U_1$ terms to capture the spin 1 signal as
\begin{equation}
\begin{pmatrix}I\\Z^Q_{1}\\Z^U_{1}\\Q\\U\end{pmatrix}
=
\begin{pmatrix}
1&\langle \cos(\psi_{i}) \rangle&\langle \sin(\psi_{i}) \rangle&\langle \cos(2\psi_{i}) \rangle&\langle \sin(2\psi_{i}) \rangle
\\\langle \cos(\psi_{i}) \rangle&\langle \cos^{2}(\psi_{i}) \rangle&\langle \cos(\psi_{i})\sin(\psi_{i}) \rangle&\langle \cos(\psi_{i})\cos(2\psi_{i}) \rangle&\langle \cos(\psi_{i})\sin(2\psi_{i}) \rangle
\\\langle \sin(\psi_{i}) \rangle&\langle \sin(\psi_{i})\cos(\psi_{i}) \rangle&\langle \sin^{2}(\psi_{i}) \rangle&\langle \sin(\psi_{i})\cos(2\psi_{i}) \rangle&\langle \sin(\psi_{i})\sin(2\psi_{i}) \rangle
\\\langle \cos(2\psi_{i}) \rangle&\langle \cos(2\psi_{i})\cos(\psi_{i}) \rangle&\langle \cos(2\psi_{i})\sin(\psi_{i}) \rangle&\langle \cos^{2}(2\psi_{i}) \rangle&\langle \cos(2\psi_{i})\sin(2\psi_{i}) \rangle
\\\langle \sin(2\psi_{i}) \rangle&\langle \sin(2\psi_{i})\cos(\psi_{i}) \rangle&\langle \sin(2\psi_{i})\sin(\psi_{i}) \rangle&\langle \sin(2\psi_{i})\cos(2\psi_{i}) \rangle&\langle \sin^{2}(2\psi_{i}) \rangle
\end{pmatrix}^{-1}
\begin{pmatrix}\langle d_{i} \rangle
\\\langle d_{i}\cos(\psi_{i}) \rangle
\\\langle d_{i}\sin(\psi_{i}) \rangle
\\\langle d_{i}\cos(2\psi_{i}) \rangle
\\\langle d_{i}\sin(2\psi_{i}) \rangle\end{pmatrix}.
\label{eq:5x5 map making}
\end{equation}
An example of a spin-1 systematic is the differential pointing terms including the $\eth T(\Omega)$ and $\bar{\eth} T(\Omega)$ fields as in equation \ref{eq:Differential Pointing}. In order to remove this contaminant from the other spin signals, they can be solved for during map making using the approach with the 5x5 matrix inversion as in equation \ref{eq:5x5 map making}. This will separate the spin 1 systematic signals from the spin-0 (temperature) and spin-2 (polarisation) signals.

\begin{figure}
  \centering
 \includegraphics[width=5.0in]{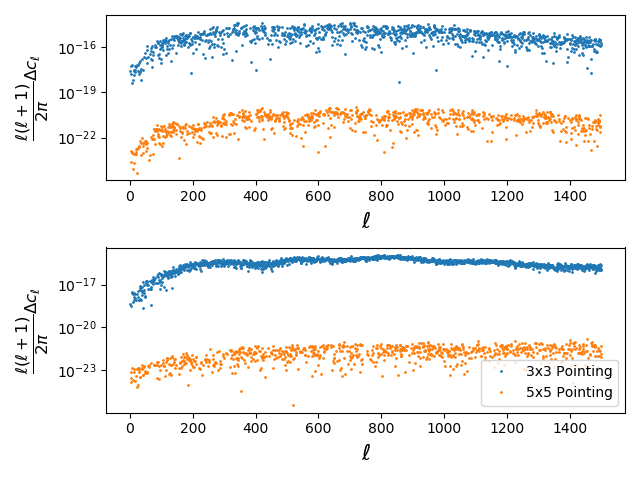}
\caption{The effect of the differential pointing on the $E$-mode (top panel) and $B$-mode (bottom panel) polarisation pseudo spectra for a simulation including a spin-1 systematic. The blue and orange dots show the systematic spectrum output from the simulation for the 3x3 and 5x5 map-making techniques respectively. These have been isolated by plotting the residual between the simulation output including the systematic and a simulation output including no systematic. A differential pointing of $\rho_1=\rho_2=0.1'$ between the two detectors in a pair is used for the simulations following equation \ref{eq:Differential Pointing}, but note that for simplicity only the spin-1 temperature leakage has been included. The systematic leakage is essentially removed entirely from both the $E$- and $B$-modes when using the 5x5 map making that incorporates the spin-1 rows. This is because the spin-1 rows have reconstructed the spin-1 pointing systematic, preventing it from leaking into the spin-2 polarisation. 
}
\label{fig:DiffPointTemp}
\end{figure}

In figure \ref{fig:DiffPointTemp} we show the efficacy of this approach using some simple simulations. We used the same ``synthetic scan" approach as in Section \ref{section:Ground Based} for the wide scan-strategy, but with $N_{\rm side} = 512$ and the parameter values $N_{\psi}=4$ and $R=0.169$ (as the 5x5 map making is not well conditioned for the original wide scan parameters\footnote{We also excluded some ill-conditioned pixels using the condition number of the matrix in the map-making process.}). As we are only concerned with demonstrating that simple binning map-making can remove spin-coupled systematics by the inclusion of additional spins, the synthetic scan approach is sufficient. In addition to the usual $I$, $Q$ and $U$ terms, in these simulations we include only the spin-1 temperature leakage systematic from differential pointing. We extend this in the next section. The simulations presented in figure \ref{fig:DiffPointTemp} show that including the spin-1 terms in map making essentially completely removes the leakage of the spin-1 systematic into the polarisation spectra, making this a promising method to explore for upcoming CMB surveys.

\subsubsection{Multiple Spin Systematics}
It is possible to extend this method to any number of spins provided there are sufficient redundancies in the scanning strategy. Note that the less optimal the scan strategy, the less well conditioned the matrix inversion will be, as such necessitating cuts of ill-conditioned pixels \citep{2009A&A...506.1511K,2017A&A...600A..60P}.

When considering how the differential pointing systematic manifests, we see that its interaction with the Stokes parameters leads to both spin-1 and spin-3 systematic signals being present as shown in equation \ref{eq:Differential Pointing}. This is due to coupling of the scan with the derivative of both the intensity and polarisation signals. As such these signals can be separated from the signals of interest by including additional rows in the map making process to solve for both spin-1 and spin-3 signals as
\footnotesize
\begin{equation}
\begin{pmatrix}I\\Z^Q_{1}\\Z^U_{1}\\Q\\U\\Z^Q_{3}\\Z^U_{3}\end{pmatrix}
=
M^{-1}
\begin{pmatrix}\langle d_{i} \rangle
\\\langle d_{i}\cos(\psi_{i}) \rangle
\\\langle d_{i}\sin(\psi_{i}) \rangle
\\\langle d_{i}\cos(2\psi_{i}) \rangle
\\\langle d_{i}\sin(2\psi_{i}) \rangle
\\\langle d_{i}\cos(3\psi_{i}) \rangle
\\\langle d_{i}\sin(3\psi_{i}) \rangle\end{pmatrix} \text{,}
\end{equation}
\normalsize
where
\footnotesize
\begin{equation}
    M = \begin{pmatrix}
1&\langle \cos(\psi_{i}) \rangle&\langle \sin(\psi_{i}) \rangle&\langle \cos(2\psi_{i}) \rangle&\langle \sin(2\psi_{i}) \rangle&\langle \cos(3\psi_{i}) \rangle&\langle \sin(3\psi_{i}) \rangle
\\\langle \cos(\psi_{i}) \rangle&\langle \cos^{2}(\psi_{i}) \rangle&\langle \cos(\psi_{i})\sin(\psi_{i}) \rangle&\langle \cos(\psi_{i})\cos(2\psi_{i}) \rangle&\langle \cos(\psi_{i})\sin(2\psi_{i}) \rangle&\langle \cos(\psi_{i})\cos(3\psi_{i}) \rangle&\langle \cos(\psi_{i})\sin(3\psi_{i}) \rangle
\\\langle \sin(\psi_{i}) \rangle&\langle \sin(\psi_{i})\cos(\psi_{i}) \rangle&\langle \sin^{2}(\psi_{i}) \rangle&\langle \sin(\psi_{i})\cos(2\psi_{i}) \rangle&\langle \sin(\psi_{i})\sin(2\psi_{i}) \rangle&\langle \sin(\psi_{i})\cos(3\psi_{i}) \rangle&\langle \sin(\psi_{i})\sin(3\psi_{i}) \rangle
\\\langle \cos(2\psi_{i}) \rangle&\langle \cos(2\psi_{i})\cos(\psi_{i}) \rangle&\langle \cos(2\psi_{i})\sin(\psi_{i}) \rangle&\langle \cos^{2}(2\psi_{i}) \rangle&\langle \cos(2\psi_{i})\sin(2\psi_{i}) \rangle&\langle \cos(2\psi_{i})\cos(3\psi_{i}) \rangle&\langle \cos(2\psi_{i})\sin(3\psi_{i}) \rangle
\\\langle \sin(2\psi_{i}) \rangle&\langle \sin(2\psi_{i})\cos(\psi_{i}) \rangle&\langle \sin(2\psi_{i})\sin(\psi_{i}) \rangle&\langle \sin(2\psi_{i})\cos(2\psi_{i}) \rangle&\langle \sin^{2}(2\psi_{i}) \rangle&\langle \sin(2\psi_{i})\cos(3\psi_{i}) \rangle&\langle \sin(2\psi_{i})\sin(3\psi_{i}) \rangle
\\\langle \cos(3\psi_{i}) \rangle&\langle \cos(3\psi_{i})\cos(\psi_{i}) \rangle&\langle \cos(3\psi_{i})\sin(\psi_{i}) \rangle&\langle \cos(3\psi_{i})\cos(2\psi_{i}) \rangle&\langle \cos(3\psi_{i})\sin(2\psi_{i}) \rangle&\langle \cos^{2}(3\psi_{i}) \rangle&\langle \cos(3\psi_{i})\sin(3\psi_{i}) \rangle
\\\langle \sin(2\psi_{i}) \rangle&\langle \sin(3\psi_{i})\cos(\psi_{i}) \rangle&\langle \sin(3\psi_{i})\sin(\psi_{i}) \rangle&\langle \sin(3\psi_{i})\cos(2\psi_{i}) \rangle&\langle \sin(3\psi_{i})\sin(2\psi_{i}) \rangle&\langle \sin(3\psi_{i})\cos(3\psi_{i}) \rangle&\langle \sin^{2}(3\psi_{i}) \rangle
\end{pmatrix}\text{,}
\end{equation}
\normalsize
and $Z^Q_1$ and $Z^U_1$ are contributed to by both temperature and polarisation spin-1 systematic terms, while the $Z^Q_3$ and $Z^U_3$ signals only contain polarisation spin-3 systematic signals.

\begin{figure}
  \centering
 \includegraphics[width=5.0in]{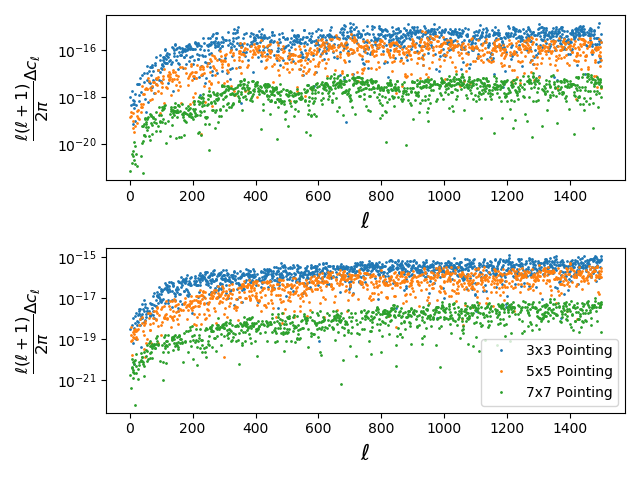}
\caption{The effect of the differential pointing on the $E$-mode (top panel) and $B$-mode (bottom panel) polarisation pseudo spectra for a simulation including spin-1 and spin-3 systematics. The blue, orange, and green dots show the systematic spectrum output from the simulation for the 3x3, 5x5, and 7x7 map-making techniques respectively. These have been isolated by plotting the residual between the simulation output including the systematic and a simulation output including no systematic. A differential pointing of $\rho_1=\rho_2=0.1'$ between the two detectors in a pair is used for the simulations following equation \ref{eq:Differential Pointing}, with both the temperature and polarisation leakage terms included. When using a 3x3 map making scheme we see that leakage from the spin-1 and spin-3 pointing signals will occur; this is reduced by the 5x5 map making which separates the spin-1 contribution, but not removed due to the spin-3 terms. The 7x7 map making approach separates both the spin-1 and spin-3 parts from the spin-2 signal and thus almost completely reduces the leakage due to the differential pointing systematic.}
\label{fig:DiffPointAll}
\end{figure}

The simulations for figure \ref{fig:DiffPointAll} use the same synthetic scan approach as in Section \ref{section:spin1mapmaking}, with $N_{\rm side} = 512$, $N_{\psi}=4$ and $R=0.169$, and we again make appropriate pixel cuts based on the condition number of the inversion of the matrix involved in the map-making. In this case we include all of the differential pointing systematics that cause both temperature and polarisation leakage as in equation \ref{eq:Differential Pointing}.

Figure \ref{fig:DiffPointAll} shows that the 7x7 approach works well in removing both the spin 1 and 3 pointing systematics from the polarisation spectra for a stable pointing systematic. It also shows that the 5x5 approach succeeds in removing the spin-1 component, but as expected the spin-3 pointing component survives this process.

\subsubsection{Unstable Systematics}
\label{section:Unstable Systematics}
We have shown that stable spin-coupled systematics of well defined spin can be solved for by extending simple binning map making. As mentioned earlier, if we consider systematics that do not vary solely with crossing angle, further difficulties arise, and the inclusion of additional rows to solve for spin-$s$ signals in the map-making may no longer remove these systematic signals. Nonetheless, if the variation within a pixel is small, then there will still be a strong spin-$s$ component to the field. We will examine this issue quantitatively with a ``toy'' systematic inspired by our differential pointing error and considerations of an absolute pointing error.

Consider a detector that has some absolute pointing offset. We may write this as an incorrect alignment of the beams of the detector by an angle $\rho(X)$ in a direction on the sky given by the angle $\chi(X)$ with respect to the orientation of the telescope with respect to North, $\psi$. Here, the $(X)$ denotes that unlike in previous work, both $\rho$ and $\chi$ can differ between measurements due to changes over time, in the atmosphere or in the environment of the telescope. The observed signal by a single detector including an absolute pointing offset may be written (where we may assume flat sky co-ordinates $\{x,y \}$ for a sufficiently small deviation and we shall only consider the temperature leakage for simplicity) as
\begin{equation}
d_i = I(x-\rho(X) \sin (\psi+\chi(X)),y-\rho(X)\cos(\psi +\chi(X)) ) 
    + Q(x,y)\cos(2\psi)) - U(x,y)\sin(2\psi) \text{,}
\end{equation}
which provided the pointing offset is small can be approximated using a Taylor expansion as
\begin{equation}
d = I(x,y)-\frac{\partial I}{\partial x}\rho(X) \sin (\psi+\chi(X))-\frac{\partial I}{\partial y}\rho(X)\cos(\psi +\chi(X)) + Q(x,y)\cos(2\psi) - U(x,y)\sin(2\psi).
\label{eq:PointingSyst}
\end{equation}

For simplicity in our demonstration we shall set the angle $\chi(X)=0$ for all measurements, and only consider a variation of $\rho(X)$ with each measurement. This means that the observations have some systematic described by $\rho(X)$ which appears to be spin-1, but any variation of $\rho$ between measurements in a pixel will mean that the average systematic signal is no longer purely spin-1, and is possibly not even well described by a finite set of spins.

To see to what extent this ``toy'' unstable systematic can be mitigated by the map-making process, we ran simulations as in section \ref{section:spin1mapmaking}, adjusted to use full sky and with parameters $N_{\psi}=10$ and $R=\pi$. The random pointing offset for each measurement is drawn from a uniform distribution centred on $\rho = 1.0$ arcmin. The levels of scatter introduced to the pointing offset were 0.01 arcmin, 0.1 arcmin, and 1.0 arcmin. These are not necessarily realistic levels, but have been chosen in order to test whether the inclusion of additional spins in a simple binning map-making approach can help suppress a systematic of unstable spin.

\begin{figure}
  \centering
 \includegraphics[width=4.5in]{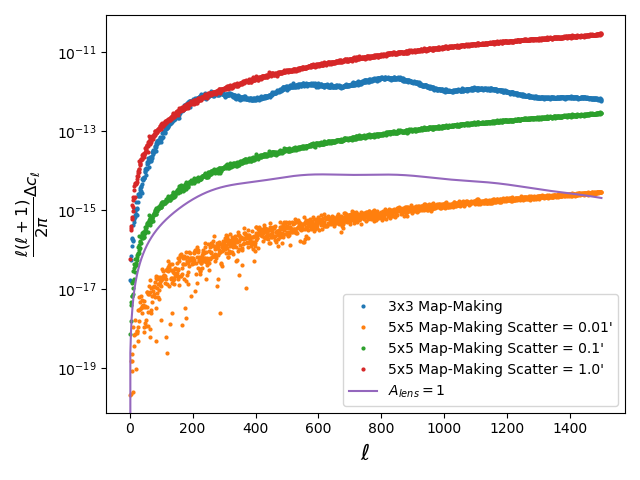}
\caption{The effect of the pointing offset on the $B$-mode polarisation pseudo spectrum for a simulation including an unstable systematic. The blue dots show the systematic spectrum output from the simulation for the 3x3 map-making technique for a pointing offset of $1.0'$ (following equation \ref{eq:PointingSyst}) and a uniform scatter of $0.01'$ (the $0.1'$, and $1.0'$ do not differ much for the 3x3 map-making so we plot only one). The orange, green and red points show the systematic spectrum output from the simulation for the 5x5 map-making technique for a pointing offset of $1.0'$, with a uniform scatter of $0.01'$, $0.1'$, and $1.0'$ respectively. These have been isolated by plotting the residual between the simulation output including the systematic and a simulation output including no systematic. For the smaller levels of scatter, the 5x5 map-making approach suppresses the systematic levels well since there is still a strong spin-1 component to the systematic signal. However, as the scatter increases, we see that the spin-1 component is no longer the dominant feature of the contamination and the 5x5 map making no longer suppresses the systematic. The theory lensing $B$-mode spectrum is also plotted for $A_{\rm lens}=1$ (with no primordial signal included) in order to compare to the levels to the systematic signals.}
\label{fig:UniformScatter}
\end{figure}

In figure \ref{fig:UniformScatter} we have plotted the simulation outputs using 3x3 and 5x5 map making.  We see that for smaller scatters there is still a strong spin-1 component to the pointing systematic, and as such the 5x5 map making does remove a considerable amount of the contaminant. As the scatter increases the 5x5 map making no longer does a good job of removing the systematic, as the spin-1 nature of the signal has been lost to the scatter. This is the first step in considering the question of how valuable this approach is if systematics are not solely functions of the crossing angle; we leave it to future work to examine this further.

\subsection{Pair Differencing - Differential Gain Issues}
\label{section:2x2 vs 3x3}
It is clear from equation \ref{eq:Differential Gain} that spin-0 signals can contribute to the spin-2 signal through a $\tilde{h}_2$ scan coupling term. The spin-0 temperature field is nominally zeroed through pair differencing. However systematics that couple the temperature directly to polarisation such as gain mismatch will still contribute a non-negligible spin-0 signal. Since the spin-0 signal has nominally been removed by the differencing process, it is common to remove the spin-0 row in the map making, reducing the 3x3 matrix inversion (see equation \ref{eq:3x3 map making}), to a less expensive 2x2 inversion,
\begin{equation}
\begin{pmatrix}Q\\U\end{pmatrix}
=
\begin{pmatrix}\langle \cos^{2}(2\psi_{i}) \rangle&\langle \cos(2\psi_{i})\sin(2\psi_{i}) \rangle\\\langle \sin(2\psi_{i})\cos(2\psi_{i}) \rangle&\langle \sin^{2}(2\psi_{i}) \rangle\end{pmatrix}^{-1}
\begin{pmatrix}\langle d_{i}\cos(2\psi_{i}) \rangle\\\langle d_{i}\sin(2\psi_{i}) \rangle\end{pmatrix}.
\label{eq:2x2 map making}
\end{equation}
This is essentially the reverse process of the extensions introduced into the map-making process earlier in this appendix.

Equation \ref{eq:Differential Gain} shows that in the presence of differential gain, the spin-2 signal has systematic contributions from a combination of leakage from the temperature signal and amplification of the polarisation signals. Following the logic of our formalism, it is clear that any gain mismatch that results in leakage of the temperature signal into the polarisation, would be prevented if the matrix was not reduced in size. A similar argument is presented in \cite{2014ApJ...794..171P}.

\begin{figure}
  \centering
 \includegraphics[width=5.0in]{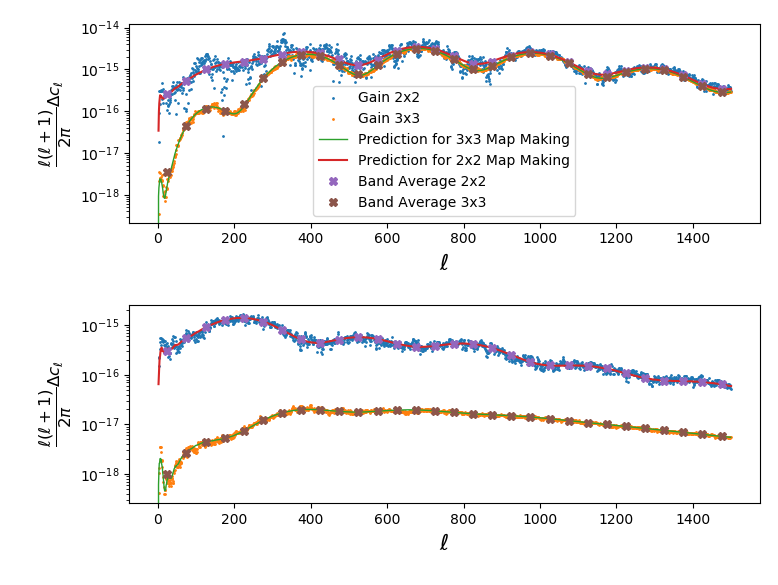}
\caption{The effect of the differential gain on the $E$-mode (top panel) and $B$-mode (bottom panel) polarisation pseudo spectra. The blue and orange dots show the systematic spectrum output from the simulation for the 2x2 and 3x3 map-making techniques respectively. These have been isolated by plotting the residual between the simulation output including the systematic and a simulation output including no systematic. A 1\% differential gain between the two detectors in a pair is used for the simulations following equation \ref{eq:Differential Gain}; see text for further details. The expected level of the systematic according to the spin decomposition in our formalism is shown as red and green lines for the 2x2 and 3x3 map-making cases respectively. The leakage that occurs in the 2x2 map-making case contains both leakage from the differential gain, and an amplification term from the absolute gain effects that is present in the cross term (equations \ref{eq:Term 3 Gain} and \ref{eq:Term 3 BB Gain}). The temperature leakage from the differential gain clearly dominates for the $B$-mode spectrum for the whole $\ell$ range in this case, while the cross term dominates for the $E$-mode for $\ell \gtrsim 400$. Using a 3x3 map making scheme removes the leakage occurring between the spin-0 temperature and spin-2 polarisation signal, however an overall amplification of each signal by the gain remains present, so the systematic is still non-zero. The predictions from the formalism work well in both cases. To further emphasise the agreement of the analytic predictions with the simulation, band averages have been overplotted as brown crosses (3x3 map-making) and purple crosses (2x2 map-making), which are sourced from the average of 50 realisations and have been smoothed into bins of width $\Delta \ell = 50$; the band averages are in agreement with the analytic predictions.}
\label{fig:GainAll}
\end{figure}

To examine this, we have run simple simulations as in section \ref{section:spin1mapmaking}, except that this time a $1\%$ gain systematic has been included rather than a pointing systematic, and the deep scan strategy of Section \ref{section:Ground Based} has been used. The results of these tests are presented in figure \ref{fig:GainAll}, clearly showing that when using a 2x2 map making scheme the leakage of the temperature to polarisation will occur if a gain mismatch is present between the detectors. In the case of the $B$-mode spectrum the leakage from the temperature is the dominant source of contamination as is evident in the lower panel of figure \ref{fig:GainAll}. The $E$-mode systematic is dominated by the temperature leakage at low multipoles ($\ell \lesssim 400$) while the polarisation amplification from the absolute gain increase is the dominant source at higher multipoles. When employing a 3x3 map making scheme we find that, as expected, there is no longer leakage from the spin-0 signal into the spin-2 polarisation signal; only the overall amplification of each term by the absolute gain remains present.

\FloatBarrier

\section{Pseudo Power Spectra Terms}
\subsection{Cross Term}
\label{section:Cross Term}
The cross term between the sky signal and the systematic of equations \ref{eq:Cross Term} and \ref{eq:Cross Term BB} will vanish under certain conditions. In the case of a full sky survey the window function applied to the survey equals 1 everywhere and as such only contributes as a monopole. This means equations \ref{eq:Cross Term Mask} and \ref{eq:Cross Term Mask BB} may be rewritten with $l_1'=m_1'=0$ and $\sum_{mm_{2}}\threej{\ell,-m}{0,0}{l_{2},m_{2}}\threej{\ell,-m}{l_1,m_1}{l_{2},m_{2}} = \delta_{0l_1}\delta_{0m_1}$ giving
\begin{equation}
\begin{split}
    \frac{1}{2 \, (2\ell + 1)} \sum_{m} \bigl\langle \tilde{E}_{\ell m} \, {}^{}_2\tilde{\Delta}_{\ell m}^* \bigr\rangle
    &= \sum_{\substack{k'm\\l_2m_2}} \bigl\langle
    \bigl\{ E_{l_2m_2} \, \bigg(\threej{l_{2},2}{0,0}{\ell,-2}
    + \threej{l_{2},-2}{0,0}{\ell,2}\bigg) + iB_{l_2m_2} \, \bigg(\threej{l_{2},2}{0,0}{\ell,-2}
    - \threej{l_{2},-2}{0,0}{\ell,2}\bigg) \bigr\} \, {}_0h_{00}
    \\
    &\bigg\{\frac{1}{2} {}_{(2-k')}h^{*}_{00} \threej{\ell,2}{0,-(2-k')}{l_2,-k'} \, {}_{k'}S^{*}_{l_2m_2}\bigg\}
    (-1)^{\ell+l_1+l_2} \tfrac{(2l_2 + 1)}{16\pi} \rangle
\end{split}
\end{equation}
and
\begin{equation}
\begin{split}
    \frac{1}{2 \, (2\ell + 1)} \sum_{m} \bigl\langle \tilde{B}_{\ell m} \, {}^{}_2\tilde{\Delta}_{\ell m}^* \bigr\rangle
    &= \sum_{\substack{k'm\\l_2m_2}} \bigl\langle
    \bigl\{ B_{l_2m_2} \, \bigg(\threej{l_{2},2}{0,0}{\ell,-2}
    - \threej{l_{2},-2}{0,0}{\ell,2}\bigg) + iE_{l_2m_2} \, \bigg(\threej{l_{2},2}{0,0}{\ell,-2}
    - \threej{l_{2},-2}{0,0}{\ell,2}\bigg) \bigr\} \, {}_0h_{00}
    \\
    &\bigg\{\frac{1}{2} {}_{(2-k')}h^{*}_{00} \threej{\ell,2}{0,-(2-k')}{l_2,-k'} \, {}_{k'}S^{*}_{l_2m_2}\bigg\}
    (-1)^{\ell+l_1+l_2} \tfrac{(2l_2 + 1)}{16\pi} \rangle.
\end{split}
\end{equation}
However, applying the rule of spherical harmonics and Wigner-3j symbols that $m \in \{-\ell,-\ell+1,-\ell+2,...,\ell\}$, the only instances when $\threej{\ell,2}{0,-(2-k')}{l_2,-k'} \neq 0$ is when $k' = 2$. As such, the cross term between the sky signal and the systematic disappears on the full sky for the cases where $k' \neq 2$. As a result, the differential pointing cross term is zero in the full sky case since $k' \neq 2$ for all its terms. Equally, this is why there is a non-zero contribution to the cross term on full sky in the case of differential gain, as there is a $k' = 2$ signal involved as seen in equation \ref{eq:Term 3 Gain}.

\subsection{Importance of Terms}
\FloatBarrier
\label{section:Importance of Terms}

We define the spectra of the $\tilde{h}_k$ fields characterising the scan strategy as
\begin{equation}
    \tilde{C}_{\ell}^{{}_{k}h{}_{k'}h} = \frac{1}{2\ell +1}\sum_{m} {}_{k}h^{}_{\ell m} {}_{k'}h^{*}_{\ell m} \text{,}
\end{equation}
where the spins $k$ and $k'$ are set to the value required to couple a spin-coupled systematic to the signal of interest. The scan terms where the spins of the scan fields satisfy $k=k'$ will dominate over terms where $k \neq k'$ since the scan auto-spectra will in general be larger than the cross-spectra. However, this statement does not necessarily translate to the systematic terms themselves being larger, as they will include a systematic dependent pre-factor.

\begin{figure}
  \centering
 \includegraphics[width=5.0in]{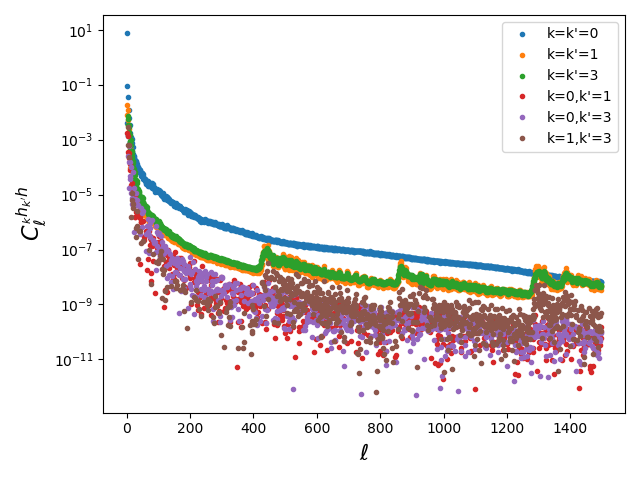}
\caption{Scan spectra of the EPIC satellite scan strategy in the presence of a galactic mask. Various auto and cross-spectra of the spin $k$ and $k'$ scan fields are plotted, showing that in general the auto-spectra are larger than the cross-spectra. The cross-spectra can take both negative and positive values so the absolute values have been plotted for easier comparison of the magnitudes. Note that the $k=k'=0$  spectrum is that of the window function, i.e. it shows the spectrum of the galactic mask. We note the presence of a slight spikes in the spectra, particularly evident in the $k=k'=1$ and $3$ spectra. These correspond to structure present in the scanning strategy of the EPIC satellite.}
\label{fig:Scan Spectra}
\end{figure}

Figure \ref{fig:Scan Spectra} shows various auto and cross-spectra of the spin $k'$ and $k''$ scan fields of the EPIC satellite scan strategy in the presence of a galactic mask. These spectra demonstrate that in general the scan spectra with spins where $k' \neq k''$ are relatively small compared to $k' = k''$.

We note the presence of slight spikes in the spectra of figure \ref{fig:Scan Spectra}, particularly evident in the $k=k'=1$ and $3$ spectra. These correspond to the structure present in the $\tilde{h}_{k}$ fields resulting from the EPIC scanning strategy, as is evident figure 2 of \cite{Wallisetal2016}. In turn those structures in the map are a consequence of the boresight angle, precession angle, spin period, and precession period that have been selected for the EPIC scanning strategy. For more details see \cite{2008arXiv0805.4207B}.

\subsection{Additional Spectra}
\label{section:Additional Spectra}
In section \ref{section:Power Spectrum Derivation} we concentrated on analysing the effect of a systematic bias on the polarisation power spectra. Here we shall also present the effects on the temperature spectrum and cross spectra without the laborious detail.

To include the effect on the intensity signal we must first define the spin-0 temperature signal using equation \ref{eq:Observed Signal}, and isolating the $k'=0$ terms as
\begin{equation}
    \tilde{T}_{\ell m}
    = \sum_{\substack{l_1m_1\\l_2m_2}} {}_0h_{l_1m_1} \, T_{l_2m_2} \, (-1)^m \, \sqrt{\tfrac{(2\ell + 1)(2l_1 + 1)(2l_2 + 1)}{4\pi}}
    \threej{\ell,-m}{l_1,m_1}{l_2,m_2} \, \threej{\ell,0}{l_1,0}{l_2,0} \text{,}
\label{eq:True Temperature}
\end{equation}
where $T_{l_2m_2}$ are the true full sky spin-0 intensity modes. The presence of the window function ${}_0h_{l_1m_1}$ means that this can be applied to both full and cut sky. The terms from the other sources, such as systematics, may be contributed to by any spin $k'$. These terms contributing to the spin-0 intensity may be written using equation \ref{eq:Observed Signal} as
\begin{align}
    \tilde{\Delta}^{T}_{\ell m} = {}_0\tilde{\Delta}_{\ell m} = (-1)^{m} \, \sum_{\substack{k'\\l_1m_1\\l_2m_2}} {}_{(0-k')}h_{l_1m_1} \, {}_{k'}\Delta_{l_2m_2}  \, \sqrt{\tfrac{(2\ell + 1)(2l_1 + 1)(2l_2 + 1)}{4\pi}}
    \threej{\ell,-m}{l_1,m_1}{l_2,m_2} \, \threej{\ell,0}{l_1,-(0-k')}{l_2,-k'}.
    \label{eq:Spin-0 General Systematic alm}
\end{align}
We may write the contributions to the total signals (signified by a hat), as the sum of the true on sky signal and the systematic $\hat{X}_{\ell m} = \tilde{X}_{\ell m}+\tilde{\Delta}^{X}_{\ell m}$, where $X \in \{T,E,B\}$, and the tilde denotes pseudo quantities i.e. those that apply on partial skies. The effect of a systematic on the power spectra may then be analytically modelled following the prescriptions demonstrated in Section \ref{section:Power Spectrum Derivation} for the polarisation spectra. In table \ref{tab:Systematic Spectra} we provide the contributions of a general systematic to the six pseudo power spectra of interest. It is also important to point out that the contaminant to the spin-0 intensity signal given by equation \ref{eq:Spin-0 General Systematic alm} and the $TT$ row of table \ref{tab:Systematic Spectra} in the power spectra case would be of importance to consider in intensity mapping experiments.

\begin{table}
\centering
 \begin{tabular}{||c c||} 
 \hline
 Spectrum & Systematic Contribution \\ [0.5ex] 
 \hline\hline
 $TT$ & $\frac{1}{(2\ell + 1)} \sum_{m} \Bigl\{ \bigl\langle {}_0\tilde{\Delta}_{\ell m} \, {}_0\tilde{\Delta}_{\ell m}^*\bigr\rangle + 2 \, {\Re{}} \langle \tilde{T}_{\ell m} \, {}_0\tilde{\Delta}_{\ell m}^* \rangle \Bigr\}$
    \\ 
 \hline
 $TE$ & $\frac{1}{(2\ell + 1)} \sum_{m} \Bigl\{ {\Re{}}\bigl\langle  {}_0\tilde{\Delta}_{\ell m} \, {}^{}_2\tilde{\Delta}_{\ell m}^*\bigr\rangle + {\Re{}} \langle \tilde{T}_{\ell m} \, {}^{}_2\tilde{\Delta}_{\ell m}^* \rangle +  \langle {}_0\tilde{\Delta}_{\ell m} \tilde{E}_{\ell m}^* \rangle \Bigr\}$
 \\
 \hline
 $TB$ & $\frac{1}{(2\ell + 1)} \sum_{m} \Bigl\{ {\Re{}}\bigl\langle i{}_0\tilde{\Delta}_{\ell m} \, {}^{}_2\tilde{\Delta}_{\ell m}^*\bigr\rangle +  {\Re{}} \langle i\tilde{T}_{\ell m} \, {}^{}_2\tilde{\Delta}_{\ell m}^* \rangle + \langle {}_0\tilde{\Delta}_{\ell m} \tilde{B}_{\ell m}^* \rangle \Bigr\}$
 \\
 \hline
 $EE$ & $\frac{1}{2 \, (2\ell + 1)} \sum_{m} \Bigl\{ \bigl\langle {}_2\tilde{\Delta}_{\ell m} \, {}_2\tilde{\Delta}_{\ell m}^*\bigr\rangle + (-1)^m \, {\Re{}} \bigl\langle {}_2\tilde{\Delta}_{\ell m} \, {}_2\tilde{\Delta}_{\ell -m}\bigr\rangle + 4 \, {\Re{}} \langle \tilde{E}_{\ell m} \, {}_2\tilde{\Delta}_{\ell m}^* \rangle \Bigr\}$
 \\ 
 \hline
 $EB$ & $\frac{1}{2 \, (2\ell + 1)} \sum_{m} \Bigl\{ (-1)^m \, {\Re{}} \bigl\langle i {}_2\tilde{\Delta}_{\ell m} \, {}_2\tilde{\Delta}_{\ell -m}\bigr\rangle + 2{\Re{}} \bigl\langle i \tilde{E}_{\ell m} {}^{}_2\tilde{\Delta}_{\ell m}^*\bigl\rangle + 2{\Re{}} \bigl\langle {}^{}_2\tilde{\Delta}_{\ell m}\tilde{B}_{\ell m}^* \bigl\rangle \Bigr\}$
 \\
 \hline
 $BB$ & $\frac{1}{2 \, (2\ell + 1)} \sum_{m} \Bigl\{ \bigl\langle{}_2\tilde{\Delta}_{\ell m} \, {}^{}_2\tilde{\Delta}_{\ell m}^*\bigr\rangle - (-1)^m \, {\Re{}} \bigl\langle{}_2\tilde{\Delta}_{\ell m} \, {}_2\tilde{\Delta}_{\ell -m}\bigr\rangle - 4 {\Im{}} \bigl\langle  \tilde{B}_{\ell m} {}^{}_2\tilde{\Delta}_{\ell m}^*\bigl\rangle \Bigr\}$
 \\ [1ex]
 \hline
\end{tabular}
\caption{\label{tab:Systematic Spectra} The systematic contributions to the 6 pseudo power spectra independent of the form of systematic and experimental set up. The appropriate substitutions of equations
\ref{eq:True Polarisation}, \ref{eq:Spin2 General Systematic alm}, \ref{eq:True Temperature}, and \ref{eq:Spin-0 General Systematic alm} must be made following the prescriptions outlined in Section \ref{section:Power Spectrum Derivation}.}
\end{table}

\clearpage
\subsection{Pointing Terms}
\label{section:Pointing Terms}
Here we present the contributions of the differential pointing systematic to the terms in the observed pseudo power spectra of equations \ref{eq:ClBB} and \ref{eq:ClEE}:
\tiny
\begin{equation}
\begin{split}
    \frac{1}{2 \, (2\ell + 1)} \sum_{m} & \bigl\langle{}_2\tilde{\Delta}_{\ell m} {}^{}_2\tilde{\Delta}_{\ell m}^*\bigr\rangle
    = \sum_{\substack{l_1l_2}}
    \tfrac{(2l_1 + 1)(2l_2 + 1)}{128\pi}
    \\
    & \Bigg\{\tilde{C}_{l_1}^{{}_1h{}_1h}l_2(l_2+1)\bigl|\rho_1 \, e^{i \, \chi_1} + \rho_2 \, e^{i \, (\chi_2 - \pi/4)}\bigr|^2 \threej{\ell,2}{l_1,-1}{l_2,-1}^2 C_{l_2}^{TT}
    \\
    &-2\tilde{C}_{l_1}^{{}_1h{}_3h}l_2(l_2+1){\Re{}}\bigg\{\bigl(\rho_1 \, e^{i \, \chi_1} + \rho_2 \, e^{i \, (\chi_2 - \pi/4)}\bigr) \bigl(\rho_1 \, e^{-i \, \chi_1} - \rho_2 \, e^{-i \, (\chi_2 - \pi/4)}\bigr)^*\bigg\} \threej{\ell,2}{l_1,-1}{l_2,-1}\threej{\ell,2}{l_1,-3}{l_2,1} C_{l_2}^{TT}
    \\
    &+\tilde{C}^{{}_1h{}_1h}_{l_1}\sqrt{l_2(l_2+1)}\sqrt{(l_2+2) \, (l_2-2+1)} \threej{\ell,2}{l_1,-1}{l_2,-1}^2   {\Re{}}\bigg\{\bigl(\rho_1 \, e^{i \, \chi_1} + \rho_2 \, e^{i \, (\chi_2 - \pi/4)}\bigr) \bigl(\rho_1 \, e^{-i \, \chi_1} + \rho_2 \, e^{-i \, (\chi_2 + \pi/4)}\bigr)^* (C^{TE}_{l_2}-iC^{TB}_{l_2})\bigg\} 
    \\
    &+\tilde{C}_{l_1}^{{}_1h{}_5h} \sqrt{l_2(l_2+1)}\sqrt{(l_2+2) \, (l_2-2+1)} \threej{\ell,2}{l_1,-1}{l_2,-1}\threej{\ell,2}{l_1,-5}{l_2,3} {\Re{}}\bigg\{\bigl(\rho_1 \, e^{i \, \chi_1} + \rho_2 \, e^{i \, (\chi_2 - \pi/4)}\bigr) \bigl(\rho_1 \, e^{-i \, \chi_1} - \rho_2 \, e^{-i \, (\chi_2 + \pi/4)}\bigr)^*
    (C_{l_2}^{TE}+iC_{l_2}^{TB})\bigg\}
    \\
    &-\tilde{C}^{{}_1h{}_3h}_{l_1}\sqrt{l_2(l_2+1)}\sqrt{(l_2-2) \, (l_2+2+1)} \threej{\ell,2}{l_1,-1}{l_2,-1}\threej{\ell,2}{l_1,-3}{l_2,1} {\Re{}}\bigg\{\bigl(\rho_1 \, e^{i \, \chi_1} + \rho_2 \, e^{i \, (\chi_2 - \pi/4)}\bigr) \bigl(\rho_1 \, e^{i \, \chi_1} - \rho_2 \, e^{i \, (\chi_2 + \pi/4)}\bigr)^* (C_{l_2}^{TE}+iC_{l_2}^{TB})\bigg\}
    \\
    &-\tilde{C}_{l_1}^{{}_1h{}_{-1}h} \sqrt{l_2(l_2+1)}\sqrt{(l_2-2) \, (l_2+2+1)} \threej{\ell,2}{l_1,-1}{l_2,-1}\threej{\ell,2}{l_1,1}{l_2,-3} {\Re{}} \bigg\{\bigl(\rho_1 \, e^{i \, \chi_1} + \rho_2 \, e^{i \, (\chi_2 - \pi/4)}\bigr) \bigl(\rho_1 \, e^{i \, \chi_1} + \rho_2 \, e^{i \, (\chi_2 + \pi/4)}\bigr)^*  (C_{l_2}^{TE}-iC_{l_2}^{TB})\bigg\}
    \\
    &+\tilde{C}_{l_1}^{{}_3h{}_3h}l_2(l_2+1)\bigl|\rho_1 \, e^{-i \, \chi_1} - \rho_2 \, e^{-i \, (\chi_2 - \pi/4)}\bigr|^2 \threej{\ell,2}{l_1,-3}{l_2,1}^2 C_{l_2}^{TT}
    \\
    &-\tilde{C}^{{}_3h{}_1h}_{l_1}\sqrt{l_2(l_2+1)}\sqrt{(l_2+2) \, (l_2-2+1)} \threej{\ell,2}{l_1,-3}{l_2,1}\threej{\ell,2}{l_1,-1}{l_2,-1} {\Re{}} \bigg\{\bigl(\rho_1 \, e^{-i \, \chi_1} - \rho_2 \, e^{-i \, (\chi_2 - \pi/4)}\bigr) \bigl(\rho_1 \, e^{-i \, \chi_1} + \rho_2 \, e^{-i \, (\chi_2 + \pi/4)}\bigr)^*  (C_{l_2}^{TE}-iC_{l_2}^{TB})\bigg\}
    \\
    &-\tilde{C}_{l_1}^{{}_3h{}_5h} \sqrt{l_2(l_2+1)}\sqrt{(l_2+2) \, (l_2-2+1)} \threej{\ell,2}{l_1,-3}{l_2,1}\threej{\ell,2}{l_1,-5}{l_2,3} {\Re{}} \bigg\{\bigl(\rho_1 \, e^{-i \, \chi_1} - \rho_2 \, e^{-i \, (\chi_2 - \pi/4)}\bigr) \bigl(\rho_1 \, e^{-i \, \chi_1} - \rho_2 \, e^{-i \, (\chi_2 + \pi/4)}\bigr)^* (C_{l_2}^{TE}+iC_{l_2}^{TB})\bigg\}
    \\
    &+\tilde{C}^{{}_3h{}_3h}_{l_1}\sqrt{l_2(l_2+1)}\sqrt{(l_2-2) \, (l_2+2+1)} \threej{\ell,2}{l_1,-3}{l_2,1}^2 {\Re{}} \bigg\{\bigl(\rho_1 \, e^{-i \, \chi_1} - \rho_2 \, e^{-i \, (\chi_2 - \pi/4)}\bigr) \bigl(\rho_1 \, e^{i \, \chi_1} - \rho_2 \, e^{i \, (\chi_2 + \pi/4)}\bigr)^* (C_{l_2}^{TE}+iC_{l_2}^{TB})\bigg\}
    \\
    &+\tilde{C}_{l_1}^{{}_3h{}_{-1}h} \sqrt{l_2(l_2+1)}\sqrt{(l_2-2) \, (l_2+2+1)} \threej{\ell,2}{l_1,-3}{l_2,1}\threej{\ell,2}{l_1,1}{l_2,-3} {\Re{}} \bigg\{\bigl(\rho_1 \, e^{-i \, \chi_1} - \rho_2 \, e^{-i \, (\chi_2 - \pi/4)}\bigr) \bigl(\rho_1 \, e^{i \, \chi_1} + \rho_2 \, e^{i \, (\chi_2 + \pi/4)}\bigr)^* (C_{l_2}^{TE}-iC_{l_2}^{TB})\bigg\}
    \\
    &+\frac{1}{4}\tilde{C}^{{}_1h{}_1h}_{l_1} (l_2+2) \, (l_2-2+1) 
    \bigl|\rho_1 \, e^{-i \, \chi_1} + \rho_2 \, e^{-i \, (\chi_2 + \pi/4)}\bigr|^2
    \threej{\ell,2}{l_1,-1}{l_2,-1}^2 (C_{l_2}^{EE}+C_{l_2}^{BB})
    \\
    &+\frac{1}{2}\tilde{C}^{{}_1h{}_5h}_{l_1} (l_2+2) \, (l_2-2+1) \threej{\ell,2}{l_1,-1}{l_2,-1}\threej{\ell,2}{l_1,-5}{l_2,3} {\Re{}} \bigg\{\bigl(\rho_1 \, e^{-i \, \chi_1} + \rho_2 \, e^{-i \, (\chi_2 + \pi/4)}\bigr) \bigl(\rho_1 \, e^{-i \, \chi_1} - \rho_2 \, e^{-i \, (\chi_2 + \pi/4)}\bigr)^* (C_{l_2}^{EE}-C_{l_2}^{BB}+2iC_{l_2}^{EB})\bigg\}
    \\
    &-\frac{1}{2}\tilde{C}^{{}_1h{}_3h}_{l_1} \sqrt{(l_2+2) \, (l_2-2+1)} \sqrt{(l_2-2) \, (l_2+2+1)} \threej{\ell,2}{l_1,-1}{l_2,-1}\threej{\ell,2}{l_1,-3}{l_2,1} {\Re{}} \bigg\{\bigl(\rho_1 \, e^{-i \, \chi_1} + \rho_2 \, e^{-i \, (\chi_2 + \pi/4)}\bigr) \bigl(\rho_1 \, e^{i \, \chi_1} - \rho_2 \, e^{i \, (\chi_2 + \pi/4)}\bigr)^* (C_{l_2}^{EE}-C_{l_2}^{BB}+2iC_{l_2}^{EB})\bigg\}
    \\
    &-\frac{1}{2}\tilde{C}^{{}_1h{}_{-1}h}_{l_1} \sqrt{(l_2+2) \, (l_2-2+1)} \sqrt{(l_2-2) \, (l_2+2+1)}  {\Re{}} \bigg\{\bigl(\rho_1 \, e^{-i \, \chi_1} + \rho_2 \, e^{-i \, (\chi_2 + \pi/4)}\bigr)^2\bigg\} \threej{\ell,2}{l_1,-1}{l_2,-1}\threej{\ell,2}{l_1,1}{l_2,-3}(C_{l_2}^{EE}+C_{l_2}^{BB})
    \\
    &+\frac{1}{4}\tilde{C}_{l_1}^{{}_5h{}_5h} (l_2+2) \, (l_2-2+1) \bigl|\rho_1 \, e^{-i \, \chi_1} - \rho_2 \, e^{-i \, (\chi_2 + \pi/4)}\bigr|^2 \threej{\ell,2}{l_1,-5}{l_2,3}^2 (C_{l_2}^{EE}+C_{l_2}^{BB})
    \\
    &-\frac{1}{2}\tilde{C}^{{}_5h{}_3h}_{l_1}\sqrt{(l_2+2) \, (l_2-2+1)}\sqrt{(l_2-2) \, (l_2+2+1)} {\Re{}} \bigg\{\bigl(\rho_1 \, e^{-i \, \chi_1} - \rho_2 \, e^{-i \, (\chi_2 + \pi/4)}\bigr)^2\bigg\} \threej{\ell,2}{l_1,-5}{l_2,3}\threej{\ell,2}{l_1,-3}{l_2,1} (C_{l_2}^{EE}+C_{l_2}^{BB})
    \\
    &-\frac{1}{2}\tilde{C}_{l_1}^{{}_5h{}_{-1}h} \sqrt{(l_2+2) \, (l_2-2+1)}\sqrt{(l_2-2) \, (l_2+2+1)} \threej{\ell,2}{l_1,-5}{l_2,3}\threej{\ell,2}{l_1,1}{l_2,-3} {\Re{}} \bigg\{\bigl(\rho_1 \, e^{-i \, \chi_1} - \rho_2 \, e^{-i \, (\chi_2 + \pi/4)}\bigr) \bigl(\rho_1 \, e^{i \, \chi_1} + \rho_2 \, e^{i \, (\chi_2 + \pi/4)}\bigr)^* (C_{l_2}^{EE}-C_{l_2}^{BB}-2iC_{l_2}^{EB})\bigg\}
    \\
    &+\frac{1}{4}\tilde{C}^{{}_3h{}_3h}_{l_1} (l_2-2) \, (l_2+2+1)
    \bigl|\rho_1 \, e^{i \, \chi_1} - \rho_2 \, e^{i \, (\chi_2 + \pi/4)}\bigr|^2 \threej{\ell,2}{l_1,-3}{l_2,1}^2 (C_{l_2}^{EE}+C_{l_2}^{BB})
    \\
    &+\frac{1}{2}\tilde{C}^{{}_3h{}_{-1}h}_{l_1} (l_2-2) \, (l_2+2+1) \threej{\ell,2}{l_1,-3}{l_2,1}\threej{\ell,2}{l_1,1}{l_2,-3} {\Re{}} \bigg\{\bigl(\rho_1 \, e^{i \, \chi_1} - \rho_2 \, e^{i \, (\chi_2 + \pi/4)}\bigr) \bigl(\rho_1 \, e^{i \, \chi_1} + \rho_2 \, e^{i \, (\chi_2 + \pi/4)}\bigr)^* (C_{l_2}^{EE}-C_{l_2}^{BB}-2iC_{l_2}^{EB})\bigg\}
    \\
    &+\frac{1}{4}\tilde{C}_{l_1}^{{}_{-1}h{}_{-1}h} (l_2-2) \, (l_2+2+1) \bigl|\rho_1 \, e^{i \, \chi_1} + \rho_2 \, e^{i \, (\chi_2 + \pi/4)}\bigr|^2 \threej{\ell,2}{l_1,1}{l_2,-3}^2 (C_{l_2}^{EE}+C_{l_2}^{BB})
    \\\\\\\\\\\\\\\\\\\\\\\\\\\\\\\\\\\\\\\\\\\\\\\\\\\\\\\
\end{split}
\label{eq:Term 1 Pointing}
\end{equation}

\begin{equation}
\begin{split}
    \frac{1}{2 \, (2\ell + 1)} & \sum_{m} (-1)^m \, \Re{} \bigl\langle{}_2\tilde{\Delta}_{\ell m} \, {}_{2}\tilde{\Delta}_{\ell -m}\bigr\rangle = \sum_{l_1l_2}
    \tfrac{(2l_1 + 1)(2l_2 + 1)}{128\pi} \Re{}
    \\
    & \Bigg\{-\tilde{C}_{l_1}^{{}_1h{}_{-1}h}l_2(l_2+1)\bigl(\rho_1 \, e^{i \, \chi_1} + \rho_2 \, e^{i \, (\chi_2 - \pi/4)}\bigr)^2 \threej{\ell,2}{l_1,-1}{l_2,-1}\threej{\ell,-2}{l_1,1}{l_2,1} C_{l_2}^{TT}
    \\
    &+\tilde{C}_{l_1}^{{}_1h{}_{-3}h}l_2(l_2+1)\bigl(\rho_1 \, e^{i \, \chi_1} + \rho_2 \, e^{i \, (\chi_2 - \pi/4)}\bigr) \bigl(\rho_1 \, e^{i \, \chi_1} - \rho_2 \, e^{i \, (\chi_2 - \pi/4)}\bigr)^* \threej{\ell,2}{l_1,-1}{l_2,-1}\threej{\ell,-2}{l_1,3}{l_2,-1} C_{l_2}^{TT}
    \\
    &-\frac{1}{2}\tilde{C}_{l_1}^{{}_1h{}_{-1}h} \sqrt{l_2(l_2+1)}\sqrt{(l_2-2) \, (l_2+2+1)} \bigl(\rho_1 \, e^{i \, \chi_1} + \rho_2 \, e^{i \, (\chi_2 - \pi/4)}\bigr) \bigl(\rho_1 \, e^{i \, \chi_1} + \rho_2 \, e^{i \, (\chi_2 + \pi/4)}\bigr)^* \threej{\ell,2}{l_1,-1}{l_2,-1} \threej{\ell,-2}{l_1,1}{l_2,1} (C_{l_2}^{TE} +i C_{l_2}^{TB})
    \\
    &-\frac{1}{2}\tilde{C}_{l_1}^{{}_1h{}_{-5}h} \sqrt{l_2(l_2+1)}\sqrt{(l_2-2) \, (l_2+2+1)} \bigl(\rho_1 \, e^{i \, \chi_1} + \rho_2 \, e^{i \, (\chi_2 - \pi/4)}\bigr) \bigl(\rho_1 \, e^{i \, \chi_1} - \rho_2 \, e^{i \, (\chi_2 + \pi/4)}\bigr)^* \threej{\ell,2}{l_1,-1}{l_2,-1}\threej{\ell,-2}{l_1,5}{l_2,-3} (C_{l_2}^{TE} -i C_{l_2}^{TB})
    \\
    &+\frac{1}{2}\tilde{C}_{l_1}^{{}_1h{}_{-3}h} \sqrt{l_2(l_2+1)}\sqrt{(l_2+2) \, (l_2-2+1)} \bigl(\rho_1 \, e^{i \, \chi_1} + \rho_2 \, e^{i \, (\chi_2 - \pi/4)}\bigr) \bigl(\rho_1 \, e^{-i \, \chi_1} - \rho_2 \, e^{-i \, (\chi_2 + \pi/4)}\bigr)^* \threej{\ell,2}{l_1,-1}{l_2,-1}\threej{\ell,-2}{l_1,3}{l_2,-1} (C_{l_2}^{TE} -i C_{l_2}^{TB})
    \\
    &+\frac{1}{2}\tilde{C}_{l_1}^{{}_1h{}_{1}h} \sqrt{l_2(l_2+1)}\sqrt{(l_2+2) \, (l_2-2+1)} \bigl(\rho_1 \, e^{i \, \chi_1} + \rho_2 \, e^{i \, (\chi_2 - \pi/4)}\bigr) \bigl(\rho_1 \, e^{-i \, \chi_1} + \rho_2 \, e^{-i \, (\chi_2 + \pi/4)}\bigr)^* \threej{\ell,2}{l_1,-1}{l_2,-1}\threej{\ell,-2}{l_1,-1}{l_2,3} (C_{l_2}^{TE} +i C_{l_2}^{TB})
    \\
    &+\tilde{C}_{l_1}^{{}_3h{}_{-1}h} l_2(l_2+1) \bigl(\rho_1 \, e^{-i \, \chi_1} - \rho_2 \, e^{-i \, (\chi_2 - \pi/4)}\bigr) \bigl(\rho_1 \, e^{-i \, \chi_1} + \rho_2 \, e^{-i \, (\chi_2 - \pi/4)}\bigr)^* \threej{\ell,2}{l_1,-3}{l_2,1} \threej{\ell,-2}{l_1,1}{l_2,1} C_{l_2}^{TT}
    \\
    &-\tilde{C}_{l_1}^{{}_3h{}_{-3}h} l_2(l_2+1)\bigl(\rho_1 \, e^{-i \, \chi_1} - \rho_2 \, e^{-i \, (\chi_2 - \pi/4)}\bigr)^2 \threej{\ell,2}{l_1,-3}{l_2,1}\threej{\ell,-2}{l_1,3}{l_2,-1} C_{l_2}^{TT}
    \\
    &+\frac{1}{2}\tilde{C}_{l_1}^{{}_3h{}_{-1}h} \sqrt{l_2(l_2+1)}\sqrt{(l_2-2) \, (l_2+2+1)} \bigl(\rho_1 \, e^{-i \, \chi_1} - \rho_2 \, e^{-i \, (\chi_2 - \pi/4)}\bigr)
    \bigl(\rho_1 \, e^{i \, \chi_1} + \rho_2 \, e^{i \, (\chi_2 + \pi/4)}\bigr)^*
    \threej{\ell,2}{l_1,-3}{l_2,1} \threej{\ell,-2}{l_1,1}{l_2,1} (C_{l_2}^{TE} +i C_{l_2}^{TB})
    \\
    &+\frac{1}{2}\tilde{C}_{l_1}^{{}_3h{}_{-5}h} \sqrt{l_2(l_2+1)}\sqrt{(l_2-2) \, (l_2+2+1)} \bigl(\rho_1 \, e^{-i \, \chi_1} - \rho_2 \, e^{-i \, (\chi_2 - \pi/4)}\bigr) \bigl(\rho_1 \, e^{i \, \chi_1} - \rho_2 \, e^{i \, (\chi_2 + \pi/4)}\bigr)^* \threej{\ell,2}{l_1,-3}{l_2,1}\threej{\ell,-2}{l_1,5}{l_2,-3} (C_{l_2}^{TE} -i C_{l_2}^{TB})
    \\
    &-\frac{1}{2}\tilde{C}_{l_1}^{{}_3h{}_{-3}h} \sqrt{l_2(l_2+1)}\sqrt{(l_2+2) \, (l_2-2+1)} \bigl(\rho_1 \, e^{-i \, \chi_1} - \rho_2 \, e^{-i \, (\chi_2 - \pi/4)}\bigr) \bigl(\rho_1 \, e^{-i \, \chi_1} - \rho_2 \, e^{-i \, (\chi_2 + \pi/4)}\bigr)^* \threej{\ell,2}{l_1,-3}{l_2,1}\threej{\ell,-2}{l_1,3}{l_2,-1} (C_{l_2}^{TE} -i C_{l_2}^{TB})
    \\
    &-\frac{1}{2}\tilde{C}_{l_1}^{{}_3h{}_{1}h} \sqrt{l_2(l_2+1)}\sqrt{(l_2+2) \, (l_2-2+1)} \bigl(\rho_1 \, e^{-i \, \chi_1} - \rho_2 \, e^{-i \, (\chi_2 - \pi/4)}\bigr) \bigl(\rho_1 \, e^{-i \, \chi_1} + \rho_2 \, e^{-i \, (\chi_2 + \pi/4)}\bigr)^* \threej{\ell,2}{l_1,-3}{l_2,1}\threej{\ell,-2}{l_1,-1}{l_2,3} (C_{l_2}^{TE} +i C_{l_2}^{TB})
    \\
    &-\frac{1}{2}\tilde{C}_{l_1}^{{}_1h{}_{-1}h} \sqrt{(l_2+2) \, (l_2-2+1)}\sqrt{l_2(l_2+1)} \bigl(\rho_1 \, e^{-i \, \chi_1} + \rho_2 \, e^{-i \, (\chi_2 + \pi/4)}\bigr) \bigl(\rho_1 \, e^{-i \, \chi_1} + \rho_2 \, e^{-i \, (\chi_2 - \pi/4)}\bigr)^* \threej{\ell,2}{l_1,-1}{l_2,-1} \threej{\ell,-2}{l_1,1}{l_2,1} (C_{l_2}^{TE} +i C_{l_2}^{TB})
    \\
    &+\frac{1}{2}\tilde{C}_{l_1}^{{}_1h{}_{-3}h} \sqrt{(l_2+2) \, (l_2-2+1)} \sqrt{l_2(l_2+1)} \bigl(\rho_1 \, e^{-i \, \chi_1} + \rho_2 \, e^{-i \, (\chi_2 + \pi/4)}\bigr) \bigl(\rho_1 \, e^{i \, \chi_1} - \rho_2 \, e^{i \, (\chi_2 - \pi/4)}\bigr)^* \threej{\ell,2}{l_1,-1}{l_2,-1}\threej{\ell,-2}{l_1,3}{l_2,-1} (C_{l_2}^{TE} +i C_{l_2}^{TB})
    \\
    &-\frac{1}{4}\tilde{C}_{l_1}^{{}_1h{}_{-1}h} \sqrt{(l_2+2) \, (l_2-2+1)} \sqrt{(l_2-2) \, (l_2+2+1)} \bigl(\rho_1 \, e^{-i \, \chi_1} + \rho_2 \, e^{-i \, (\chi_2 + \pi/4)}\bigr)^2 \threej{\ell,2}{l_1,-1}{l_2,-1} \threej{\ell,-2}{l_1,1}{l_2,1} (C_{l_2}^{EE} - C_{l_2}^{BB} +2iC_{l_2}^{EB})
    \\
    &-\frac{1}{4}\tilde{C}_{l_1}^{{}_1h{}_{-5}h} \sqrt{(l_2+2) \, (l_2-2+1)} \sqrt{(l_2-2) \, (l_2+2+1)} \bigl(\rho_1 \, e^{-i \, \chi_1} + \rho_2 \, e^{-i \, (\chi_2 + \pi/4)}\bigr) \bigl(\rho_1 \, e^{i \, \chi_1} - \rho_2 \, e^{i \, (\chi_2 + \pi/4)}\bigr)^* \threej{\ell,2}{l_1,-1}{l_2,-1}\threej{\ell,-2}{l_1,5}{l_2,-3} (C_{l_2}^{EE} + C_{l_2}^{BB})
    \\
    &+\frac{1}{4}\tilde{C}_{l_1}^{{}_1h{}_{-3}h} (l_2+2) \, (l_2-2+1) \bigl(\rho_1 \, e^{-i \, \chi_1} + \rho_2 \, e^{-i \, (\chi_2 + \pi/4)}\bigr) \bigl(\rho_1 \, e^{-i \, \chi_1} - \rho_2 \, e^{-i \, (\chi_2 + \pi/4)}\bigr)^* \threej{\ell,2}{l_1,-1}{l_2,-1}\threej{\ell,-2}{l_1,3}{l_2,-1} (C_{l_2}^{EE} + C_{l_2}^{BB})
    \\
    &+\frac{1}{4}\tilde{C}_{l_1}^{{}_1h{}_{1}h} (l_2+2) \, (l_2-2+1) \bigl|\rho_1 \, e^{-i \, \chi_1} + \rho_2 \, e^{-i \, (\chi_2 + \pi/4)}\bigr|^2 \threej{\ell,2}{l_1,-1}{l_2,-1} \threej{\ell,-2}{l_1,-1}{l_2,3} (C_{l_2}^{EE} - C_{l_2}^{BB} +2iC_{l_2}^{EB})
    \\
    &-\frac{1}{2}\tilde{C}_{l_1}^{{}_5h{}_{-1}h} \sqrt{(l_2+2) \, (l_2-2+1)}\sqrt{l_2(l_2+1)} \bigl(\rho_1 \, e^{-i \, \chi_1} - \rho_2 \, e^{-i \, (\chi_2 + \pi/4)}\bigr) \bigl(\rho_1 \, e^{-i \, \chi_1} + \rho_2 \, e^{-i \, (\chi_2 - \pi/4)}\bigr)^*  \threej{\ell,2}{l_1,-5}{l_2,3} \threej{\ell,-2}{l_1,1}{l_2,1} (C_{l_2}^{TE} - iC_{l_2}^{TB})
    \\
    &+\frac{1}{2}\tilde{C}_{l_1}^{{}_5h{}_{-3}h} \sqrt{(l_2+2) \, (l_2-2+1)}\sqrt{l_2(l_2+1)} \bigl(\rho_1 \, e^{-i \, \chi_1} - \rho_2 \, e^{-i \, (\chi_2 + \pi/4)}\bigr) \bigl(\rho_1 \, e^{i \, \chi_1} - \rho_2 \, e^{i \, (\chi_2 - \pi/4)}\bigr)^* \threej{\ell,2}{l_1,-5}{l_2,3}\threej{\ell,-2}{l_1,3}{l_2,-1} (C_{l_2}^{TE} - iC_{l_2}^{TB})
    \\
    &-\frac{1}{4}\tilde{C}_{l_1}^{{}_5h{}_{-1}h} \sqrt{(l_2+2) \, (l_2-2+1)}\sqrt{(l_2-2) \, (l_2+2+1)} \bigl(\rho_1 \, e^{-i \, \chi_1} - \rho_2 \, e^{-i \, (\chi_2 + \pi/4)}\bigr) \bigl(\rho_1 \, e^{i \, \chi_1} + \rho_2 \, e^{i \, (\chi_2 + \pi/4)}\bigr)^* \threej{\ell,2}{l_1,-5}{l_2,3} \threej{\ell,-2}{l_1,1}{l_2,1} (C_{l_2}^{EE} + C_{l_2}^{BB})
    \\
    &-\frac{1}{4}\tilde{C}_{l_1}^{{}_5h{}_{-5}h} \sqrt{(l_2+2) \, (l_2-2+1)}\sqrt{(l_2-2) \, (l_2+2+1)} \bigl(\rho_1 \, e^{-i \, \chi_1} - \rho_2 \, e^{-i \, (\chi_2 + \pi/4)}\bigr)^2 \threej{\ell,2}{l_1,-5}{l_2,3}\threej{\ell,-2}{l_1,5}{l_2,-3} (C_{l_2}^{EE} - C_{l_2}^{BB} -2iC_{l_2}^{EB})
    \\
    &+\frac{1}{4}\tilde{C}_{l_1}^{{}_5h{}_{-3}h} (l_2+2) \, (l_2-2+1) \bigl|\rho_1 \, e^{-i \, \chi_1} - \rho_2 \, e^{-i \, (\chi_2 + \pi/4)}\bigr|^2 \threej{\ell,2}{l_1,-5}{l_2,3}\threej{\ell,-2}{l_1,3}{l_2,-1} (C_{l_2}^{EE} - C_{l_2}^{BB} -2iC_{l_2}^{EB})
    \\
    &+\frac{1}{4}\tilde{C}_{l_1}^{{}_5h{}_{1}h} (l_2+2) \, (l_2-2+1) \bigl(\rho_1 \, e^{-i \, \chi_1} - \rho_2 \, e^{-i \, (\chi_2 + \pi/4)}\bigr) \bigl(\rho_1 \, e^{-i \, \chi_1} + \rho_2 \, e^{-i \, (\chi_2 + \pi/4)}\bigr)^* \threej{\ell,2}{l_1,-5}{l_2,3}\threej{\ell,-2}{l_1,-1}{l_2,3} (C_{l_2}^{EE} + C_{l_2}^{BB})
    \\
    &+\frac{1}{2}\tilde{C}_{l_1}^{{}_3h{}_{-1}h} \sqrt{(l_2-2) \, (l_2+2+1)} \sqrt{l_2(l_2+1)} \bigl(\rho_1 \, e^{i \, \chi_1} - \rho_2 \, e^{i \, (\chi_2 + \pi/4)}\bigr) \bigl(\rho_1 \, e^{-i \, \chi_1} + \rho_2 \, e^{-i \, (\chi_2 - \pi/4)}\bigr)^* \threej{\ell,2}{l_1,-3}{l_2,1} \threej{\ell,-2}{l_1,1}{l_2,1} (C_{l_2}^{TE} -i C_{l_2}^{TB})
    \\
    &-\frac{1}{2}\tilde{C}_{l_1}^{{}_3h{}_{-3}h} \sqrt{(l_2-2) \, (l_2+2+1)} \sqrt{l_2(l_2+1)} \bigl(\rho_1 \, e^{i \, \chi_1} - \rho_2 \, e^{i \, (\chi_2 + \pi/4)}\bigr) \bigl(\rho_1 \, e^{i \, \chi_1} - \rho_2 \, e^{i \, (\chi_2 - \pi/4)}\bigr)^* \threej{\ell,2}{l_1,-3}{l_2,1}\threej{\ell,-2}{l_1,3}{l_2,-1} (C_{l_2}^{TE} -i C_{l_2}^{TB})
    \\
    &+\frac{1}{4}\tilde{C}_{l_1}^{{}_3h{}_{-1}h} (l_2-2) \, (l_2+2+1) \bigl(\rho_1 \, e^{i \, \chi_1} - \rho_2 \, e^{i \, (\chi_2 + \pi/4)}\bigr) \bigl(\rho_1 \, e^{i \, \chi_1} + \rho_2 \, e^{i \, (\chi_2 + \pi/4)}\bigr)^* \threej{\ell,2}{l_1,-3}{l_2,1} \threej{\ell,-2}{l_1,1}{l_2,1} (C_{l_2}^{EE} + C_{l_2}^{BB})
    \\
    &+\frac{1}{4}\tilde{C}_{l_1}^{{}_3h{}_{-5}h} (l_2-2) \, (l_2+2+1) \bigl|\rho_1 \, e^{i \, \chi_1} - \rho_2 \, e^{i \, (\chi_2 + \pi/4)}\bigr|^2 \threej{\ell,2}{l_1,-3}{l_2,1}\threej{\ell,-2}{l_1,5}{l_2,-3} (C_{l_2}^{EE} - C_{l_2}^{BB} -2iC_{l_2}^{EB})
    \\
    &-\frac{1}{4}\tilde{C}_{l_1}^{{}_3h{}_{-3}h} \sqrt{(l_2-2) \, (l_2+2+1)} \sqrt{(l_2+2) \, (l_2-2+1)} \bigl(\rho_1 \, e^{i \, \chi_1} - \rho_2 \, e^{i \, (\chi_2 + \pi/4)}\bigr)^2 \threej{\ell,2}{l_1,-3}{l_2,1}\threej{\ell,-2}{l_1,3}{l_2,-1} (C_{l_2}^{EE} - C_{l_2}^{BB} -2iC_{l_2}^{EB})
    \\
    &-\frac{1}{4}\tilde{C}_{l_1}^{{}_3h{}_{1}h} \sqrt{(l_2-2) \, (l_2+2+1)} \sqrt{(l_2+2) \, (l_2-2+1)} \bigl(\rho_1 \, e^{i \, \chi_1} - \rho_2 \, e^{i \, (\chi_2 + \pi/4)}\bigr) \bigl(\rho_1 \, e^{-i \, \chi_1} + \rho_2 \, e^{-i \, (\chi_2 + \pi/4)}\bigr)^*  \threej{\ell,2}{l_1,-3}{l_2,1} \threej{\ell,-2}{l_1,-1}{l_2,3} (C_{l_2}^{EE} + C_{l_2}^{BB})
    \\
    &+\frac{1}{2}\tilde{C}_{l_1}^{{}_{-1}h{}_{-1}h} \sqrt{(l_2-2) \, (l_2+2+1)}\sqrt{l_2(l_2+1)} \bigl(\rho_1 \, e^{i \, \chi_1} + \rho_2 \, e^{i \, (\chi_2 + \pi/4)}\bigr) \bigl(\rho_1 \, e^{-i \, \chi_1} + \rho_2 \, e^{-i \, (\chi_2 - \pi/4)}\bigr)^* \threej{\ell,2}{l_1,1}{l_2,-3} \threej{\ell,-2}{l_1,1}{l_2,1} (C_{l_2}^{TE} + iC_{l_2}^{TB})
    \\
    &-\frac{1}{2}\tilde{C}_{l_1}^{{}_{-1}h{}_{-3}h} \sqrt{(l_2-2) \, (l_2+2+1)}\sqrt{l_2(l_2+1)} \bigl(\rho_1 \, e^{i \, \chi_1} + \rho_2 \, e^{i \, (\chi_2 + \pi/4)}\bigr) \bigl(\rho_1 \, e^{i \, \chi_1} - \rho_2 \, e^{i \, (\chi_2 - \pi/4)}\bigr)^* \threej{\ell,2}{l_1,1}{l_2,-3}\threej{\ell,-2}{l_1,3}{l_2,-1} (C_{l_2}^{TE} + iC_{l_2}^{TB})
    \\
    &+\frac{1}{4}\tilde{C}_{l_1}^{{}_{-1}h{}_{-1}h} (l_2-2) \, (l_2+2+1) \bigl|\rho_1 \, e^{i \, \chi_1} + \rho_2 \, e^{i \, (\chi_2 + \pi/4)}\bigr|^2 \threej{\ell,2}{l_1,1}{l_2,-3} \threej{\ell,-2}{l_1,1}{l_2,1} (C_{l_2}^{EE} - C_{l_2}^{BB} +2iC_{l_2}^{EB})
    \\
    &+\frac{1}{4}\tilde{C}_{l_1}^{{}_{-1}h{}_{-5}h} (l_2-2) \, (l_2+2+1) \bigl(\rho_1 \, e^{i \, \chi_1} + \rho_2 \, e^{i \, (\chi_2 + \pi/4)}\bigr) \bigl(\rho_1 \, e^{i \, \chi_1} - \rho_2 \, e^{i \, (\chi_2 + \pi/4)}\bigr)^* \threej{\ell,2}{l_1,1}{l_2,-3}\threej{\ell,-2}{l_1,5}{l_2,-3} (C_{l_2}^{EE} + C_{l_2}^{BB})
    \\
    &-\frac{1}{4}\tilde{C}_{l_1}^{{}_{-1}h{}_{-3}h} \sqrt{(l_2-2) \, (l_2+2+1)} \sqrt{(l_2+2) \, (l_2-2+1)} \bigl(\rho_1 \, e^{i \, \chi_1} + \rho_2 \, e^{i \, (\chi_2 + \pi/4)}\bigr) \bigl(\rho_1 \, e^{-i \, \chi_1} - \rho_2 \, e^{-i \, (\chi_2 + \pi/4)}\bigr)^* \threej{\ell,2}{l_1,1}{l_2,-3}\threej{\ell,-2}{l_1,3}{l_2,-1} (C_{l_2}^{EE} + C_{l_2}^{BB})
    \\
    &-\frac{1}{4}\tilde{C}_{l_1}^{{}_{-1}h{}_{1}h} \sqrt{(l_2-2) \, (l_2+2+1)} \sqrt{(l_2+2) \, (l_2-2+1)} \bigl(\rho_1 \, e^{i \, \chi_1} + \rho_2 \, e^{i \, (\chi_2 + \pi/4)}\bigr)^2 \threej{\ell,2}{l_1,1}{l_2,-3}\threej{\ell,-2}{l_1,-1}{l_2,3} (C_{l_2}^{EE} - C_{l_2}^{BB} +2iC_{l_2}^{EB})
    \Bigg\}
    \\\\\\\\\\\\\\
\end{split}
\label{eq:Term 2 Pointing}
\end{equation}

\begin{equation}
\begin{split}
    \frac{1}{2 \, (2\ell + 1)} \sum_{m} {\Re{}} &\bigl\langle \tilde{E}_{\ell m} \, {}^{}_2\tilde{\Delta}_{\ell m}^* \bigr\rangle
    = \sum_{l_1l_2}{\Re{}} (-1)^{\ell+l_1+l_2}
    \bigg\{\tilde{C}^{{}_0h{}_1h}_{l_1} \, \sqrt{l_2 \, (l_2+1)} \, \bigl(\rho_1 \, e^{i \, \chi_1} + \rho_2 \, e^{i \, (\chi_2 - \pi/4)}\bigr)^* \, \threej{\ell,2}{l_1,-1}{l_2,-1}
    \bigl\{C^{TE}_{l_2} \, \bigg(\threej{l_{2},2}{l_1,0}{\ell,-2}
    + \threej{l_{2},-2}{l_1,0}{\ell,2}\bigg) + iC^{TB}_{l_2} \, \bigg(\threej{l_{2},2}{l_1,0}{\ell,-2}
    - \threej{l_{2},-2}{l_1,0}{\ell,2}\bigg) \bigr\}
    \\
    &- \tilde{C}^{{}_0h{}_3h}_{l_1} \, \sqrt{l_2 \, (l_2+1)} \, \bigl(\rho_1 \, e^{-i \, \chi_1} - \rho_2 \, e^{-i \, (\chi_2 - \pi/4)}\bigr)^* \, \threej{\ell,2}{l_1,-3}{l_2,1} \bigl\{C^{TE}_{l_2} \, \bigg(\threej{l_{2},2}{l_1,0}{\ell,-2}
    + \threej{l_{2},-2}{l_1,0}{\ell,2}\bigg) + iC^{TB}_{l_2} \, \bigg(\threej{l_{2},2}{l_1,0}{\ell,-2}
    - \threej{l_{2},-2}{l_1,0}{\ell,2}\bigg) \bigr\}
    \\
    &+\frac{1}{2} \tilde{C}^{{}_0h{}_1h}_{l_1} \, \sqrt{(l_2+2) \, (l_2-2+1)}
    \, \bigl(\rho_1 \, e^{-i \, \chi_1} + \rho_2 \, e^{-i
    \, (\chi_2 + \pi/4)}\bigr)^* \, \threej{\ell,2}{l_1,-1}{l_2,-1}
    \bigl\{(C^{EE}_{l_2}-iC^{EB}_{l_2}) \, \bigg(\threej{l_{2},2}{l_1,0}{\ell,-2}
    + \threej{l_{2},-2}{l_1,0}{\ell,2}\bigg)
    + (C^{BB}_{l_2}+iC^{BE}_{l_2}) \, \bigg(\threej{l_{2},2}{l_1,0}{\ell,-2}
    - \threej{l_{2},-2}{l_1,0}{\ell,2}\bigg) \bigr\}
    \\
    &+\frac{1}{2} \tilde{C}^{{}_0h{}_5h}_{l_1} \, \sqrt{(l_2+2) \, (l_2-2+1)}
    \, \bigl(\rho_1 \, e^{-i \, \chi_1} - \rho_2 \, e^{-i
    \, (\chi_2 + \pi/4)}\bigr)^* \, \threej{\ell,2}{l_1,-5}{l_2,3}
    \bigl\{(C^{EE}_{l_2}+iC^{EB}_{l_2}) \, \bigg(\threej{l_{2},2}{l_1,0}{\ell,-2}
    + \threej{l_{2},-2}{l_1,0}{\ell,2}\bigg) 
    + (iC^{BE}_{l_2}-C^{BB}_{l_2}) \, \bigg(\threej{l_{2},2}{l_1,0}{\ell,-2}
    - \threej{l_{2},-2}{l_1,0}{\ell,2}\bigg) \bigr\}
    \\
    &-\frac{1}{2} \tilde{C}^{{}_0h{}_3h}_{l_1} \, \sqrt{(l_2-2) \, (l_2+2+1)}
    \bigl(\rho_1 \, e^{i \, \chi_1} - \rho_2 \, e^{i \, (\chi_2 + \pi/4)}\bigr)^* \,
    \threej{\ell,2}{l_1,-3}{l_2,1}
    \bigl\{(C^{EE}_{l_2}+iC^{EB}_{l_2}) \, \bigg(\threej{l_{2},2}{l_1,0}{\ell,-2}
    + \threej{l_{2},-2}{l_1,0}{\ell,2}\bigg) + (iC^{BE}_{l_2}-C^{BB}_{l_2}) \, \bigg(\threej{l_{2},2}{l_1,0}{\ell,-2}
    - \threej{l_{2},-2}{l_1,0}{\ell,2}\bigg) \bigr\}
    \\
    &-\frac{1}{2} \tilde{C}^{{}_0h{}_{-1}h}_{l_1} \, \sqrt{(l_2-2) \, (l_2+2+1)} \, \bigl(\rho_1 \, e^{i \, \chi_1} + \rho_2 \, e^{i \, (\chi_2 + \pi/4)}\bigr)^* \,
    \threej{\ell,2}{l_1,1}{l_2,-3}
    \bigl\{(C^{EE}_{l_2}-iC^{EB}_{l_2}) \, \bigg(\threej{l_{2},2}{l_1,0}{\ell,-2}
    + \threej{l_{2},-2}{l_1,0}{\ell,2}\bigg) + (C^{BB}_{l_2}+iC^{BE}_{l_2}) \, 
    \bigg(\threej{l_{2},2}{l_1,0}{\ell,-2}
    - \threej{l_{2},-2}{l_1,0}{\ell,2}\bigg) \bigr\}\bigg\}
    \tfrac{(2l_1 + 1)(2l_2 + 1)}{64\pi}
\end{split}
\label{eq:Term 3 Pointing}
\end{equation}

\begin{equation}
\begin{split}
    \frac{1}{2 \, (2\ell + 1)} \sum_{m} {\Im{}} &\bigl\langle  \tilde{B}_{\ell m} \, {}^{}_2\tilde{\Delta}_{\ell m}^* \bigr\rangle
    = \sum_{l_1l_2} {\Im{}} (-1)^{\ell+l_1+l_2}
    \bigg[\tilde{C}^{{}_0h{}_1h}_{l_1} \, \sqrt{l_2 \, (l_2+1)} \, \bigl(\rho_1 \, e^{i \, \chi_1} + \rho_2 \, e^{i \, (\chi_2 - \pi/4)}\bigr)^* \, \threej{\ell,2}{l_1,-1}{l_2,-1}
    \bigl\{C^{TB}_{l_2} \, \bigg(\threej{l_{2},2}{l_1,0}{\ell,-2}
    + \threej{l_{2},-2}{l_1,0}{\ell,2}\bigg) - iC^{TE}_{l_2} \, \bigg(\threej{l_{2},2}{l_1,0}{\ell,-2}
    - \threej{l_{2},-2}{l_1,0}{\ell,2}\bigg) \bigr\}
    \\
    &- \tilde{C}^{{}_0h{}_3h}_{l_1} \, \sqrt{l_2 \, (l_2+1)} \, \bigl(\rho_1 \, e^{-i \, \chi_1} - \rho_2 \, e^{-i \, (\chi_2 - \pi/4)}\bigr)^* \, \threej{\ell,2}{l_1,-3}{l_2,1} \bigl\{C^{TB}_{l_2} \, \bigg(\threej{l_{2},2}{l_1,0}{\ell,-2}
    + \threej{l_{2},-2}{l_1,0}{\ell,2}\bigg) - iC^{TE}_{l_2} \, \bigg(\threej{l_{2},2}{l_1,0}{\ell,-2}
    - \threej{l_{2},-2}{l_1,0}{\ell,2}\bigg) \bigr\}
    \\
    &+\frac{1}{2} \tilde{C}^{{}_0h{}_1h}_{l_1} \, \sqrt{(l_2+2) \, (l_2-2+1)}
    \, \bigl(\rho_1 \, e^{-i \, \chi_1} + \rho_2 \, e^{-i
    \, (\chi_2 + \pi/4)}\bigr)^* \, \threej{\ell,2}{l_1,-1}{l_2,-1}
    \bigl\{(C^{BE}_{l_2}-iC^{BB}_{l_2}) \, \bigg(\threej{l_{2},2}{l_1,0}{\ell,-2}
    + \threej{l_{2},-2}{l_1,0}{\ell,2}\bigg)
    - (C^{EB}_{l_2}+iC^{EE}_{l_2}) \, \bigg(\threej{l_{2},2}{l_1,0}{\ell,-2}
    - \threej{l_{2},-2}{l_1,0}{\ell,2}\bigg) \bigr\}
    \\
    &+\frac{1}{2} \tilde{C}^{{}_0h{}_5h}_{l_1} \, \sqrt{(l_2+2) \, (l_2-2+1)}
    \, \bigl(\rho_1 \, e^{-i \, \chi_1} - \rho_2 \, e^{-i
    \, (\chi_2 + \pi/4)}\bigr)^* \, \threej{\ell,2}{l_1,-5}{l_2,3}
    \bigl\{(C^{BE}_{l_2}+iC^{BB}_{l_2}) \, \bigg(\threej{l_{2},2}{l_1,0}{\ell,-2}
    + \threej{l_{2},-2}{l_1,0}{\ell,2}\bigg) 
    - (iC^{EE}_{l_2}-C^{EB}_{l_2}) \, \bigg(\threej{l_{2},2}{l_1,0}{\ell,-2}
    - \threej{l_{2},-2}{l_1,0}{\ell,2}\bigg) \bigr\}
    \\
    &-\frac{1}{2} \tilde{C}^{{}_0h{}_3h}_{l_1} \, \sqrt{(l_2-2) \, (l_2+2+1)}
    \bigl(\rho_1 \, e^{i \, \chi_1} - \rho_2 \, e^{i \, (\chi_2 + \pi/4)}\bigr)^* \,
    \threej{\ell,2}{l_1,-3}{l_2,1}
    \bigl\{(C^{BE}_{l_2}+iC^{BB}_{l_2}) \, \bigg(\threej{l_{2},2}{l_1,0}{\ell,-2}
    + \threej{l_{2},-2}{l_1,0}{\ell,2}\bigg) 
    - (iC^{EE}_{l_2}-C^{EB}_{l_2}) \, \bigg(\threej{l_{2},2}{l_1,0}{\ell,-2}
    - \threej{l_{2},-2}{l_1,0}{\ell,2}\bigg) \bigr\}
    \\
    &-\frac{1}{2} \tilde{C}^{{}_0h{}_{-1}h}_{l_1} \, \sqrt{(l_2-2) \, (l_2+2+1)} \, \bigl(\rho_1 \, e^{i \, \chi_1} + \rho_2 \, e^{i \, (\chi_2 + \pi/4)}\bigr)^* \,
    \threej{\ell,2}{l_1,1}{l_2,-3}
    \bigl\{(C^{BE}_{l_2}-iC^{BB}_{l_2}) \, \bigg(\threej{l_{2},2}{l_1,0}{\ell,-2}
    + \threej{l_{2},-2}{l_1,0}{\ell,2}\bigg) 
    - (C^{EB}_{l_2}+iC^{EE}_{l_2}) \, 
    \bigg(\threej{l_{2},2}{l_1,0}{\ell,-2}
    - \threej{l_{2},-2}{l_1,0}{\ell,2}\bigg) \bigr\}\bigg]
    \tfrac{(2l_1 + 1)(2l_2 + 1)}{64\pi} \text{.}
\end{split}
\label{eq:Term 3 Pointing BB}
\end{equation}
\normalsize
These equations provide a complete characterisation of the effects of differential pointing for the experimental setup of two sets of detector pairs.


\bsp	
\label{lastpage}
\end{document}